\documentclass[11pt,a4paper]{article}

\usepackage[english]{babel}
\usepackage{authblk}
\usepackage{graphicx}
\usepackage{hyperref} 
\usepackage{geometry}
\usepackage{amsmath}
\usepackage{amssymb}
\usepackage{amsfonts}
\usepackage{multicol}
\usepackage{slashed}
\usepackage{float}
\usepackage{caption}
\usepackage{subcaption}
\usepackage{cite}
\usepackage{amsthm}
\usepackage{xcolor}
\usepackage{musicography}
\usepackage{bbm}
\usepackage{enumitem} 
\usepackage{csquotes} 
\numberwithin{equation}{section}

\newtheorem*{mydef}{Definition}

\newtheorem*{theorem}{Theorem}
\newtheorem*{corollary}{Corollary}
\newtheorem*{lemma}{Lemma}

\newcommand{\ri}{{\rm i}}
\newcommand{\M}{\mathcal{M}}
\newcommand{\R}{\mathbb{R}}
\newcommand{\E}{\mathbb{E}}
\newcommand{\p}{\partial}
\newcommand{\indep}{\perp \!\!\! \perp}
\newcommand{\cev}[1]{\reflectbox{\ensuremath{\vec{\reflectbox{\ensuremath{#1}}}}}}
\def\Xint#1{\mathchoice
	{\XXint\displaystyle\textstyle{#1}}%
	{\XXint\textstyle\scriptstyle{#1}}%
	{\XXint\scriptstyle\scriptscriptstyle{#1}}%
	{\XXint\scriptscriptstyle\scriptscriptstyle{#1}}%
	\!\int}
\def\XXint#1#2#3{{\setbox0=\hbox{$#1{#2#3}{\int}$ }
		\vcenter{\hbox{$#2#3$ }}\kern-.6\wd0}}

\def\dashint{\Xint-}

\def\upint{\mkern13mu\overline{\vphantom{\intop}\mkern10mu}\mkern-20mu\int}
\def\lowint{\mkern3mu\underline{\vphantom{\intop}\mkern10mu}\mkern-10mu\int}

\title{\bf Stochastic Mechanics and the Unification of Quantum Mechanics with Brownian Motion}
\author[1]{Folkert~Kuipers\thanks{E-mail: Kuipers@na.infn.it}}
\affil[1]{\em INFN, Sezione di Napoli\authorcr \em Complesso Universitario di Monte S. Angelo\authorcr \em Via Cintia Edificio 6, 80126 Napoli, Italy}

\begin{document}

\maketitle

\begin{abstract}
We unify Brownian motion and quantum mechanics in a single mathematical framework. In particular, we show that non-relativistic quantum mechanics of a single spinless particle on a flat space can be described by a Wiener process that is rotated in the complex plane.
We then extend this theory to relativistic stochastic theories on manifolds using the framework of second order geometry. 
As a byproduct, our results show that a consistent path integral based formulation of a quantum theory on a Lorentzian (Riemannian) manifold requires an It\^o deformation of the Poincar\'e (Galilean) symmetry, arising due to the coupling of the quadratic variation to the affine connection.
\end{abstract}
\vspace{1cm}

\fbox{%
	\parbox{\textwidth}{ 
		This is a preprint of the following work:\\
		Folkert Kuipers, ``Stochastic Mechanics: the Unification of Quantum Mechanics with Brownian Motion'', SpringerBriefs in Physics, Springer (2023),\\[0.1cm]
		reproduced with permission of Springer Nature Switzerland AG. \\[0.2cm]
		The final authenticated version is available online at:\\
		\href{https://link.springer.com/book/10.1007/978-3-031-31448-3}{http://dx.doi.org/10.1007/978-3-031-31448-3}
	}
}

\pagenumbering{roman}
\thispagestyle{empty}
\clearpage

\tableofcontents

\clearpage

\setcounter{page}{1}
\pagenumbering{arabic}

\section{Introduction}
The quantum theory that has been developed since the early 20th century has greatly improved our understanding of fundamental physics. It lies at the heart of all our understanding of both particle physics and condensed matter theory. Moreover, it forms the basis of a large fraction of the technological advancement of the last century.
However, despite all these advances, several fundamental aspects of quantum theory are not yet understood. In this regard, the most pressing question is perhaps how to reconciliate quantum theory with the theory of gravity that, in its current form, also stems from the early 20th century.
\par

It is widely known that there exist various equivalent formulations of quantum mechanics. Perhaps the best known is Dirac's canonical formalism, which incorporates both Heisenberg's matrix mechanics and Schr\"odinger's complex diffusion theory. A second widely used approach is the path integral formulation that was later developed by Feynman.
It is less known, however, that there exists another formulation of quantum mechanics, which was pioneered by Edward Nelson and is better known as stochastic mechanics. 
This research program aims to resolve the two major issues that are left by the other two formulations of quantum mechanics.
\par 

The first issue is that the canonical formulation of quantum mechanics leaves a rather unsatisfactory picture of physical reality. An orthodox interpretation of the canonical formulation forces one to give up on the notion of a classical configuration space. This classical configuration space only appears, if the probability interpretation is imposed. However, this probability interpretation is ad hoc, as it is not properly defined within modern probability theory. For this reason the foundations of quantum mechanics have been the subject of many debates. 
\par 

Feynman's path integral formulation, on the other hand, provides a rather intuitive picture of quantum mechanics in which the classical configuration space is retained. However, the path integral formulation has the major drawback that the path integral is a heuristic object and that a precise mathematical definition that applies to all physical theories is still absent.
\par 

Stochastic mechanics tries to resolve these two issues by interpreting the paths in Feynman's path integral as the sample paths of a stochastic process. If this process exists, Feynman's path integral is well defined as an It\^o integral and can be studied using the tools from stochastic analysis. Moreover, it provides a physical picture in which the quantum fluctuations are similar to the statistical fluctuations encountered in the theory of Brownian motion. Stochastic mechanics thus provides a natural interpretation of quantum mechanics in which the classical configuration space can be retained and the probability interpretation is well defined.
\par 

The main challenge faced by stochastic mechanics is to prove the equivalence between stochastic mechanics and quantum theories. Although this has not yet been achieved for all quantum theories, this equivalence has been established for a single spinless particle on a smooth manifold subjected to any potential, and there are strong indications that this equivalence can be generalized to any other quantum theory.
\par 

In this book, we review the stochastic theory of a single spinless particle in detail. More precisely, we generalize the stochastic quantization procedure, that was originally formulated by Nelson, and apply this procedure to a single (non-)relativistic spinless particle moving on a pseudo-Riemannian manifold subjected to a vector and scalar potential. As we use a generalized stochastic quantization procedure, we obtain a class of stochastic processes. In a certain limit, this class reduces to the Wiener process that describes Brownian motion, while in another limit it reduces to processes that are similar to the ones studied in stochastic mechanics and describe quantum mechanics.

\subsection{A Brief History}
The history of Brownian motion can be traced back until at least ca. 60 B.C., when Lucretius hypothesized in his book `De Rerum Natura' that the existence of atoms should, due to their collisions with larger particles, induce a jiggly motion for the larger particles. It would take until the 18th century, however, that, thanks to developments in microscopy, such a phenomenon was observed in nature by many scientists who studied particles suspended in a liquid \cite{Nelson:1967}.
\par 

Knowledge about the existence of this phenomenon would become more widespread after Brown published his observation of jiggly motion of pollen suspended in water through a microscope \cite{Brown}. Due to the increased interest in this phenomenon at the time, this jiggly motion became known as Brownian motion.
Since then Brownian motion has become a topic of considerable interest in many branches of science.
\par

In the early 20th century, interest for the phenomenon arose in the physics community, due to the mathematical models developed by Einstein and Smoluchowski \cite{EinsteinBM,Smoluchowski}. These models were put forward as a way to confirm the existence of atoms and molecules introduced in statistical physics. Soon after, their models were confirmed experimentally by Perrin \cite{Perrin}, who thus proved the atomic nature of matter. Later, the models developed by Einstein and Smoluchowski were improved by various other physicists, most notably by Ornstein and Uhlenbeck \cite{Ornstein}, who built on the theory of stochastic differential equations developed by Langevin \cite{Langevin}.
\par 

The first mathematically rigorous description of Brownian motion was developed in the framework of functional analysis by Wiener \cite{Wiener}. Due to this work, the stochastic process describing Brownian motion is also known as the Wiener process. Further progress in the mathematical study of Brownian motion was made after the formulation of the axioms of probability theory by Kolmogorov \cite{Kolmogorov}. This allowed to rigorously describe Brownian motion in the framework of stochastic analysis \cite{Doob}, which, in turn, allowed for generalizations to much larger classes of stochastic processes \cite{Levy}, and for the development of stochastic calculus by It\^o \cite{Ito}, Fisk \cite{Fisk} and Stratonovich \cite{Stratonovich}. This calculus provides a rigorous foundation and generalization of the stochastic calculus that was developed earlier by Langevin.
\par 

Quantum mechanics arose in the early 20th century, starting from Planck's explanation of the black body spectrum \cite{Planck:1901tja} and Einstein's explanation of the photo-electric effect \cite{Einstein:1905cc}. Both these works served to explain observed phenomena that could not be reconciled with classical physics and quickly gave rise to a radically new theory known as quantum mechanics. Many of the elementary building blocks of quantum mechanics were developed in the early 20th century by Born, Dirac, Heisenberg, Jordan, Von Neumann, Pauli, Schr\"odinger, Weyl, Wigner and others. This led to two formulations of quantum theory: one based on Schr\"odinger's complex diffusion equation and one based on Heisenberg's matrix mechanics. These two approaches were later synthesized by Dirac \cite{Dirac}, which led to the Dirac-Von Neumann axioms \cite{VonNeumann}. These axioms form, together with various other postulates, the mathematical basis of quantum mechanics of non-relativistic particles.
\par 

After these initial developments, quantum mechanics was extended to quantum field theory, which provided a quantum description of both relativistic and non-relativistic classical field theories. This led to the formulation of the Wightman axioms of quantum field theory \cite{Wightman}. Around the same time, Feynman developed a new formalism of quantum mechanics based on path integrals \cite{Feynman:1948ur}, which was shown to be mathematically equivalent to the canonical formulation of quantum mechanics that was developed earlier by Dirac. This reformulation by Feynman provided a first rigorous connection between the theories of Brownian motion and quantum mechanics, due to its similarities with the Wiener integral that governs the dynamics of a Brownian motion. 
A more direct connection between quantum mechanics and stochastic theories, such as Brownian motion, was later established by Nelson~\cite{Nelson:1966sp,Nelson:1967,Nelson}, who introduced another equivalent reformulation of quantum mechanics, better known as stochastic mechanics.

\subsection{Diffusion Theories}
Given a configuration space $\M$ and an interval $\mathcal{T}\subseteq\R$, a complex diffusion equation is an equation of the form 
\begin{equation}\label{eq:DiffEqDef}
	\alpha \, \frac{\p}{\p t} \Psi(x,t) = H(x,\hat{p},t) \, \Psi(x,t) \, ,
\end{equation}
where $\alpha\in \mathbb{C}$, $t\in\mathcal{T}$, $x\in\M$. Moreover, $\Psi$ is a bounded complex valued function and $H$ is a Hamiltonian or diffusion operator. $H$ is second order in $\hat{p}$ and $\hat{p}$ is an operator acting on the space of complex functions on $\M$.
\par

The typical example is the setting where $\M=\R^d$, $\Psi$ is (square-)integrable and ${\hat{p}=\pm \alpha \, \p_x}$ is a differential operator, such that $H$ is of the form
\begin{equation}\label{eq:HamExample}
	H = \frac{\alpha^2}{2} \, \delta^{ij} \left( \frac{\p}{\p x^i} + A_i(x,t) \right)  \left( \frac{\p}{\p x^j} + A_j(x,t) \right) + \mathfrak{U}(x,t) \, ,
\end{equation}
where $\mathfrak{U}$ is a scalar field and $A$ is a covector field.
\par 

If $\alpha\in(0,\infty)$, the diffusion equation is called the heat equation. The heat equation presents a special limit within all diffusion equations, since it reduces to a real diffusion equation for real valued functions $\Psi$. A characteristic feature of real diffusion equations is that their solutions are dissipative, meaning that solutions $\Psi$ spread out in the space $\M$, when the time is evolved. This dissipation introduces an asymmetry under time reversal in the theory.
\par 

If $\alpha\in\ri \times (0,\infty)$, the diffusion equation is called the Schr\"odinger equation\footnote{The Schr\"odinger equation is sometimes erroneously referred to as a wave equation, due to its wave-like solutions. We emphasize, therefore, that, by definition, a diffusion equation contains a first order time derivative, while a wave equation contains a second order time derivative, i.e., a wave equation is of the form $\alpha^2 \p_t^2 \Psi = H \, \Psi$.}, which is encountered in quantum mechanics. The Schr\"odinger equation also presents a special limit within all diffusion equations, as it is a conservative diffusion equation \cite{Nelson}, which means that the time evolution operator is unitary. As a consequence, the dissipative behavior, that is characteristic to real diffusions, is completely absent in the solutions of the Schr\"odinger equation. Therefore, the solutions of the Schr\"odinger equation are symmetric under time reversal.
\par

The heat equation is intimately related to the theory of Brownian motion, because of the Kolmogorov equations \cite{KolmogorovEq} and the Feynman-Kac theorem \cite{FKac}. The Kolmogorov equations describe how the probability density of a continuous-time Markov process evolves over time. There are two such equations. The Kolmogorov forward equations describe the evolution of the probability that a process is in a certain state, given the probability density at an earlier time. The Kolmogorov backward equations describe the evolution of the probability that a process is in a certain state, given the probability density at a later time.
\par 

The Feynman-Kac theorem states that, given the diffusion equation \eqref{eq:DiffEqDef} with ${\alpha\in(-\infty,0)}$, $\mathcal{T}=[0,T]$, $\Psi:\R\times\R^d\rightarrow\R$ and $H$ as given in eq.~\eqref{eq:HamExample}, there exists an $\R^d$-valued stochastic process $X=C+M$, where $M$ is a Wiener process and $C$ a finite variation process, such that 
\begin{equation}\label{eq:FKacThm}
	\Psi(x,t) = \E\left[ \Psi(X_T,T) \,\exp\left(-\int_t^T \mathfrak{U}(X_s,s) \, ds \right) \, \Big| \, X_t = x\right]
\end{equation}
for all $t\in\mathcal{T}$. Conversely, given an $\R^d$-valued stochastic process $X=C+M$, where $M$ is a Wiener process and $C$ a finite variation process, there exists a probability measure, such that \eqref{eq:FKacThm} satisfies the diffusion equation \eqref{eq:DiffEqDef} with $\alpha\in(-\infty,0)$, $\mathcal{T}=[0,T]$, $\Psi:\R\times\R^d\rightarrow\R$ and $H$ as given in eq.~\eqref{eq:HamExample}.
We note that a similar result follows for $\alpha\in(0,\infty)$ by considering a stochastic process that evolves backward in time.
\par 

The Kolomogorov backward equations can be derived from the Feynman-Kac theorem by noting that the probability that the process $X$ will evolve to the state $X_T$, given that at time $t$ it is in state $X_t$, is given by
\begin{equation}
	\rho(X_T,T;X_t,t) = \E\left[ \Psi(X_T,T) \,\exp\left(-\int_t^T \mathfrak{U}(X_s,s) \, ds \right) \, \Big| \, X_t \right] .
\end{equation}
\par 

Therefore, the time evolution of the probability density for a Brownian motion is governed by the heat equation. Conversely, for any normalized solution $\Psi$ of the heat equation, one can construct an $\R^d$-valued stochastic process $X=C+M$, where $C$ is a finite variation process and $M$ is a Wiener process, such that $\Psi$ describes the probability density of $X$.

\subsection{The Wick Rotation}
The similarities between the Schr\"odinger equation and the heat equation have been exploited in many ways in quantum physics. At the heart of any such exploitation lies the Wick rotation \cite{Wick:1954eu}, which can be regarded as a combination of two different transformations:
\begin{itemize}
	\item the Wick rotation maps a theory on a pseudo-Riemannian manifold onto a theory on a Riemannian manifold;
	\item the Wick rotation maps a quantum theory onto a statistical theory by mapping the imaginary diffusion constant onto a real diffusion constant.
\end{itemize}
The Wick rotation thus allows to study quantum theories on pseudo-Riemannian manifolds by mapping them onto statistical theories on Riemannian manifolds. These statistical theories can be studied using the tools from stochastic analysis and the results can be translated back to the original theory by inverting the Wick rotation.
\par 

The Wick rotation plays a crucial role in Feynman's reformulation of quantum theories using path integrals. The path integral is an object of the form
\begin{equation}\label{eq:pathintegral}
	W(X_f,t_f;X_0,t_0) = \int e^{\frac{\ri}{\hbar} \int_{t_0}^{t_f} L(X,\dot{X},t) \, dt} \, DX
\end{equation}
that integrates over all possible paths that the system $X$ can take in evolving from state $X(t_0)=X_0$ to the state $X(t_f)=X_f$. As these paths are weighted with their probability $e^{\frac{\ri}{\hbar}S(X)}$, this integral represents the probability amplitude for the system $X$ of evolving from state $X(t_0)=X_0$ to a state $X(t_f)=X_f$.
\par 
 
The path integral \eqref{eq:pathintegral} is a heuristic object, which can be given a precise meaning by mapping it to other functional integrals or to stochastic integrals that have been given a rigorous definition. For the general case this has not yet been established cf. e.g. Ref.~\cite{Albeverio} for a review. However, if one applies a Wick rotation to \eqref{eq:pathintegral}, such that the probability of the sample paths is given by $e^{-\frac{1}{\hbar}S_{\rm Eucl.}}$, the path integral becomes a Wiener integral, which has a well defined meaning as a functional integral and as an It\^o integral. This observation inspired Kac in formulating and proving the Feynman-Kac theorem \cite{FKac}.
\par 

The path integral method and the Wick rotation have led to major progress in quantum field theory. Moreover, the Feynman-Kac theorem allows to define Euclidean quantum field theories in mathematically rigorous way \cite{NelsonPath,Osterwalder:1973dx,Osterwalder:1974tc,Glimm:1987ng,Schwinger:1958mma}. Therefore, the path integral approach and the Wick rotation also serve as important tools in constructive quantum field theory, which aims to axiomatize quantum field theory in such a way that all quantum field theories are mathematically well defined.
\par 

The similarities between statistical theories and quantum theories have also been exploited in the framework of stochastic quantization developed by Parisi and Wu~\cite{Parisi:1980ys,Damgaard:1987rr}. In this framework, Euclidean quantum theories are studied by introducing an additional fictious time parameter $\tau$. One then imposes that the theory evolves as a standard Brownian motion with respect to this fictious parameter and studies the equilibrium limit $\tau\rightarrow\infty$. In this equilibrium limit, the statistical averages of the extended field theory become identical to the vacuum expectation values of the original Euclidean quantum theory. This allows to employ the methods from statistical physics to study Euclidean quantum field theories. 
\par 

We emphasize that, although there are various similarities, the stochastic quantization method as proposed by Parisi and Wu is different from the stochastic quantization procedure that is employed in stochastic mechanics and was developed by Nelson and Yasue. Stochastic quantization in the sense of Parisi and Wu will not be studied in this work, whereas the stochastic quantization procedure, as developed by Nelson and Yasue, is the central object of study in this book. Therefore, in the remainder of this book, stochastic quantization will solely refer to the quantization procedure employed in stochastic mechanics, which will be reviewed in the next section.

\subsection{Stochastic Mechanics}\label{sec:introStochMech}
Stochastic analysis plays a central role in the study of Euclidean quantum theories, which has led to major progress in both rigorous mathematical studies of quantum theories and in numerical simulation of quantum theories. The obvious downside of this use of stochastic tools in the study of quantum theories is that they are applied directly to Euclidean quantum theories, while their application to Lorentzian quantum theories remains indirect, as it relies on the Wick rotation.
\par 

Stochastic mechanics is a theoretical framework that aims to apply stochastic analysis to all quantum theories in a direct manner by explicitly constructing stochastic processes that describe the quantum systems. If such a construction can be performed, this has two major advantages:
\begin{itemize}
	\item it provides a natural interpretation of quantum mechanics as a stochastic theory;
	\item it allows to apply the stochastic tools that have been very successful in the study of Euclidean quantum theories directly to Lorentzian quantum theories.
\end{itemize}
\par 

Since Euclidean quantum theories of spinless particles are in essence stochastic theories, the major challenge of stochastic mechanics is to construct processes for Lorentzian quantum theories and for theories with spin. In this book, we will focus on the extension to Lorentzian theories of spinless particles. In order to construct stochastic processes that describe such theories, stochastic mechanics must overcome the two issues that are usually resolved by the Wick rotation and were discussed in previous section.
\par

We stress that this provides a major challenge, already for non-relativistic theories, since a straightforward generalization of the Feyman-Kac theorem, involving a single real Wiener process, to complex diffusion equations such as the Schr\"odinger equation does not exist. More precisely, it has been shown that the complex measure, which is necessary to construct such an equivalence will have an infinite total variation \cite{Cameron,Daletskii,Albeverio}. Therefore, a successful theory of stochastic mechanics must involve a generalization of the standard Wiener process. 
\par 

In early formulations of stochastic mechanics, this generalization is achieved by considering linear combinations of a future directed and a past directed Wiener process \cite{Nelson:1966sp,Nelson:1967,Nelson,Guerra:1981ie}. In more recent formulations \cite{Pavon:1995April,Pavon:1995June,Pavon:1996}, these two processes have been reformulated as a single complex process. Such processes avoid the issues related to the infinite variation of the complex measure \cite{Pavon:2000}. In this book, we discuss a further generalization of this complex theory suggested in Ref.~\cite{Kuipers:2021ylr}, which allows to describe both stochastic processes associated to quantum mechanics and the ordinary Wiener process in a single framework.
\par 

The first theory of stochastic mechanics was developed by F\'enyes \cite{Fenyes}, who constructed a stochastic process for which the diffusion equation is given by the Schr\"odinger equation. A few years later, a similar derivation was developed independently by Kershaw \cite{Kershaw} and by Nelson \cite{Nelson:1966sp}. These derivations of the Schr\"odinger equation in stochastic mechanics share strong resemblance with the hydrodynamical theory of quantum mechanics that was developed earlier by Madelung \cite{MadelungI,MadelungII}. However, in stochastic mechanics the Madelung equations are derived from an underlying theory of stochastic processes.
\par 

The initial works by F\'enyes and Nelson consider a stochastic particle subjected to a stochastic version of Newton's second law, but, soon after, it was realized that this theory could be reformulated in a Lagrangian or Hamiltonian formulation using stochastic variational calculus \cite{Nelson,Yasue:1979rf,Yasue:1981,Yasue:1981wu,Zambrini,Yasue:1981hq,Zambrini:1982,Yasue:1978Ann}. This reformulation allowed to axiomatize stochastic mechanics by means of a stochastic quantization condition. This condition plays a similar role as the canonical quantization condition and can be formulated as follows:
\begin{itemize}
	\item the trajectory of a quantum particle can be described by a future directed stochastic process $X(t)=C_+(t)+M_+(t)$;
	\item the time reversed process is well defined and can be described by a past directed stochastic process $X(t)=C_-(t)+M_-(t)$;
	\item the processes $C_\pm(t)$ represent a drift that minimizes a stochastic action;
	\item the processes $M_\pm(t)$ are Wiener processes, i.e. they represent a Gaussian noise with covariance matrix ${\E\left[M_\pm^i(t) \, M_\pm^j(t) \, | \, M_\pm(t_0)\right]= \frac{\hbar}{m} \, \delta^{ij} \, |t-t_0|}$.
\end{itemize}
\par 

The theory of stochastic mechanics, as formulated by Nelson, only applies to a single non-relativistic massive spinless particle subjected to a scalar potential and charged under a vector potential in flat spaces $\R^d$ of arbitrary dimension $d\in\mathbb{N}$. Nevertheless, this theory already allows to give a mathematically consistent description of elementary quantum theories such as the free particle and the harmonic oscillator in terms of stochastic processes \cite{Nelson,Guerra:1981ie}. Moreover, this stochastic theory provides an interpretation of the single and double slit experiments \cite{Nelson}. In this stochastic interpretation of the double slit experiment, the quantum particle passes only one slit. Nonetheless, the stochastic theory correctly produces an interference pattern, as this is the unique solution of the unitary stochastic process that is studied in stochastic mechanics.
\par 

Stochastic mechanics has been extended in various ways. In particular, the extension of stochastic mechanics to a single relativistic particle is now well understood~\cite{Guerra:1973ck,GuerraRuggiero,Dohrn:1985iu,Marra:1989bi,Guerra:1981ie,Morato:1995ty,Garbaczewski:1995fr,Pavon:2001,Kuipers:2021aok}. Also, the generalization of stochastic mechanics to particles moving on Riemannian manifolds\cite{Dankel,DohrnGuerraI,DohrnGuerraII,Guerra:1982fn,Nelson,Koide:2019} is well established. Recently \cite{Kuipers:2021jlh,Kuipers:2022wpy}, it was shown that the techniques developed in the stochastic mechanics literature to achieve such extensions can be defined in a rigorous way in the framework of second order geometry \cite{Schwartz,Meyer,Emery,Huang:2022}.
\par 

Furthermore, it has been pointed out that a description of spin lies within the scope of stochastic mechanics~\cite{Nelson,Guerra:1981ie,Dankel,Faris:1982,DeAngelis:1985hp,Dohrn:1978gd}, and stochastic mechanics has been extended to describe various multi-particle systems and field theories~\cite{Nelson,Guerra:1973ck,Guerra:1980sa,Guerra:1981ie,Kodama:2014dba,Morato:1995ty,Garbaczewski:1995fr,Koide:2015,NelsonPath,Nelson:2014exa,Guerra:1973ck,Yasue:1978JMP,Davidson:1980,Koide:2014zkj,DeSiena:1983bx,DeSiena:1986nc,Davidson:1980df}. In addition, various other aspects of stochastic mechanics have been studied, cf.~e.g. Refs.~\cite{Falco:1983,Golin:1985,Pena,Olavo,Petroni_2000,Gazeau:2019amk,Koide:2012ya,Rosenbrock:1986,Rosenbrock:1999}. 
Finally, as we will explain in section \ref{sec:Qfoam}, stochastic mechanics provides a mathematically clean description of the quantum foam that is hypothesized in quantum gravity. It is, therefore, not surprising that the tools from stochastic mechanics have also found their way into models of quantum gravity \cite{Santos:1998jb,Smolin:2001hh,Markopoulou:2003ps,Erlich:2018qfc,Erlich:2022eku}.
\par 

Despite the fact that stochastic mechanics provides a fully consistent description of quantum mechanics of a single particle, and has been used to explain many of the peculiar properties of quantum mechanics, the theory of stochastic mechanics has never become widely known. This may in part be ascribed to various unjustified criticisms of the theory \cite{Guerra:1981ie}. Two of those will be discussed in more detail in sections \ref{sec:TimeRev} and \ref{sec:IntroHiddenVar}.
\par 

A third criticism was formulated by Wallstrom \cite{Wallstrom:1988zf,WallstromII}, which is discussed in more detail in section \ref{sec:diffusionreal}. Various answers to this criticism have been formulated, cf. e.g. \cite{Derakhshani:2017vls,Schmelzer}, which require a new assumption in the theory of stochastic mechanics. In section \ref{sec:diffusionreal}, we will show, however, that Wallstrom's criticism results from an incomplete analysis of stochastic mechanics. Therefore, the criticism is invalid and no new assumptions are necessary to resolve the criticism.
\par 

A fourth criticism was raised by Nelson \cite{Nelson:1986,Nelson:2011}, and states that the processes that are studied in stochastic mechanics appear to fail to recover multi-time correlations predicted by quantum theories. A possible resolution of this criticism within the traditional framework of stochastic mechanics is discussed in Refs.~\cite{Blanchard:1986zd,Derakhshani:2022esz}. In this book, we adopt a different approach to this problem, as we modify the stochastic processes that are being studied. More precisely, we rotate the Wiener process in the complex plane and study its real projection. One can  verify that these new processes correctly reproduce the multi-time correlations of quantum theories, and thus resolve Nelson's criticism. Using these new processes, the stochastic quantization condition is modified to the following statement:
\begin{itemize}
	\item the trajectory of a stochastic particle can be described by a future and a past directed stochastic process $X(t)=C_\pm(t)+{\rm Re}[M(t)]$;
	\item the processes $C_\pm(t)$ represent a drift that minimizes a stochastic action;
	\item the process $M(t)$ is a complex stochastic processes, such that ${\rm Re}[M]$ and ${\rm Im}[M]$ are correlated Wiener processes, i.e. it represents a Gaussian noise with covariance matrix ${\E\left[M^i(t) \, M^j(t) \, | \, M(t_0)\right]= \alpha \, \frac{\hbar}{m} \, \delta^{ij} \, |t-t_0|}$, where $\alpha\in\mathbb{C}$.
\end{itemize}
\par

We conclude this section by emphasizing that stochastic mechanics is not in conflict with ordinary quantum theory. Instead, stochastic mechanics aims to derive the Dirac-Von Neumann axioms and the postulates of quantum mechanics from an underlying stochastic theory. Hence, any viable formulation of stochastic mechanics is consistent with all the results from standard quantum theory. The major difference is that stochastic mechanics is by construction compatible with the Kolmogorov axioms of probability theory. Therefore, in stochastic mechanics, the probability interpretation of quantum mechanics is naturally embedded in the theory, whereas in ordinary quantum theory the probability interpretation is imposed ad hoc by means of the Born rule \cite{Born:1926yhp}.
Moreover, in the stochastic theory, the Hilbert space structure, that is central to the Dirac-Von Neumann axioms, arises in a natural way. Here, we will explain this in a qualitative fashion, making use of notions from measure theoretic probability theory that are reviewed in appendices \ref{ap:ReviewStochProb} and \ref{ap:ReviewStochProc}.
\par 

Stochastic mechanics studies stochastic processes. By definition, a stochastic process is a family of random variables $\{X_t\,|\,t\in[t_0,t_f]\}$, where, for any $t\in[t_0,t_f]$, the random variable $X_t$ is a map from a probability space $(\Omega,\Sigma,\mathbb{P})$ to the measurable configuration space $(\M,\mathcal{B}(\M))$. Hence, for any time $t\in[t_0,t_f]$, the stochastic process induces a measure $\mu_{X_t}=\mathbb{P}\circ X_t^{-1}$ on the configuration space $(\M,\mathcal{B}(\M))$. Then, in order to do analysis, one must introduce an $L^p$-norm that turns the space of all random variables into an $L^p$-space denoted by  $L^p_t(\Omega,\Sigma,\mathbb{P})$.
\par 

The construction of the $L^p$-space of random variables allows to study the dynamics of the stochastic process, but does not provide the observables of the process. In a classical theory, observables are obtained by applying smooth functions $f\in C^\infty(T^\ast\M)$ to the trajectory of the particle. In the stochastic theory, the observables are given by the expectation value of real valued measurable functions ${f\in L^p(\M,\mathcal{B}(\M),\mu_{X_t})}$ on the space $(\M,\mathcal{B}(\M))$ with induced measure $\mu_{X_t}$. If one fixes the norm by setting $p=2$, one finds that $L^2(\M,\mathcal{B}(\M),\mu_{X_t})$ is a Hilbert space. Therefore, in the stochastic theory, observables are elements of the real Hilbert space $L_{\R}^2(\M,\mathcal{B}(\M),\mu_{X_t})$.
\par 

This Hilbert space is different from the static Hilbert space $L_\mathbb{C}^2(\M,\mathcal{B}(\M),V_R)$, which is introduced in the Dirac-Von Neumann axioms, as it is real and time-dependent. However, by changing the measure from $\mu_{X_t}$ to the Riemann measure $V_R$, one obtains the real Hilbert space $L_\R^2(\M,\mathcal{B}(\M),V_R)$, and the Radon-Nykod\'ym derivative associated to this change of measure is the probability density $\rho_{X_t}$. The complex Hilbert space $L_\mathbb{C}^2(\M,\mathcal{B}(\M),V_R)$ can then be obtained by complexifying $L_\R^2(\M,\mathcal{B}(\M),V_R)$. On this complex Hilbert space, the observables $f$ act as self-adjoint operators and the elements of this Hilbert space are complex functions $\psi$, such that $|\psi|^2\propto\rho_{X_t}$, where the proportionality denotes equality up to the scalar multiplication of the Hilbert space.

\subsection{Time Reversibility}\label{sec:TimeRev}
It is often believed that any stochastic process is inherently time irreversible. This idea has given rise to a criticism of any stochastic formulation of quantum mechanics that can be formulated as follows:
\begin{quote}
	``Stochastic processes are often used to describe dissipative diffusion phenomena. These are non-unitary, thus time irreversible. Quantum mechanics, on the other hand, is a unitary theory, implying a degree of time reversibility. Therefore, quantum mechanics cannot be described by stochastic processes''
\end{quote}
At the heart of this criticism lies the incorrect assumption that any diffusion theory is non-unitary, hence inherently dissipative and time irreversible.\footnote{As pointed out in Ref.~\cite{Nelson}, the idea that dissipation is intrinsic is historically reminiscent to the Aristotelean school of thought. In Aristotelean dynamics, friction is a fundamental property of all physical systems, as it is believed that all objects tend to their rest position. For a long time, it was believed that this dynamical principle was correct, as it was compatible with the observations of the time, since most physical systems are subjected to friction. Galileo showed, however, that friction is not a fundamental property of deterministic dynamics. This led to Galilean relativity as a new dynamical principle in which friction is no longer fundamental to deterministic theories. This principle can be generalized to stochastic theories: although many stochastic systems are dissipative, dissipation is not a fundamental property of stochastic dynamics.} This assumption is itself based on a typical misunderstanding of the notion of time reversibility in stochastic theories, as we will explain below in a qualitative fashion.
\par

The motion of a deterministic particle is governed by deterministic laws of motion. These are typically encoded in a Lagrangian or Hamiltonian formalism by means of a stationary action principle. Using this stationary action principle, one can derive equations of motion for the particle. These equations define an initial value problem, which allows to determine the trajectory $(X,V)(t)$ for all $t\in\mathcal{T}\subseteq\R$, when the state $(X,V)(t_0)$ is given at some $t_0\in\mathcal{T}$.
\par 

A crucial aspect of deterministic theories is the principle of time reversal invariance, which states that the physical laws that govern the system must be invariant under time reversal. For deterministic theories, this time reversal invariance of the laws of motion implies that the solutions of the equations of motion have the same shape under a time reversal operation.
\par

The notion of time reversibility is more subtle in stochastic theories. First of all, stochastic trajectories cannot be characterized completely by the state $(X,V)$, due to the fact that the objects $\E[X^k]-\E[X]^k$, where $\E$ denotes expectation value and $k\in\mathbb{N}$, are non-vanishing in stochastic theories. Therefore, the state of a stochastic process is described by a state $(X,V_1,V_2,V_3,...)$, where $V_k$ denotes a velocity associated to the moment $\E[X^k]$. Only in deterministic theories, $\E[X^k]=\E[X]^k$, which implies that all velocities $V_k$ are completely determined by $V_1$. This enables the phase space reduction to the state $(X,V)$.
\par 

In order to simplify our further discussion, we will focus on processes for which all the moments $\E[X^k]$ are completely determined by the moments $\E[X]$ and $\E[X^2]$. For these processes, the state is described by a trajectory $(X,V,V_2)$ in a higher order phase space, where $V$ determines the drift velocity, while $V_2$ is the velocity of the variance ${\rm Var}(X) = \E[X^2]-\E[X]^2$.
\par 

As is the case in a deterministic theory, the motion of the stochastic particle is governed by physical laws of motion. In particular, the deterministic laws, encoded in the stationary action principle, can be generalized to the stochastic theory by the construction of a stochastic Lagrangian or Hamiltonian. However, these laws will not completely determine the trajectory $(X,V,V_2)(t)$. Roughly speaking, these laws will only provide $(X,V)$, while $V_2$ remains unknown. In order to obtain the full trajectory, one must introduce a new stochastic law of motion that fixes the velocity $V_2$. 
\par 

To further simplify our discussion, we will now focus on the Wiener process. For the Wiener process localized at $(x_0,t_0)$, the stochastic law of motion states that its evolution is governed by the heat kernel
\begin{equation}\label{eq:heatkernel}
	\Phi(x,t;x_0,t_0) = \big[4 \, \pi \, \alpha \, (t-t_0)\big]^{-\frac{d}{2}}\,  \exp\left[- \frac{||x - x_0||^2}{4 \, \alpha \, (t-t_0)}\right] ,
\end{equation}
where $\alpha>0$ is the diffusion constant and $d$ is the spatial dimension. This stochastic law only defines the forward evolution of the system, i.e. it defines the evolution for any $t,t_0\in\mathcal{T}$ provided that $t\geq t_0$. In order to specify the backward evolution, we must impose another stochastic law, for which there exist infinitely many choices. There is, however, a unique choice that is compatible with time reversal invariance, which is given by  the kernel
\begin{equation}\label{eq:heatkernel2}
	\Phi(x,t;x_0,t_0) = \big[4 \, \pi \, \alpha \, |t-t_0|\big]^{-\frac{d}{2}} \, \exp\left[- \frac{||x - x_0||^2}{4 \, \alpha \, |t-t_0|}\right] .
\end{equation}
This new kernel defines the evolution of the process for any $t,t_0\in\mathcal{T}$, and processes governed by this kernel are called two-sided Wiener processes \cite{Nelson:1967}.
\par 

By L\'evy's characterization of the Wiener process \cite{Levy}, cf. appendix \ref{ap:Wiener}, imposing the heat kernel \eqref{eq:heatkernel} is equivalent to imposing that the second order velocity is given by
\begin{equation}\label{stochasticLaw}
	V_2^{ij}= \alpha \, \delta^{ij} \, .
\end{equation}
Similarly, imposing the kernel \eqref{eq:heatkernel2} is equivalent to imposing 
\begin{alignat}{2}\label{stochasticlaw}
	V_{2,+}^{ij} &=  V_2^{ij} &&= \alpha \, \delta^{ij} \, , \nonumber\\
	V_{2,-}^{ij} &= -V_2^{ij} &&= - \alpha \, \delta^{ij},
\end{alignat}
where the $+$ solution corresponds to processes evolving forward in time and the $-$ solution to processes evolving backward in time. 
\par 

We emphasize that, by construction, the laws of motion, i.e. the stationary action principle and the stochastic law \eqref{stochasticlaw}, are time reversal invariant. However, in contrast to the deterministic theory, this does not imply that the solutions of the equations of motion are also time reversal invariant. In fact, they are not, as the forward solution is described by $(X,V_+,+V_2)$, while a backward solution is given by $(X,V_-,-V_2)$, where\footnote{The presence of two different velocities $V_+$ and $V_-$ reflects that a Wiener process is almost surely not differentiable. This can be interpreted as follows: at every time $t\in\mathcal{T}$, the Brownian particle gets a kick from a microscopic particle, which induces a discontinuous change of velocity from $V_-$ to $V_+$.} $V_+\neq V_-$.
\par

Often, in the study of stochastic processes, one only cares about the forward evolution of the system, such that one has to worry only about the stochastic law \eqref{stochasticLaw}. Stochastic mechanics, on the other hand, studies processes for which the stochastic law is time reversible, which is imposed by setting $V_{2,-}=-V_{2,+}$. As a consequence, stochastic mechanics studies both the forward solutions $(X,V_+,V_2)$ and the backward solutions $(X,V_-,-V_2)$.
\par 

Up to this point, we have discussed the notion of time reversibility in stochastic theories. This discussion does not resolve the criticism formulated at the beginning of this section, as the two-sided Wiener process governed by the kernel \eqref{eq:heatkernel2} remains a dissipative process, albeit with respect to both the future and the past. However, things change drastically, when one considers superpositions of processes. Stochastic mechanics proves that such superpositions can be governed by a unitary time evolution, as encountered in quantum mechanics. 
\par 

The traditional approach in the stochastic mechanics literature, cf. e.g. Refs.~\cite{Nelson:1966sp,Nelson:1967,Nelson,Guerra:1981ie,Zambrini}, for obtaining such processes with a unitary time evolution is to consider a superposition of the forward and backward solutions. As mentioned in section \ref{sec:introStochMech}, we adopt a different approach in this book: we consider two correlated two-sided Wiener processes $M_x$ and $M_y$, and study the complex process $M = M_x + \ri \, M_y$. Then, we show that the evolution of the real projection ${\rm Re}[M]=M_x$ is governed by the heat kernel \eqref{eq:heatkernel2}, if the processes $M_x$ and $M_y$ are uncorrelated, and that the evolution is governed by a unitary Schr\"odinger kernel, if the processes are maximally correlated.

\subsection{Hidden Variables}\label{sec:IntroHiddenVar}
Another criticism that is often faced by stochastic interpretations of quantum mechanics is that stochastic theories open the door to hidden variable theories. This criticism can be formulated as follows
\begin{quote}
	``Stochastic theories often arise as an effective theory that replaces a more fundamental hidden background field of microscopic particles. On the other hand, the Bell theorems exclude any locally real hidden variable theory. Therefore, quantum theories must be fundamentally different from stochastic theories.''
\end{quote}
In the physical models of Brownian motion and other stochastic processes, the theory contains a background field and expectation values are obtained by calculating ensemble averages. This background field is governed by deterministic laws of motion, but, due to the enormous computational complexity of this deterministic theory, one introduces an effective stochastic theory. This stochastic theory allows to determine the behavior of the macroscopic particle in a much more efficient manner.
\par 

In the mathematical theory of stochastic processes, the background field is replaced by a probability space and expectation values are defined as Lebesgue integrals over this probability space. Stochastic mechanics is built entirely within this mathematical framework. Therefore, stochastic mechanics should be interpreted as a mathematically rigorous implementation of the statement that `God plays dice'.
\par

Stochastic mechanics remains agnostic about the question whether the stochastic theory based on the idea of `God playing dice' may be replaced with a statistical theory containing some background field.
Moreover, if such a background field is introduced, stochastic mechanics only imposes a condition on the stochastic law that is obtained in the continuum limit.\footnote{The condition is that the ensemble averages $\langle.\rangle$ calculated in the statistical theory converge to the expectation values $\E[.]$ calculated in the stochastic theory, i.e., for any observable $A$, $\langle A \rangle\rightarrow \E[A]$ in the limit where the number of particles in the background field goes to infinity, while their size goes to zero.} It does not impose any conditions on the (non-)deterministic laws that govern the background field on a microscopic level.
\par 

We point out that the Bell theorems do not rule out the replacement of the probability space in stochastic mechanics by some background field that is governed by deterministic laws of motion. Such deterministic hidden variable theories generically imply the presence of uncertainty principles in the corresponding stochastic theory, while the derivation of the Bell inequalities relies on the absence of such uncertainty relations. This argument is worked out in more detail in section \ref{sec:Bell}.

\subsection{Outline for the book}
In this book, we study a spinless stochastic particle subjected to a scalar potential $\mathfrak{U}$ and a vector potential $A$. We derive the stochastic equations of motion for this particle and show that the (proper) time evolution of the probability density of this particle is subjected to the diffusion equation \eqref{eq:DiffEqDef} with $\M$ a Riemannian or Lorentzian manifold, $\Psi\in L^2(\M)$ complex valued, and $H$ given by the covariant generalization of eq.~\eqref{eq:HamExample}.
\par 

Conversely, we show, by explicit construction, that for any solution of the diffusion equation \eqref{eq:DiffEqDef} with $\M$ a Riemannian or Lorentzian manifold, $\Psi\in L^2(\M)$, and $H$ given by the covariant generalization of eq.~\eqref{eq:HamExample}, there exists a stochastic process $X$ such that $|\Psi|^2$ describes the probability density of $X$.
\par

The presentation in this book is self-contained: the results review, reformulate and build on results from the stochastic mechanics literature, but no prior knowledge of stochastic mechanics is assumed.
\par 

The book is organized as follows: in chapter \ref{sec:Class}, we review the classical dynamics of a deterministic spinless non-relativistic particle on $\R^d$. In chapter \ref{sec:StochDynR}, we superimpose a Brownian motion onto this particle and derive the Feynman-Kac theorem using the tools from stochastic mechanics. In chapter \ref{sec:StochDynC}, we complexify the Wiener process and perform the same analysis, which allows to extend the results from real to complex diffusion equations. In chapter \ref{sec:RLT}, we extend this complex stochastic theory to relativistic particles on $\R^{d,1}$. In chapter \ref{sec:Manifolds}, we extend the results to Riemannian and Lorentzian manifolds.
In chapter \ref{sec:HiddenVariables}, we discuss some important aspects of the stochastic interpretation of quantum mechanics that is implied by the results. Finally, in chapter \ref{sec:Conclusion}, we conclude.
\par 

Our analysis heavily relies on standard results from stochastic analysis, which are reviewed in appendices \ref{ap:ReviewStochProb}, \ref{ap:ReviewStochProc} and \ref{ap:ReviewStochCalc}. Moreover, the extension of the theory to manifolds requires the framework of second order geometry, which is reviewed in appendix \ref{Ap:2Geometry}. Finally, in appendices \ref{sec:ItoLagrangian} and \ref{ap:VariationalCalculus}, the equations of motion for the stochastic particle are derived using stochastic variational calculus.

\clearpage

\section{Classical Dynamics on $\R^d$}\label{sec:Class}
We consider a particle with mass $m$ moving in the $d$-dimensional real space $\R^d$. We assume the particle to be charged under a vector potential $A_i(x,t)$ with charge $q$ and a scalar potential $\mathfrak{U}(x,t)$. We are interested in its trajectory $\{X_t\,|\,t\in\mathcal{T}\}$ with ${X_t = X(t) : \mathcal{T} \rightarrow \R^d}$ parameterized by the time $t\in \mathcal{T}=[t_0,t_f]$ with ${t_0<t_f\in\R}$. 
\par

The motion of this particle is governed by a Lagrangian $L:T\R^{d} \times \mathcal{T}\rightarrow \R$ that is defined on the tangent bundle (phase space) $T\R^{d}\cong\R^{2d}$ and is given by 
\begin{equation}\label{eq:ClassLagrangian}
	L(x,v,t) =\frac{m}{2} \, \delta_{ij} \, v^i v^j + q \, A_i(x,t) \, v^i - \mathfrak{U}(x,t) \, .
\end{equation}
The corresponding action is given by the integral
\begin{equation}
	S(X) = \int_{t_0}^{t_f} L(X_t,V_t,t) \, dt \, ,
\end{equation}
where $(X,V):\mathcal{T}\rightarrow T\R^d$ describes a trajectory on the tangent bundle.
By extremizing this action one finds the Euler-Lagrange equations. These are given by
\begin{equation}
	\frac{d}{dt} \frac{\p L}{\p v^i} = \frac{\p L}{\p x^i} \Big|_{(x,v)=(X_t,V_t) \, .}
\end{equation}
For the Lagrangian \eqref{eq:ClassLagrangian}, the Euler-Lagrange equations yield
\begin{equation}
	\frac{d}{dt} \left( m \, \delta_{ij} V^j + q \, A_i \right) = q \, V^j \p_i A_j - \p_i \mathfrak{U} \, .
\end{equation}
Then, using that the velocity satisfies
\begin{equation}\label{eq:VelocityELClass}
	V^i_t = \frac{d X^i_t}{dt}\, ,
\end{equation}
one finds
\begin{equation}\label{eq:ELFirstOrder}
	m \, \delta_{ij} \frac{dV^j}{dt}
	= q \, F_{ij} \,  V^j - q \, \p_t A_i - \p_i \mathfrak{U} \, ,
\end{equation}
where the field strength is defined by
\begin{equation}
	F_{ij} := \p_i A_j - \p_j A_i \, .
\end{equation}
The trajectory $(X,V):\mathcal{T}\rightarrow T\R^d$ is now uniquely determined by eqs.~\eqref{eq:VelocityELClass} and \eqref{eq:ELFirstOrder} supplemented with an initial condition, i.e. by the initial value problem
\begin{equation}\label{eq:ELFirstOrderset}
	\begin{cases}
		\frac{d X_t^i}{dt} &= V_t^i \, , \\
		m \, \delta_{ij} \frac{dV_t^j}{dt}
		&= q \, F_{ij}(X_t,t) \,  V_t^j - q \, \p_t A_i(X_t,t) - \p_i \mathfrak{U}(X_t,t) \, , \\
		(X_{t_0},V_{t_0}) &= (x_0,v_0)\, .
	\end{cases}
\end{equation}
\par 

Alternatively, the trajectory of the particle can be derived in the Hamiltonian formulation of classical mechanics. In this equivalent description, the motion is governed by a Hamiltonian $H:T^\ast\R^d \times \mathcal{T} \rightarrow\R$ defined on the cotangent bundle (phase space) $T^\ast\R^d\cong\R^{2d}$, which can be obtained from the Lagrangian by a Legendre transform
\begin{equation}
	H(x,p,t) = p_i v^i - L(x,v,t) \, ,
\end{equation}
where $p_i$ is the canonical momentum defined by
\begin{equation}
	p_i := \frac{\p L}{\p v^i} \, .
\end{equation}
Inversely, the Lagrangian can be obtained from the Hamiltonian through a Legendre transform
\begin{equation}
	L(x,v,t) = p_i v^i - H(x,p,t)
\end{equation}
with the canonical velocity
\begin{equation}
	v^i := \frac{\p H}{\p p_i} \, .
\end{equation}
Using the Hamiltonian, one can derive the Hamilton equations, which is a set of ordinary differential equations. These are equivalent to the Euler-Lagrange equations and given by
\begin{equation}
	\left.
	\begin{cases}
		\frac{d x^i}{dt} = \frac{\p H}{\p p_i}\\
		\frac{d p_i}{dt} = - \frac{\p H}{\p x^i}\\
		\frac{\p H}{\p t} = - \frac{\p L}{\p t} 
	\end{cases}
	\right|_{(x,p)=(X_t,P_t) \, .} 
\end{equation}
\par

There exists a third equivalent formulation of classical mechanics, which is the Hamilton-Jacobi formalism. In this formulation, one defines Hamilton's principal function $S:\R^d \times \mathcal{T} \rightarrow \R$, such that
\begin{equation}
	S(x,t) = S(x,t;x_0,t_0) = \int_{t_0}^{t} L(X_s,V_s,s)\, ds \, ,
\end{equation}
where $(X_s,V_s)$ is a solution of the Euler-Lagrange equations passing through $(x_0,t_0)$ and $(x,t)$ with $t_0 \leq t \leq t_f$. Using Hamilton's principal function, one can derive another equivalent set of equations of motion. These are the Hamilton-Jacobi equations:
\begin{equation}\label{eq:HamJacClass}
	\left.
	\begin{cases}
		\frac{\p S(x,t)}{\p x^i} = p_i\\
		\frac{\p S(x,t)}{\p t} = - H(x,p,t) 
	\end{cases}
	\right|_{(x,p)=(X_t,p(X_t,t))\,.}
\end{equation}
For the Lagrangian \eqref{eq:ClassLagrangian}, this yields
\begin{equation}\label{eq:ClassHamJacobi}
	\left.
	\begin{cases}
		\frac{\p S}{\p x^i} = m \, \delta_{ij} v^j + q \, A_i\\
		\frac{\p S}{\p t} = - \frac{m}{2} \, \delta_{ij} v^i v^j - \mathfrak{U}
	\end{cases}
	\right|_{(x,v)=(X_t,v(X_t,t))\,.}
\end{equation}
Here, $(x,v)$ and $(x,p)$ are coordinates on the (co)tangent bundle, $v(x,t)$ and $p(x,t)$ are (co)vector fields over $\R^d$, $X_t$ is a trajectory on $\R^d$ and $(X_t,V_t)$, $(X_t,P_t)$ are trajectories on the (co)tangent bundle.
\par

One can explicitly show the equivalence between the Hamilton-Jacobi equations \eqref{eq:ClassHamJacobi} and the Euler-Lagrange equations \eqref{eq:ELFirstOrder} by taking a spatial derivative of the second Hamilton-Jacobi equation and plugging in the first equation. This leads to
\begin{equation}\label{eq:HamJacIntermediate}
	m \, \delta_{ij} \, \frac{\p}{\p t} v^j + q \, \frac{\p}{\p t} A_i = - m \, \delta_{jk} v^k \p_i v^j  - \p_i \mathfrak{U} \, .
\end{equation}
Then, using the first Hamilton-Jacobi equation, one finds
\begin{align}\label{eq:pivjpjvi}
	m \, \delta_{jk} \, \p_i v^j 
	&= \p_i \left(\p_k S - q \, A_k  \right)\nonumber\\
	&= \p_k \left(\p_i S - q \, A_i  \right) - q \left(\p_i A_k - \p_k A_i \right) \nonumber\\	
	&= m \, \delta_{ij}\, \p_k v^j - q \, F_{ik} \, .
\end{align}
Plugging this relation into eq.~\eqref{eq:HamJacIntermediate} yields
\begin{equation}\label{eq:velocityFieldClassical}
	\left[ m \, \delta_{ij} \left(\frac{\p}{\p t} + v^k \p_k\right) - q \, F_{ij} \right] v^j  
	= 
	- q \, \frac{\p}{\p t} A_i - \p_i \mathfrak{U}\,.
\end{equation}
Finally, using that
\begin{equation}\label{eq:velocity}
	\frac{dX_t}{dt} = v(X_t,t),
\end{equation}
one finds that this is equivalent to eq.~\eqref{eq:ELFirstOrder} with $V_t$ replaced by $v(X_t,t)$.
\par

We conclude this chapter with two remarks. First, we note that one can write down a partial differential equation for Hamilton's principal function by combining the expressions in eq.~\eqref{eq:ClassHamJacobi}. This yields
\begin{equation}\label{eq:ClassDiffEq}
	- 2\, m\, \frac{\p S}{\p t} = \p^i S \, \p_i S - 2 \, q \, A^i \, \p_i S  + q^2 \, A^i A_i + 2 \, m \, \mathfrak{U} \, .
\end{equation}
Secondly, we point out that Hamilton's principal function can also be defined as
\begin{equation}
	S(x,t) = S(x,t;x_f,t_f) = -\int_{t}^{t_f} L(X_s,V_s,s)\, ds\, ,
\end{equation}
where $(X_s,V_s)$ is a solution of the Euler-Lagrange equations passing through $(x,t)$ and $(x_f,t_f)$ with $t_0 \leq t \leq t_f$. The Hamilton-Jacobi equations for this principal function are also given by eq.~\eqref{eq:HamJacClass}.

\clearpage
\section{Stochastic Dynamics on $\R^d$}\label{sec:StochDynR}
In previous chapter, we discussed the equations of motion that govern a deterministic theory. In this chapter, we introduce a notion of randomness in the theory by promoting the deterministic trajectories to stochastic processes. We will be interested in processes $X$ of the form
\begin{equation}\label{eq:procInitial}
	X_t = C_{t} + M_t \, ,
\end{equation}
where $C$ represents a deterministic trajectory, while $M$ represents a stochastic noise. Our aim is to derive equations of motion for such a process by minimizing an action, as was done in the previous chapter. 
\par

We can make this idea precise in the language of stochastic analysis, which we review in appendices \ref{ap:ReviewStochProb} and \ref{ap:ReviewStochProc}. In this language, a stochastic process that can be decomposed as in eq.~\eqref{eq:procInitial} is called a semi-martingale, the process $C$ is a c\`adl\`ag process\footnote{A c\`adl\`ag process is right-continuous with left limits. In this book, starting from eq.~\eqref{eq:StructRelation}, we only consider processes that are continuous, which is a stronger assumption than being c\`adl\`ag.} of finite variation\footnote{Roughly speaking, the requirement of finite variation ensures that $C$ is deterministic, bounded and that it does not oscillate with an infinite frequency.} and $M$ is a local martingale process\footnote{A martingale process is a process that does not drift, i.e. its expectation value is constant in time: $\E[M_t|\{M_r:r\in[t_0,s]\}]=M_s$ for all $t>s\in\mathcal{T}$. For local martingales, this  martingale property is only required to hold locally, i.e. if $s$ and $t$ are close to each other..}.
\par 

In addition, we will impose a notion of time reversibility to the process by requiring that the laws of motion are invariant under time reversal. As discussed in section \ref{sec:TimeRev}, the time reversal symmetry of the laws of motion does not imply time reversal symmetry of the solutions of the equations of motion. Therefore, solutions will be a two-sided\footnote{Being two-sided means that the stochastic evolution of the process is well defined and governed by the same stochastic law with respect to both the future and past directed evolution.} semi-martingale of the form
\begin{equation}\label{eq:DMdecomp}
	X_t^i = C_{\pm,t}^i + M_t^i \, ,
\end{equation}
where $C_+$ is c\`adl\`ag process of finite variation that evolves forward in time and $C_-$ is c\`agl\`ad process\footnote{A c\`agl\`ad process is left-continuous with right limits.} of finite variation that evolves backward in time. Moreover, $M$ is a two-sided local martingale.
\par

In the following chapters, it will prove to be useful to let the processes $X$ and $M$ evolve in different copies of the configuration space $\R^d$. This can be achieved by rewriting eq.~\eqref{eq:DMdecomp} as
\begin{equation}\label{eq:DMdecompImproved}
	X^i_t = C^i_{\pm,t} + \delta_a^i \, M^a_t \, .
\end{equation}
Here, we have promoted the configuration space $\M=\R^d$ to the frame bundle $(E,\pi,\M)$ with base space $\M=\R^d$ and fibers $F=\R^d$, such that $X$ is a two-sided process on the base space $\M$, $C_\pm$ are one-sided processes on the base space $\M$, and $M$ is a two-sided martingale on the fibers $F$. Moreover, the Kronecker delta $\delta_a^i$ defines an orthonormal frame spanning $F$ at every point $x\in\M$ and is commonly referred to as a vielbein or polyad.
\par

In the remainder of this chapter, we will discuss the derivation of the equations of motion for the stochastic trajectory by generalizing the principle of stationary action to a stochastic theory. Such equations of motion will only fix the deterministic part contained in $C$, but not the noise contained in $M$. Therefore, we must impose additional  laws of motion to fix the stochastic behavior of $X$.

\subsection{The Stochastic Law}
The aim of this section is to fix the stochastic law of the martingale process $M$. In this chapter, we will fix $M$ to be a Wiener process. However, as we will generalize the theory in the next chapters, we will provide a more general discussion.
\par 

A standard way of fixing the stochastic law of a stochastic process $X$ is by fixing its characteristic functional
\begin{equation}
	\varphi_{X}(J) := \E\left[ e^{\ri \int_\mathcal{T} J_i(t)\, X^i(t) \, dt}\right] .
\end{equation}
Our first assumption is that the moment generating functional for $X$, defined by
\begin{equation}
	M_{X}(J) := \E\left[ e^{\int_\mathcal{T} J_i(t)\, X^i(t) \, dt}\right] ,
\end{equation}
has a non-zero radius of convergence. Due to this assumption, fixing $\varphi_{X}(J)$ is equivalent to fixing $M_{X}(J)=\varphi_{X}(- \ri \, J)$. In addition, the non-zero radius of convergence of $M_{X}(J)$ implies that all moments $\E[X^k]$ for $k\in\mathbb{N}$ exist, and that, within its radius of convergence, $M_X(J)$ can be characterized completely by its moments. Therefore, we can determine the stochastic law of $X$ by specifying the moments $\E[X^k]$ for all $k\in\mathbb{N}$. Furthermore, since $X=C_\pm+M$ and $C_\pm$ has finite variation, $\E[C_\pm^k]=\E[C_\pm]^k$. Hence, we only need to specify the moments $\E[M^k]$, More precisely, we must specify all moments 
\begin{equation}\label{eq:Moments1}
	\E\left[\prod_{i=1}^k \, (M^{a_i}_{t_i}-M^{a_i}_{s_i})\right] \qquad \forall \; t_i,s_i \in \mathcal{T}, \; a_i\in \{1,...,d\}, \; k\in\mathbb{N} \, .
\end{equation}
Since $M$ is a martingale, the first moment is fixed by the martingale property \eqref{eq:MartingaleProperty}, as
\begin{align}
	\E[(M^a_t-M^a_s)] 
	&= \E[\E[(M^a_t-M^a_s)|\{M_r:t_0\leq r \leq s\}]]\nonumber\\
	&= \E[M^a_s - M^a_s] \nonumber\\
	&= 0\, .
\end{align}
\par

In order to fix the other moments, we will make the additional assumption that $M$ is a L\'evy process.\footnote{L\'evy processes can be regarded as the continuous time analogue of the random walk, cf. appendix \ref{ap:Levy}. Examples of L\'evy processes include the Wiener process and the Poisson process.} Therefore, $M$ has independent increments, and, due to this independence,\footnote{Another consequence of the independent increments is that $M$ is a Markov process.} all moments~\eqref{eq:Moments1} are completely determined by the subset of moments
\begin{equation}\label{eq:Moments2}
	\E\left[\prod_{i=1}^k \, (M^{a_i}_{t}-M^{a_i}_{s})\right] \qquad \forall \; s<t \in \mathcal{T}, \; a_i\in \{1,...,d\}, \; k\in\mathbb{N} \, .
\end{equation}
We will fix these remaining moments by imposing a structure relation, using a differential notation. In this differential notation, the decomposition \eqref{eq:DMdecompImproved} is given by the stochastic differential equation\footnote{Cf. appendix \ref{ap:ReviewStochCalc} for a review of stochastic calculus.}
\begin{equation}\label{eq:ItoEqInitial}
	d_\pm X_t^i = v_\pm^i(X_t,t) \, dt + \delta^i_a \, d_\pm M_t^a \, ,
\end{equation}
where $d_+$ is a forward It\^o differential and $d_-$ is a backward It\^o differential, i.e.
\begin{align}
	d_+ X_t &:= X_{t+dt} - X_t \, ,\nonumber\\
	d_- X_t &:= X_t - X_{t-dt} \, .
\end{align}
We introduce a bracket $[.,.]$, called the quadratic variation\footnote{Cf. appendix \ref{ap:QVar} for a definition and discussion of the properties of quadratic variation.} of $X$, which is given by
\begin{align}
	d_+[X,X]_t &:= \left[ X_{t+dt} - X_t \right]\otimes \left[ X_{t+dt} - X_t \right] \, ,\nonumber\\
	d_-[X,X]_t &:= \left[ X_t - X_{t-dt}\right] \otimes \left[ X_t - X_{t-dt}\right]\, .
\end{align}
Since $C_\pm$ has finite variation, $dC_t=\mathcal{O}(dt)$, but $dM_t=o(1)$, whence\footnote{This reflects the fact that the stochastic law of $X$ is determined by the stochastic law of $M$.}
\begin{align}
	d_\pm[X,X]_t &= d_\pm[C,C]_t + d_\pm[C,M]_t + d_\pm[M,C]_t + d_\pm[M,M]_t \nonumber\\
	&= d_\pm[M,M]_t + o(dt) \, .
\end{align}
The quadratic variation can be used to calculate the quadratic moment of $M$, since 
\begin{align}\label{eq:QuadMomInitial}
	\E\left[(M^a_t-M^a_s)(M^b_t-M^b_s)\right] 
	&= \E\left[\E\left[(M^a_t-M^a_s)(M^b_t-M^b_s) \, \Big| \, \{M_r : t_0\leq r\leq s \}\right] \right] \nonumber\\
	&= \E\left[ \int_s^t d_+M^a_{r_1}  \int_s^t d_+M^b_{r_2} \right]\nonumber\\
	&= \E\left[ \int_s^t d_+[M^a,M^b]_r  \right] ,
\end{align}
where we used that $M_t=M_s+\int_s^t d_+M_r$ for $s<t$. Furthermore, using that ${M_s=M_t+\int_t^s d_-M_r}$ for $s<t$, we find
\begin{align}\label{eq:QuadMomInitialBack}
	\E\left[(M^a_t-M^a_s)(M^b_t-M^b_s)\right] 
	&= \E\left[\E\left[(M^a_t-M^a_s)(M^b_t-M^b_s) \Big| \, \{M_r : t\leq r\leq t_f \}\right] \right] \nonumber\\
	&=
	\E\left[ \int_t^s d_-M^a_{r_1}  \int_t^s d_-M^b_{r_2} \right]\nonumber\\
	&=
	\E\left[ \int_t^s d_-[M^a,M^b]_r \right]\nonumber\\
	&= \E\left[ - \int_s^t d_-[M^a,M^b]_r \right] .
\end{align}
Now, we can introduce the notation
\begin{equation}
	d[M,M]_t := d_+[M,M]_t = - d_-[M,M]_t\, ,
\end{equation}
and impose a structure relation for the quadratic variation $d[M,M]_t$. A general structure relation takes the form
\begin{equation}\label{eq:StructRelationGen}
	d[M^a,M^b] = \frac{\alpha \, \hbar}{m} \, A^{ab} \, dt + \frac{\beta}{\kappa} \, B^{ab}_c \, dM^c_t \, ,
\end{equation}
where $A$ and $B$ are dimensionless structure constants with $A$ symmetric and positive definite. $m$, $\hbar$ and $\kappa$ are introduced to fix the physical dimensions, such that $m$ is the mass of the particle, $\hbar$ the reduced Planck constant and $\kappa$ a constant with dimension $[\kappa]=L^{-1}$. $\alpha\in[0,\infty)$ and $\beta\in\R$ are dimensionless parameters and $\alpha,\beta\rightarrow 0$ yield the deterministic limit. Furthermore, $M$ is continuous, if and only if $\beta=0$.
\par

In this book, we assume that $X$ is continuous, which fixes $\beta=0$, we set $A^{ab}=\delta^{ab}$ and work in natural units where $\hbar=1$. The structure relation then simplifies to
\begin{equation}\label{eq:StructRelation}
	m \, d[M^a,M^b]_t = \alpha \, \delta^{ab} \, dt 
\end{equation}
with\footnote{Note that the real martingale $M$ exists, if and only if $\alpha\geq0$, and is non-trivial, if and only if $\alpha\neq0$.} $\alpha\geq0$.	We note that, by the L\'evy characterization of Brownian motion, this structure relation implies that $M$ is a Wiener process, cf. appendix~\ref{ap:Wiener}.
\par

The structure relation \eqref{eq:StructRelation} fixes all the moments given in eq.~\eqref{eq:Moments2}. Indeed, using eq.~\eqref{eq:QuadMomInitial}, we find that the quadratic moment is given by
\begin{align}\label{eq:QuadMom}
	\E\left[(M^a_t-M^a_s)(M^b_t-M^b_s)\right] 
	&= \E\left[ \int_s^t d[M^a,M^b]_r \right]\nonumber\\
	&= \E\left[\frac{\alpha}{m} \, \int_s^t \delta^{ab} \, dr \right]\nonumber\\
	&= \frac{\alpha }{m} \, (t-s) \, \delta^{ab} \, .
\end{align}
Furthermore, all moments of order $k$ can be expressed as a linear combination of moments of order $(k-2)$ using that
\begin{align}
	\E\left[\prod_{i=1}^k \, (M^{a_i}_{t}-M^{a_i}_{s})\right]
	&=
	\E\left[\prod_{i=1}^k \, \int_s^t dM^{a_i}_{r_i}\right] 
	\nonumber\\
	&=
	\sum_{j=1}^{k-1} \E\left[\int_s^t d[M^{a_k},M^{a_j}]_{r_k} \, \prod_{i=1,\,i\neq j}^{k-1} \, \int_s^t dM^{a_i}_{r_i}\right]
	\nonumber\\
	&=
	\frac{\alpha}{m} \, (t-s) \, \sum_{j=1}^{k-1} \delta^{a_k a_j} \, \E\left[ \prod_{i=1,\,i\neq j}^{k-1} \,(M^{a_i}_{t}-M^{a_i}_{s})\right] .
\end{align}
Therefore,
\begin{equation}
	\E\left[\prod_{i=1}^k \, (M^{a_i}_{t}-M^{a_i}_{s})\right] = 0 \qquad {\rm if} \; k {\rm \; is\; odd}
\end{equation}
and all even moments can be expressed in terms of the quadratic moment \eqref{eq:QuadMom}. We remark that this result is known as Isserlis' theorem or Wick's probability theorem.

\subsection{Stochastic Phase Space}
Now that we have fixed the stochastic law of $X$, we would like to derive the equations of motion for $X$. In order to do so, we must construct a stochastic action using a stochastic Lagrangian and derive equations of motion by stochastically minimizing this action.
\par

Here, we encounter a difficulty: the classical Lagrangian is defined on the tangent bundle $T\R^d\cong\R^{2d}$, so we expect that the stochastic Lagrangian is also defined on a tangent bundle. However, stochastic processes are almost surely not differentiable. Therefore, there is no trivial notion of velocity. In the deterministic theory, we can define for any trajectory $X$ a velocity, as given in eq.~\eqref{eq:VelocityELClass}:
\begin{eqnarray}\label{eq:naivevelocity}
	V_t 
	&:=& \lim_{dt\rightarrow 0} \frac{X_{t+dt} - X_{t-dt}}{2 \, dt}\nonumber\\
	&=& \lim_{dt\rightarrow 0} \frac{X_{t+dt} - X_t}{dt}\nonumber\\
	&=& \lim_{dt\rightarrow 0} \frac{X_t - X_{t-dt}}{dt} \, ,
\end{eqnarray}
such that the pair $(X_t,V_t)$ is a trajectory on the tangent bundle $T\R^d\cong\R^{2d}$. However, when $X$ is a stochastic process, none of these expressions is well defined.
\par

Nevertheless, using conditional expectations, we can construct a velocity field, which is the stochastic equivalent of the velocity field given in eq.~\eqref{eq:velocity}, but there exist two inequivalent possibilities:
\begin{align}
	v_+(X_t,t) &= \lim_{dt\rightarrow 0} \E \left[\frac{ X_{t+dt} - X_{t}}{dt} \, \Big| \, X_t \right],\\
	v_-(X_t,t) &= \lim_{dt\rightarrow 0} \E \left[\frac{ X_{t} - X_{t-dt}}{dt} \, \Big| \, X_t \right],
\end{align}
which are the forward and backward It\^o velocities.
In addition, we can define a Stratonovich velocity by
\begin{align}
	v_\circ(X_t,t) &= \lim_{dt\rightarrow 0} \E \left[\frac{ X_{t+dt} - X_{t-dt}}{2 \, dt} \, \Big| \, X_t \right] \nonumber\\
	&= 
	\lim_{dt\rightarrow 0} \left\{ 
	\E \left[\frac{ X_{t+dt} - X_t}{2\,dt} \, \Big|\, X_t \right]
	+ \E \left[\frac{ X_t - X_{t-dt}}{2\, dt} \, \Big|\, X_t \right]
	\right\} \nonumber\\
	&=
	\frac{1}{2} \, \Big[ v_+(X_t,t) + v_-(X_t,t) \Big] .
\end{align}
\par 

Moreover, in contrast to the classical case, we can define another non-vanishing velocity field, which results from the non-vanishing quadratic variation and is given by
\begin{equation}
	v_2(X_t,t) 
	:= \lim_{dt\rightarrow 0} \E \left[ 
	\frac{ \left(X_{t+dt} - X_t\right)  \otimes \left(X_{t+dt} - X_t\right) }{dt} \,	\Big| \, X_t 
	\right] .
\end{equation}
One could again take the limit in various ways, yielding velocities $v_{2,+}$, $v_{2,-}$ and $v_{2,\circ}$. However, this does not lead to independent velocities, since, by eqs.~\eqref{eq:QuadMomInitial} and \eqref{eq:QuadMomInitialBack},
\begin{align}
	v_{2,+}(X_t,t) &= v_{2}(X_t,t) \, , \nonumber\\
	v_{2,-}(X_t,t) &= - v_{2}(X_t,t) \, , \nonumber\\
	v_{2,\circ}(X_t,t) &= 0 \, .
\end{align}
%,
\par

In the classical theory, the velocity fields are sections of the tangent bundle ${T\R^d\cong\R^{2d}}$. Similarly, in the stochastic theory, the fields $v_\circ$, $(v_+,v_2)$, $(v_-,-v_2)$ can be regarded as sections of the tangent bundles $T_\circ\R^d$, $T_+\R^d$ and $T_-\R^d$ respectively.\footnote{Cf. appendix \ref{Ap:2Geometry} for more detail.} For the Stratonovich bundle we have $T_\circ\R^d\cong\R^{2d}$, but the It\^o bundles $T_\pm\R^d$ have a larger dimension, due to the $\frac{d(d+1)}{2}$ additional degrees of freedom contained in the symmetric object $v_2$, such that $T_\pm\R^d\cong\R^{\frac{d(d+5)}{2}}$.
\par

We can study processes $(X_t,V_{\circ,t})$ on $T_\circ\R^d$ and processes $(X_t,V_{\pm,t},\pm V_{2,t})$ on $T_\pm\R^d$ and define a relation between $X_t$ and $V_t$, which is similar to the classical relation given in eq.~\eqref{eq:VelocityELClass}: the processes $V_\circ$, $V_+$, $V_-$ and $V_2$ are the velocity above $X$, if
\begin{align}\label{eq:VelProcess}
	\E \left[ \int f_i(X_t,t) \, V^i_{\circ,t} \, dt \right] &= \E \left[ \int f_i(X_t,t) \, d_\circ X^i_t \right] ,\nonumber\\
	\E \left[ \int f_i(X_t,t) \, V^i_{+,t} \, dt \right] &= \E \left[ \int f_i(X_t,t) \, d_+ X^i_t \right] ,\nonumber\\
	\E \left[ \int f_i(X_t,t) \, V^i_{-,t} \, dt \right] &= \E \left[ \int f_i(X_t,t) \, d_- X^i_t \right] ,\nonumber\\
	\E \left[ \int g_{ij}(X_t,t) \, V^{ij}_{2,t} \, dt \right] &= \E \left[ \int g_{ij}(X_t,t) \, d[X^i,X^j]_t \right]
\end{align}
for any Lebesgue integrable $f\in T^\ast \R^d$ and $g\in T^2(T^\ast \R^d)$. In the remainder of the book, we will shorten these expressions using the differential notation
\begin{align}
	V^i_{\circ,t} \, dt &= d_\circ X^i_t \, , \label{eq:VelProcessStrat}\\
	V^i_{+,t} \, dt &= d_+ X^i_t \, , \label{eq:VelProcessItoF}\\
	V^i_{-,t} \, dt &= d_- X^i_t \, , \label{eq:VelProcessItoB}\\
	V^{ij}_{2,t} \, dt &= d[X^i,X^j]_t \, \label{eq:VelProcessQVar}.
\end{align}
\par

The velocity field $v_2$ is fixed by the structure relation for $M$, as the structure relation \eqref{eq:StructRelation} implies	
\begin{align}\label{eq:StructureRelationX}
	m \, d[X^i,X^j]_t 
	&= \delta_a^i \, \delta_b^j \, d[M^a,M^b]_t + o(dt)\nonumber\\
	&= \alpha \, \delta_a^i \delta_b^j \delta^{ab} \, dt + o(dt)\nonumber\\
	&= \alpha \, \delta^{ij} \, dt + o(dt) \, .
\end{align}
Hence,
\begin{align}\label{eq:v2fixed}
	v_2^{ij} &= 
	\lim_{dt\rightarrow 0} \E\left[\frac{d[X^i,X^j]}{dt} \right]
	\nonumber\\
	&=
	\frac{\alpha}{m} \, \delta^{ij} \, .
\end{align}

\subsection{Stochastic Action}
In the previous section, we have constructed the stochastic phase spaces. This allows to define Lagrangian functions $L^\circ:T_\circ\R^d \times \mathcal{T} \rightarrow \R$, $L^\pm:T_\pm\R^d \times \mathcal{T}\rightarrow \R$ and action functionals
\begin{align}
	S_\circ(X) &= \E\left[\int_{t_0}^{t_f} L^\circ(x,v_\circ,t) \, dt\right],\\
	S_\pm(X) &= \E\left[\int_{t_0}^{t_f} L^\pm(x,v_\pm,v_2,t) \, dt\right].
\end{align}
We must now find the stochastic equivalent of the classical Lagrangian \eqref{eq:ClassLagrangian}. Since the Stratonovich tangent bundle is similar to the classical tangent bundle, there is a natural choice for the Stratonovich Lagrangian:
\begin{equation}
	L^\circ(x,v_\circ,t) = L(x,v_\circ,t) \,.
\end{equation}
Hence, the Stratonovich Lagrangian associated with the classical Lagrangian \eqref{eq:ClassLagrangian} is 
\begin{equation}\label{eq:StratLag}
	L^\circ(x,v_\circ,t) = \frac{m}{2} \, \delta_{ij} \, v_\circ^i v_\circ^j + q \, A_i(x,t) \, v_\circ^i - \mathfrak{U}(x,t).
\end{equation}
For the It\^o Lagrangians, on the other hand, there is no obvious choice. We can, however, construct these Lagrangians from the Stratonovich Lagrangian by imposing
\begin{equation}\label{StochasticAction}
	S(X) := S_\circ(X) = S_\pm(X) \, .
\end{equation}
In appendix \ref{sec:ItoLagrangian}, we show that this condition implies that the forward and backward It\^o Lagrangians corresponding to the Stratonovich Lagrangian \eqref{eq:StratLag} are given by
\begin{equation}
	L^\pm(x,v_\pm,v_2,t) = L_0^\pm(x,v_\pm,v_2,t) \pm L_\infty (x,v_\circ) 
\end{equation}
with finite part
\begin{equation}\label{eq:ItoLag}
	L_0^\pm(x,v_\pm,v_2,t) = \frac{m}{2} \, \delta_{ij} \, v_\pm^i v_\pm^j + q \, A_i(x,t) \, v_\pm^i \pm \frac{q}{2} \, \p_j A_i(x,t) \, v_2^{ij} - \mathfrak{U}(x,t)
\end{equation}
and a divergent part defined by the integral condition
\begin{equation}\label{eq:ItoLagDiv}
	\E\left[ \int L_\infty (x,v_\circ) \, dt  \right]
	=
	\E\left[ \int \frac{m}{2} \, \delta_{ij} \, d[x^i,v_\circ^j] \right] .
\end{equation}

\subsection{Stochastic Euler-Lagrange Equations}\label{eq:StochEQM}
By minimizing the stochastic action $S(X)$, one can derive the Euler-Lagrange equations. In appendix \ref{Ap:StratELEqs}, this is done in the Stratonovich formulation, which yields the  Stratonovich-Euler-Lagrange equations
\begin{equation}\label{eq:SELeq}
	d_\circ \frac{\p L^\circ}{\p v_\circ^i} = \frac{\p L^\circ}{\p x^i} \,  dt \, .
\end{equation}
For the Lagrangian \eqref{eq:StratLag}, this becomes
\begin{equation}\label{eq:StratEL}
	d_\circ \left( m \, \delta_{ij} V^j_{\circ} + q \, A_i \right) 
	= \left( q \, \p_i A_j \, V^j_{\circ} - \p_i \mathfrak{U} \right) dt\, .
\end{equation}
Hence, using eq.\eqref{eq:VelProcessStrat}, one finds
\begin{align}\label{StochStratEq}
	d_\circ X^i &= V^i_{\circ} \, dt \, , \nonumber\\
	m \, \delta_{ij} \, d_\circ V^j_{\circ} 
	&= q \, F_{ij} \, V_{\circ} \, dt - q \, \p_t A_i \, dt - \p_i \mathfrak{U} \, dt \, .
\end{align}
This is a set of stochastic differential equations in the sense of Stratonovich and is the stochastic equivalent of the classical equation \eqref{eq:ELFirstOrderset}.
\par

Alternatively, one can work in the It\^o formulation. We minimize the action for the It\^o Lagrangians in appendix \ref{Ap:ItoELEqs}, which yields the It\^o-Euler-Lagrange equations
\begin{equation}\label{eq:IELeq}
	d_\pm \frac{\p L^\pm}{\p v_\pm^i} = \frac{\p L^\pm}{\p x^i} \,  dt  \, .
\end{equation}
For the Lagrangian \eqref{eq:ItoLag}, the forward equations become
\begin{equation}\label{eq:ItoEL}
	d_+ \left( m \, \delta_{ij} V^j_{+} + q \, A_i \right) 
	= \left( q \, \p_i A_j \, V^j_{+} + \frac{q}{2} \, \p_i \p_j A_k \, V^{jk}_{2} - \p_i \mathfrak{U} \right) dt \, .
\end{equation}
Hence, using eqs.~\eqref{eq:VelProcessItoF} and \eqref{eq:VelProcessQVar}, one finds
\begin{align}\label{StochItoEq}
	d_+ X^i &= V^i_{+} \, dt \, , \nonumber\\
	d[X^i,X^j] &= V^{ij}_{2} \, dt \, ,\nonumber\\
	m \, \delta_{ij} \, d_+ V^j_{+} 
	&= 
	q \, F_{ij} \, V_{+}^j \, dt
	+ \frac{q}{2} \, \p_k F_{ij} \, V_2^{jk} \, dt
	- q \, \p_t A_i \, dt 
	- \p_i \mathfrak{U} \, dt \, .
\end{align}
This is a set of stochastic differential equations in the sense of It\^o, and is equivalent to the Stratonovich equation \eqref{StochStratEq}.
\par

Finally, by a similar calculation, one can derive the equivalent set of backward equations. These are given by
\begin{align}\label{StochItoEqb}
	d_- X^i &= V^i_{-} \, dt \, ,\nonumber\\
	d[X^i,X^j] &= V^{ij}_{2} \, dt \, , \nonumber\\
	m \, \delta_{ij} \, d_- V^j_{-} 
	&= 
	q \, F_{ij} \, V^j_{-} \, dt
	- \frac{q}{2} \, \p_k F_{ij} \, V_2^{jk} \, dt
	- q \, \p_t A_i \, dt 
	- \p_i \mathfrak{U} \, dt \, .
\end{align}

\subsection{Boundary Conditions}\label{sec:Boundary}
The stochastic Euler-Lagrange equations, that were derived in previous section, admit solutions. However, uniqueness of the solutions can only be achieved, if appropriate initial conditions are specified. Such initial conditions can be obtained by measuring the particle at an initial time $t_0$, but the measurement process is different for deterministic and stochastic theories.
\par 

In a deterministic theory, one can in theory perform a measurement of both the position $X_{t_0}$ and the velocity $V_{t_0}$ with an infinite precision. However, in practice, one will always have a finite uncertainty on such measurements, since any measurement device has a finite precision. Moreover, a measurement requires an interaction between the measurement device and the particle, which alters the state of the particle, and thus induces an additional uncertainty.
\par

In a stochastic theory, the position process is almost surely not differentiable. Therefore, one cannot provide an initial condition for the process $V_t$. Nevertheless, one can still obtain unique solutions to the stochastic equations of motion.\footnote{Uniqueness refers to the uniqueness of the stochastic process $X:\mathcal{T}\times(\Omega,\Sigma,\mathbb{P})\rightarrow(\M,\mathcal{B}(\M))$. When this stochastic process is evolved in time, it will follow one of all possible sample paths $X(\cdot,\omega) : \mathcal{T} \rightarrow \M$, where $\omega\in\Omega$ is chosen according to the probability measure $\mathbb{P}$. Uniqueness of the process does not imply uniqueness of the sample path.} 
The reason for this is that the initial condition that must be seeded into the stochastic differential equation is an initial probability measure $\mu_{X_{t_0}}$. This measure can be obtained from a probability density $\rho(X_{t_0})$, using that $d\mu = \rho \, d^dx$, which itself can be obtained by measuring the moments $\E[X_{t_0}^n]$ for all $n\in\mathbb{N}$.
\par

The moments $\E[X_{t_0}^n]$ can in theory be measured up to an infinite precision, but in practice there will always be a finite experimental uncertainty, as is the case in deterministic theories. In principle, one could also measure the moments $\E[V_{t_0}^n]$ of the velocity process, which would provide the initial measure for the velocity process $\mu_{V_{t_0}}$. However, this would be in contradiction with the the non-differentiability of $X$, which implied that initial measure for the process $\mu_{V_{t_0}}$ does not exist.
\par 

The resolution of this paradox is that the moments $\E[X_{t}^n]$ and $\E[V_{t}^n]$ can be measured at any time $t$, but the moments of $X$ and of $V$ cannot be measured simultaneously. Hence, if the measure $\mu_{X_{t_0}}$ is constructed, the measure $\mu_{V_{t_0}}$ does not exist and vice versa. This feature leads to a theoretical uncertainty, which is inherently different from the classical experimental uncertainties that we discussed earlier. In the next section, we show that this theoretical uncertainty, that can be formulated as an uncertainty principle, is a generic feature of stochastic theories.

\subsection{The Momentum Process}\label{sec:MomentumProcess}
In the previous sections, we have studied a stochastic process $X:\mathcal{T}\times (\Omega,\Sigma,\mathbb{P})\rightarrow(\R^d,\mathcal{B}(\R^d))$ describing the position of a particle with mass $m$ charged under a scalar potential $\mathfrak{U}$ and a vector potential $A_i$ with charge $q$. We will now construct a dual process $P$, which we call the momentum process. 
\par 

As reviewed in appendix \ref{ap:ReviewStochProc}, the stochastic process $X$ is a family of random variables $\{X_t \, | \, t\in\mathcal{T}\}$, and, for every $t\in\mathcal{T}$, $X_t\in L_t^2(\Omega,\Sigma,\mathbb{P})$. This allows to consider elements of the dual space, $\tilde{P}_t\in L_t^2(\Omega,\Sigma,\mathbb{P})^\ast$, which are maps $\tilde{P}_t:L_t^2(\Omega,\Sigma,\mathbb{P})\rightarrow \R$. Since $L^2$-spaces are self-dual, there exists a family of isomorphisms $\phi_t:(L_t^2)^\ast\rightarrow L_t^2$. Hence, we can define a momentum process $\{P_t \, | \, t\in\mathcal{T}\}$, where  $P_t=\phi_t(\tilde{P}_t):(\Omega,\Sigma,\mathbb{P})\rightarrow(\R^d,\mathcal{B}(\R^d))$ is a momentum random variable for every $t\in\mathcal{T}$.
\par 

We can consider the process $(X,P)$ on the cotangent bundle $T^\ast\R^d\cong\R^{2d}$. When studying this process, we would like to condition $(X,P)$ on its previous or future states. For this, we require the notion of a filtration, cf. appendix \ref{ap:filtration}. However, the processes $X$ and $P$ are mutually incompatible in the sense that they are not adapted to each others filtration.
\par 

We already encountered this incompatibility when studying the process $(X,V)$ on the tangent bundle $T\R^d\cong\R^{2d}$. There, we noted that the process $V$ is not differentiable, which necessitated the introduction of It\^o processes $V_\pm$ and a Stratonovich process ${V_\circ=\frac{1}{2}(V_+ + V_-)}$. In a position representation,\footnote{i.e. with respect to the to the natural filtration $\mathcal{F}^X$ of $X$.} these velocity processes are integrable, such that
\begin{equation}
	\E[V_t \, | X_t] = v(X_t,t) \qquad \forall \, t\in\mathcal{T} \, ,
\end{equation}
but not square integrable, as
\begin{equation}
	\E\left[ |V_t|^2 \, | X_t\right] = \infty \qquad \forall \, t\in\mathcal{T}.
\end{equation}
\par 

The same reasoning can be applied to the momentum processes: one can introduce It\^o processes $P_\pm$ and a Stratonovich process $P_\circ=\frac{1}{2}(P_++P_-)$, and the expected value for these processes is given by
\begin{equation}
	\E[P_{t} \, | X_t] = p(X_t,t) \qquad \forall \, t\in\mathcal{T}\, ,
\end{equation}
but
\begin{equation}\label{eq:momentumdivergence}
	\E\left[ |P_{t}|^2 \, | X_t\right] = \infty \qquad \forall \, t\in\mathcal{T}.
\end{equation}
\par 

Alternatively, one could work in a momentum representation,\footnote{i.e. with respect to the to the natural filtration $\mathcal{F}^P$ of $P$.} and introduce It\^o position processes $X_\pm$ and a Stratonovich process $X_\circ=\frac{1}{2} (X_+ + X_-)$. The expected value for these processes is given by 
\begin{equation}
	\E[X_{t} \, | P_t] = x(P_t,t) \qquad \forall \, t\in\mathcal{T}\, ,
\end{equation}
but
\begin{equation}\label{eq:positiondivergence}
	\E\left[ |X_{t}|^2 \, | P_t\right] = \infty \qquad \forall \, t\in\mathcal{T}.
\end{equation}
\par 

We can estimate the divergences appearing in the second moment of $P$ in the position representation and in the second moment of $X$ in the momentum representation.\footnote{In this book, we work in a position representation. Therefore, we only discuss the divergence appearing in \eqref{eq:momentumdivergence}, but a similar reasoning can be applied to \eqref{eq:positiondivergence}.}
\par

Suppose that we are given the state $M_0$ of a Wiener process at time $t_0$. We can calculate its quadratic variation for any time $t \geq t_0$, and find
\begin{equation}
	[M^a,M^b]_t = \int_{t_0}^t d[M^a,M^b]_s = \frac{\alpha }{m} \, (t-t_0) \, \delta^{ab}\, .
\end{equation}
This expression yields the covariance matrix of the process $M$, given by
\begin{align}
	{\rm Cov}_{t_0}(M^a_t,M^b_t) 
	&= \E [ M^a_t M^b_t \, | \, M_0 ] - \E[ M^a_t \, | \, M_0 ] \, \E[ M^b_t \, | \, M_0 ] \nonumber\\
	&= \frac{\alpha }{m} \, (t-t_0) \, \delta^{ab}\, .
\end{align}
\par 

The velocity process $\dot{M}$ is not adapted to $M$. Therefore, the quadratic variation of $\dot{M}$ is not well defined in the position representation. We can, however, write down a formal expression by inverting the time parameter. This yields
\begin{equation}\label{eq:QVarP}
	[\dot{M}^a,\dot{M}^b]_t = \frac{\alpha }{m} \, (t-t_0)^{-1} \, \delta^{ab} \, .
\end{equation}
As expected, this term contains a divergence at $t_0$, but we have now identified the order of this divergence.
Furthermore, the covariance matrix follows from this expression and is given by
\begin{align}
	{\rm Cov}_{t_0}(\dot{M}^a_t,\dot{M}^b_t) 
	&= \E [ \dot{M}^a_t \dot{M}^b_t \, | \, M_0 ] - \E[ \dot{M}^a_t \, | \, M_0 ] \, \E[ \dot{M}^b_t \, | \, M_{0} ] \nonumber\\
	&= \frac{\alpha }{m} \, (t-t_0)^{-1} \, \delta^{ab} \, .
\end{align}
Then, for any time $t$, the product of the covariance matrices is given by
\begin{equation}
	{\rm Cov}(\dot{M}^a,\dot{M}^b) \, {\rm Cov}(M^c,M^d)  
	= \frac{\alpha^2}{m^2} \, \delta^{ab} \, \delta^{cd} \, .
\end{equation}
This defines an uncertainty principle, as it provides a minimal theoretical uncertainty on the joint process $(M,\dot{M})$, which takes values in $T_\circ\R^d$. This derivation can be made more rigorous using the isomorphisms $\phi_t:(L_t^2)^\ast\rightarrow L_t^2$. These induce a map from the characteristic functional $\varphi_{M_t}$ to the probability density $\rho_{\dot{M}_t}$ implying that $\rho_{M_t}$ and $\rho_{\dot{M}_t}$ are related by a Fourier transform.
\par 

We will now consider the quadratic covariation of the processes $M$ and $\dot{M}$. As is the case for the quadratic variation of $\dot{M}$ in the position representation, this covariation is not well defined, but we can give a formal expression. For this, we first write the quadratic variation of $\dot{M}$ in its differential form:
\begin{align}
	d[\dot{M}^a,\dot{M}^b]_t 
	&= \frac{\alpha}{m} \, \delta^{ab} \, d(t-t_0)^{-1} \nonumber\\
	&= - \frac{\alpha}{m} \, (t-t_0)^{-2} \, \delta^{ab} \, dt \, .
\end{align}
In a similar fashion, we find that the covariation of $\dot{M}$ and $M$ is given by
\begin{align}\label{QVARPX}
	d[\dot{M}^a,M^b]_t 
	&= \frac{\alpha}{m} \, \delta^{ab} \, (t-t_0) \, d(t-t_0)^{-1} \, \nonumber\\
	&= - \frac{\alpha}{m} \, (t-t_0)^{-1} \, \delta^{ab} \, dt 
\end{align}
and
\begin{align}\label{QVARXP}
	d[M^a,\dot{M}^b]_t 
	&= \frac{\alpha}{m} \, \delta^{ab} \, (t-t_0)^{-1} \, d(t-t_0) \, \nonumber\\
	&= + \frac{\alpha}{m} \, (t-t_0)^{-1} \, \delta^{ab} \, dt  \, .
\end{align}
Here, we note that the covariation $[M,\dot{M}]$ is not symmetric, which reflects the mutual incompatibility of the processes $M$ and $\dot{M}$.\footnote{The limit in the definition \eqref{eq:defQVar} of the quadratic covariation does not converge to a unique value for $[M,\dot{M}]$.} Furthermore, the sign of these expressions depends on a time ordering convention, such that the opposite time ordering leads to opposite signs.
\par 

Eqs. \eqref{QVARPX} and \eqref{QVARXP} suggest a non-commutativity of $M$ and $\dot{M}$. This can be made explicit by interpreting the commutator as the difference of a time ordered product, such that
\begin{align}\label{eq:commutator}
	[M^a,\dot{M}^b]
	&= \lim_{dt\rightarrow 0} M^a_{t+dt} \, \dot{M}^b_t - \dot{M}^b_{t+dt} \, M^a_t \nonumber\\
	&= \lim_{dt\rightarrow 0} \frac{ M^a_{t+dt} \, (M^b_{t+dt}-M^b_t) - (M^b_{t+dt}-M^b_{t}) \, M^a_t }{dt} \nonumber\\
	&= \lim_{dt\rightarrow 0} \frac{(M^a_{t+dt} - M^a_t) \, (M^b_{t+dt} - M^b_t)}{dt} \, \nonumber\\
	&= \lim_{dt\rightarrow 0} \frac{d[M^a,M^b]_t}{dt} \nonumber\\
	&= \frac{\alpha}{m} \, \delta^{ab}.
\end{align}
\par 

The expressions \eqref{QVARPX} and \eqref{QVARXP} for the covariation of $M$ and $\dot{M}$ can be used to give meaning to the covariation $[X,V_\circ]$ that is encountered in the divergent part of the It\^o Lagrangian \eqref{eq:ItoLagDiv}. This covariation can be derived by identifying $P=P_\circ$ and using that the momentum $P$ of $X=C+M$ is given by
\begin{equation}
	P_i = \frac{\p L}{\p v^i} = m \, \delta_{ij} V^j + q \, A_i \, .
\end{equation}
Moreover, the commutator \eqref{eq:commutator} implies a commutation relation for the position and momentum, which is given by
\begin{align}
	[X^i,P_j] 
	&= m \, [X^i,\delta_{jk} V^k + q \, A_j(X)] \nonumber\\
	&= m \, \delta_{jk} \, [X^i, V^k]_c + m \, q \, [X^i,A_j(X)] \nonumber\\
	&= m \, \delta_{jk} \delta^i_a \delta^k_b \, [M^a, \dot{M}^b]
	+ m \, q \, \p_k A_j(X) \, \delta^i_a \, \delta^k_b [M^a,M^b] \nonumber\\
	&= \alpha \, \delta^i_j \, .
\end{align}

\subsection{Stochastic Hamilton-Jacobi Equations}
In this section, we construct the stochastic equations of motion in the Hamilton-Jacobi formalism. We define Hamilton's principal function for $L^+$ as
\begin{align}\label{eq:HamPrinFnct+}
	S^+(x,t) 
	&= S^+(x,t;x_f,t_f) \nonumber\\
	&= -\E\left[\int_{t}^{t_f} L^+(X_s,V_{+,s},V_{2,s},s)\, ds \, \Big| \, X_t = x, X_{t_f} = x_f\right] ,
\end{align}
where $(X_s,V_s)$ is a solution of the stochastic It\^o-Euler-Lagrange equations passing through $(x,t)$ and $(x_f,t_f)$. Moreover, for  $L^-$, we define the principal function as
\begin{align}\label{eq:HamPrinFnct-}
	S^-(x,t)
	&= S^-(x,t;x_0,t_0) \nonumber\\
	&= \E\left[\int_{t_0}^{t} L^-(X_s,V_{-,s},V_{2,s},s)\, ds \, \Big| \, X_t = x, X_{t_0} = x_0 \right] ,
\end{align}
where $(X_s,V_s)$ is a solution of the It\^o-Euler-Lagrange equations passing through $(x_0,t_0)$ and $(x,t)$.
\par

Using Hamilton's principal function, one can derive a set of stochastic Hamilton-Jacobi equations, which is done in appendix \ref{ap:HamJac}. We find
\begin{equation}\label{eq:HamJacStoch}
	\begin{cases}
		\frac{\p}{\p x^i} S^\pm(x,t) &= p^\pm_{i}\\
		\frac{\p}{\p t} S^\pm(x,t) &= - H_0^\pm(x,p^\pm,\p p^\pm,t)
	\end{cases}
\end{equation}
with the momentum given by
\begin{equation}
	p^\pm_{i} = \frac{\p L_0^\pm(x,v_\pm,v_2,t)}{\p v_\pm^i}
\end{equation}
and the Hamiltonian by
\begin{equation}
		H^\pm_0(x,p^\pm,\p p^\pm,t) = p^\pm_{i} v_\pm^i \pm \frac{1}{2} \p_j p^\pm_{i} \, v_2^{ij} - L^\pm_0(x,v_\pm,v_2,t) \, .
\end{equation}
For the It\^o Lagrangian \eqref{eq:ItoLag}, this yields
\begin{equation}\label{eq:StochHamJac}
	\begin{cases}
		\frac{\p S^\pm}{\p x^i} = m \, \delta_{ij} v_\pm^j + q \, A_i \, ,\\
		\frac{\p S^\pm}{\p t} = - \frac{m}{2} \, \delta_{ij} v_\pm^i v_\pm^j \mp \frac{m}{2}\, \delta_{ij} v_2^{ik} \p_k v_\pm^j - \mathfrak{U} \, .
	\end{cases}
\end{equation}
In addition, we find an integral constraint for the velocity field $v_\pm(X_t,t)$ given by
\begin{equation}\label{eq:IntConstr}
	\oint\left(  p^\pm_i \, v_\pm^i \pm \frac{1}{2} \, v_2^{ij} \, \p_j p^\pm_i \right) dt
	=
	\pm \, \E\left[ \oint L_\infty(X_s,V_{\circ,s}) \, ds \, \Big| \, X_t\right] .
\end{equation}
For the divergent Lagrangian \eqref{eq:ItoLagDiv}, the right hand side yields
\begin{align}
	\E\left[ \oint_\gamma L_\infty \, ds \, \Big| \, X_t \right]
	&=
	\E\left[ \oint_\gamma \frac{m}{2}  \, \delta_{ij} \, d[X^i,V_\circ^j]_{s} \, \Big| \, X_t \right] \nonumber\\
	&=
	\E\left[ \oint_\gamma  \frac{1}{2} \, \Big( d[X^i,P_{\circ,i}]_{s} - q \, \p_j A_i \, d[X^i,X^j]_s \Big) \Big| \, X_t \right] \nonumber\\
	&=
	\frac{\alpha}{2} \, \oint_\gamma \left( \frac{\delta^i_i}{s-t} - \frac{q}{m} \, \p_i A^i \right) ds  \nonumber\\
	&=
	\alpha \, \pi \, \ri \, k^i_i  \, ,
\end{align}
where $k^i_j\in\mathbb{Z}^{d\times d}$ is a matrix of winding numbers counting the number of times the loop $\gamma$ winds around the pole at $s=t$.
\par 

As was done in the classical theory, we can rewrite the equations \eqref{eq:StochHamJac} into a form that is equivalent to the It\^o-Euler-Lagrange equations \eqref{StochItoEq} and \eqref{StochItoEqb}. This can be achieved by taking a spatial derivative of the second equation and plugging in the first equation. This yields
\begin{equation}\label{eq:HamJacIntermediateStoch}
	m \, \delta_{ij} \, \frac{\p}{\p t} v_\pm^j + q \, \frac{\p}{\p t} A_i 
	= - m \, \delta_{jk} v_\pm^k \p_i v_\pm^j \mp \frac{m}{2} \delta_{jk} v_2^{kl} \p_l \p_i v_\pm^j - \p_i \mathfrak{U} \, .
\end{equation}
Then, using eq.~\eqref{eq:pivjpjvi}, i.e.
\begin{equation*}
	m \, \delta_{jk} \, \p_i v_\pm^j 	
	= m \, \delta_{ij}\, \p_k v_\pm^j - q \, F_{ik} \, ,
\end{equation*}
we find
\begin{equation}\label{eq:ELStochFieldNot}
	\left[
		m \, \delta_{ij} \left(
			\p_t 
			+ v_\pm^k \p_k 
			\pm \frac{1}{2} v_2^{kl} \p_l \p_k
		\right)
		- q \, F_{ij} 
	\right] v_\pm^j  
	= 
	\pm \frac{q}{2} \, v_2^{jk} \p_k F_{ij}
	- q \, \p_t A_i 
	- \p_i \mathfrak{U} \, .
\end{equation}
When supplemented with
\begin{align}
	d_\pm X^i_t &= v_\pm^i(X_t,t) \, dt + \delta^i_a \, dM^a_t \, ,\\
	d[M^a,M^b]_t &= \frac{\alpha}{m} \, \delta^{ab} \, dt \, ,
\end{align}
this set of equations is equivalent to the It\^o-Euler-Lagrange equations \eqref{StochItoEq} and \eqref{StochItoEqb} with the process $V_{\pm,t}$ replaced by the field $v_\pm(X_t,t)$.

\subsection{Diffusion Equations}\label{sec:diffusionreal}
As anticipated in chapter \ref{sec:Class}, one can derive a partial differential equation for Hamilton's principal functions $S^\pm(X_t,t)$. Combining the stochastic Hamilton-Jacobi equations \eqref{eq:StochHamJac} and plugging in the expression \eqref{eq:v2fixed} for $v_2$ yields
\begin{align}\label{eq:DtochHamJacScrod}
	- 2 \, m \, \frac{\p S^\pm}{\p t}
	&=
	\p_i S^\pm \, \p^i S^\pm
	\pm \alpha \, \p_i\p^i S^\pm
	- 2\, q \, A^i \, \p_i S^\pm
	\mp \alpha \, q \, \p_i A^i
	+ q^2 \, A_i A^i
	+ 2 \, m \, \mathfrak{U}\, .
\end{align}
This is the generalization of eq.~\eqref{eq:ClassDiffEq} to a stochastic theory.
\par

We can define the wave functions
\begin{equation}\label{WaveFunction}
	\Psi_\pm(x,t) 
	:= \exp\left( \pm \frac{ S^\pm(x,t)}{\alpha} \right) .
\end{equation}
Then, eq.~\eqref{eq:DtochHamJacScrod} implies that $\Psi_+(x,t)$ is a solution of the time reversed heat equation
\begin{equation}\label{eq:diffusionBack}
	- \alpha \, \frac{\p}{\p t} \Psi
	=
	\left[
		\frac{\delta^{ij}}{2 \, m} 
		\left(\alpha \, \frac{\p}{\p x^i} - q \, A_i \right)
		\left(\alpha \, \frac{\p}{\p x^j} - q \, A_j \right)
		+ \mathfrak{U}
	\right] \Psi
\end{equation}
subjected to a terminal condition $\Psi_+(x,t_f)=u_f(x)$.
Moreover, $\Psi_-(x,t)$ is a solution of the heat equation
\begin{equation}\label{eq:diffusionFward}
	\alpha \, \frac{\p}{\p t} \Psi
	=
	\left[
	\frac{\delta^{ij}}{2 \, m} 
	\left(\alpha \, \frac{\p}{\p x^i} + q \, A_i \right)
	\left(\alpha \, \frac{\p}{\p x^j} + q \, A_j \right)
	+ \mathfrak{U}
	\right] \Psi
\end{equation}
subjected to an initial condition $\Psi_-(x,t_0)=u_0(x)$. Thus, we can associate solutions of the heat equation to solutions of the stochastic Hamilton-Jacobi and Euler-Lagrange equations.
\par

Conversely, we would like to associate a stochastic process $X$ to any solution of the diffusion equations \eqref{eq:diffusionBack} and \eqref{eq:diffusionFward}. Here, we encounter a small caveat, as Hamilton's principal function is multi-valued: due to the integral constraint \eqref{eq:IntConstr}, Hamilton's principal function defines an equivalence class $[S^\pm]$ under the equivalence relation
\begin{equation}
	\tilde{S}^\pm \sim  S^\pm \qquad {\rm if} \quad \tilde{S}^\pm = S^\pm + \alpha \, \pi \, \ri \, \sum_{i=1}^d k_i \,  \quad {\rm with} \; k_i\in \mathbb{Z} \, . 
\end{equation}
This implies that the wave functions \eqref{WaveFunction} define equivalence classes $[\Psi_\pm]$ with equivalence relations
\begin{equation}\label{eq:EquivRel}
	\begin{cases}
		\tilde{\Psi}_+ \sim \Psi_+ \qquad {\rm if} \qquad \tilde{\Psi}_+ = \pm \Psi_+ \, ,\\
		\tilde{\Psi}_- \sim \Psi_- \qquad {\rm if} \qquad \tilde{\Psi}_- = \pm \Psi_- \, ,
	\end{cases}
\end{equation}
which must be taken into account when formulating the converse statement.
\par

Hence, we find that, for any solution $\Psi_+$ of eq.~\eqref{eq:diffusionBack} and $\Psi_-$ of eq.~\eqref{eq:diffusionFward} modulo the equivalence relation \eqref{eq:EquivRel}, one can construct velocity fields
\begin{equation}
	v_\pm^i = \frac{\delta^{ij}}{m} \, \Big( \pm \alpha \, \p_j \ln\Psi_\pm - q \, A_j \Big)\, ,
\end{equation}
such that solutions $X$ of the It\^o equation 
\begin{align}
	d_\pm X^i_t &= v_\pm^i(X_t,t) \, dt + \delta^i_a \, dM^a_t \\
	d[M^a,M^b]_t &= \frac{\alpha}{m} \, \delta^{ab} \, dt 
\end{align}
minimize the stochastic action $S(X)$ associated to Hamilton's principal function
\begin{equation}\label{WavefunctionInverse}
	S_\pm = \alpha \, \ln \Psi_\pm \, .
\end{equation}
\par

We conclude that there is a correspondence between solutions of the diffusion equations \eqref{eq:diffusionBack} and \eqref{eq:diffusionFward} with $\alpha>0$ modulo the equivalence relation \eqref{eq:EquivRel} and semi-martingale processes $X$ that minimize the stochastic action $S(X)$ and satisfy the structure relation \eqref{eq:StructureRelationX}. This correspondence is similar to the one established by the Feynman-Kac theorem \cite{FKac}, where the time reversed heat equation \eqref{eq:diffusionBack} is the Kolmogorov backward equation for the process $X$, when evolved forward in time, and the heat equation \eqref{eq:diffusionFward} is the Kolmogorov backward equation for the process $X$, when evolved backward in time.
\par

We conclude this section by pointing out that in earlier formulations of stochastic mechanics, the divergent part of the Lagrangian is discarded, as it does not contribute to the equations of motion for the process $X$. However, when this is done, one does not have an equivalence between solutions of the of the diffusion equations and solutions of the Hamilton-Jacobi equations.
\par 

Indeed, for any solution $\Psi$ of the diffusion equations, one can construct an equivalent solution $\Psi \, \exp( 2  \, \pi \, \ri \, \sum_{i=1}^d k_i)$, which imposes an equivalence relation on Hamilton's principal functions $S^\pm$. However, when the divergent part of the Lagrangian is ignored, such an equivalence relation is not present in the theory and must be imposed by hand.
\par 

This issue is known as Wallstrom's criticism of stochastic mechanics \cite{Wallstrom:1988zf,WallstromII}. Here, we have resolved this criticism, as we have shown that such an equivalence relation follows from the integral condition \eqref{eq:IntConstr}, which is part of the theory, when the divergent part of the Lagrangian is properly taken into account.

\clearpage
\section{Complex Stochastic Dynamics on $\R^d$}\label{sec:StochDynC}
In the previous chapter, we studied stochastic processes on $\R^d$. More precisely, we studied semi-martingale processes $X = C + M$ that can be decomposed in a deterministic trajectory $C$ and a noise term $M$. Then, we imposed the process $M$ to be a Wiener process using the structure relation \eqref{eq:StructRelation}, and we used the stationary action principle to derive equations of motion for the process $X$. Finally, using the equations of motion, we derived a correspondence between solutions of real diffusion equations and semi-martingale processes that minimize the stochastic action.
\par

Given the fact that the Schr\"odinger equation is a complex diffusion equation, one might wonder whether our analysis can be generalized, such that a correspondence between a generalized stochastic process and complex diffusion equations is obtained. In this chapter, we show that this can be done, when the martingale $M$ is complexified. We will do this by complexifying the fibers to $F=\mathbb{C}^d$, while keeping the base space $\M=\R^d$ real.
\par

In order to incorporate complex martingale processes, while keeping the process $X\in\R^d$ real, we will consider a semi-martingale that can be decomposed as 
\begin{align}
	X_t^i 
	&= C^{i}_{x,t} + \delta^i_a \, {\rm Re} \left[ M_t^a\right]\nonumber\\
	&= C^{i}_{x,t} + \delta^i_a \, M_{x,t}^a \, ,
\end{align}
where $C_{x,t}$ is a real valued continuous trajectory with finite variation on $\R^d$ and $M$ is a complex valued martingale on $\mathbb{C}^d$, i.e.
\begin{equation}
	M_t = M_{x,t} + \ri \, M_{y,t} \, ,
\end{equation}
where $M_x,M_y$ are real $d$-dimensional martingales.
\par

We will make the assumption that $M_x$ and $M_y$ are L\'evy processes\footnote{Note that $M_t$ is not a complex L\'evy process, as we do not assume that $M_x$ and $M_y$ are independent.} and therefore also Markov processes. As before, we will further specify $M$ using a structure relation. In particular, we will consider
\begin{align}\label{eq:ComplexStructRelation}
	m \, d[M^a,M^b]_t &=  \alpha \, \delta^{ab} \, dt \nonumber\\
	&=  |\alpha| \, e^{\ri \, \phi}\, \delta^{ab} \, dt
\end{align}
with $\alpha\in\mathbb{C}$. Hence, compared to the structure relation \eqref{eq:StructRelation} from previous chapter, we have added the complex factor $e^{\ri \, \phi}$.
As we are studying a complex process, we must also specify the structure relations for the conjugated process $\overline{M}$. A general complex structure relation is of the form
\begin{equation}
	d
	\begin{pmatrix}
		[M^a,M^b]_t & [M^a,\overline{M}{}^b]_t\\
		[\overline{M}{}^a,M^b]_t & [\overline{M}{}^a,\overline{M}{}^b]_t
	\end{pmatrix}
	=
	\frac{\delta^{ab}}{m} 
	\begin{pmatrix}
		\alpha & |\alpha| + \gamma\\
		|\alpha|+ \gamma & \overline{\alpha}
	\end{pmatrix}
	\, dt \, .
\end{equation}
with\footnote{These are necessary and sufficient conditions for the existence of the complex martingale $M$. For $\alpha=0$, the process is the complex Wiener process, which is an example of a conformal martingale.} $\alpha\in\mathbb{C}$ and $\gamma\in[0,\infty)$. Such a general process requires new degrees of freedom, as it defines a diffusion in $\mathbb{C}^d$ instead of $\R^d$. We can resolve this issue by setting $\gamma=0$, such that
\begin{equation}\label{eq:complexStrucRel}
	d
	\begin{pmatrix}
		[M^a,M^b]_t & [M^a,\overline{M}{}^b]_t\\
		[\overline{M}{}^a,M^b]_t & [\overline{M}{}^a,\overline{M}{}^b]_t
	\end{pmatrix}
	=
	\frac{\delta^{ab}}{m} 
		\begin{pmatrix}
			\alpha & |\alpha|\\
			|\alpha| & \overline{\alpha}
		\end{pmatrix}
		\, dt \, .
\end{equation}
For this choice, the structure relations are degenerate, which implies that the complex process $M\in\mathbb{C}^n$ has only $d$ degrees of freedom instead of $2d$, such that $M$ is restricted to the subspace $e^{\frac{\ri \, \phi}{2}}\times \R^d \subset \mathbb{C}^d$. Hence, the structure relation \eqref{eq:complexStrucRel} characterizes a real Wiener process that is rotated in the complex plane over an angle $\frac{\phi}{2}$.
\par

We can also define an auxiliary process that can be decomposed as
\begin{align}
	Y_t^i 
	&= C^{i}_{y,t} + \delta^i_a \, {\rm Im} \left[ M_t^a\right]\nonumber\\
	&= C^{i}_{y,t} + \delta^i_a \, M_{y,t}^{a}\, ,
\end{align}
where $C_{y,t}$ is another $\R^d$-valued continuous trajectory with finite variation. 
In addition, we write
\begin{equation}
	Z_t = X_t + {\rm i} \, Y_t 
\end{equation}
and
\begin{equation}
	C_t = C_{x,t} + {\rm i} \, C_{y,t} \,.
\end{equation}
Using eq.~\eqref{eq:complexStrucRel}, we find the quadratic covariation of the processes $X$ and $Y$, which is given by
\begin{equation}\label{eq:QuadVarComplexProjections}
	d
	\begin{pmatrix}
		[X^a,X^b]_t & [X^a,Y^b]_t\\
		[Y^a,X^b]_t & [Y^a,Y^b]_t
	\end{pmatrix}
	=
	\frac{|\alpha|}{2 \, m} \, \delta^{ab} 
	\begin{pmatrix}
		1 + \cos \phi & \sin \phi\\
		\sin \phi & 1 - \cos \phi
	\end{pmatrix}
	\, dt \, .
\end{equation}
\par

As in previous chapter, one can derive the equations of motion using a Stratonovich formulation, a forward It\^o formulation or a backward It\^o formulation. In order to do so, we must introduce the velocity fields. These are now given by the complex fields 
\begin{equation}
	w^i(X_t,t) = v^i(X_t,t) + \ri \, u^i(X_t,t).
\end{equation}
Since the stochastic motion is not differentiable, we obtain a forward and a backward velocity field
\begin{align}
	w_+^i(X_t,t) &= \lim_{h\rightarrow 0} \E \left[\frac{ Z^i_{t+dt} - Z^i_t}{dt} \, \Big| \, X_t \right],\\
	w_-^i(X_t,t) &= \lim_{h\rightarrow 0} \E \left[\frac{ Z^i_t - Z^i_{t-dt}}{dt} \, \Big| \, X_t \right].
\end{align}
In addition, we define a Stratonovich velocity field
\begin{align}
	w_\circ^i(X_t,t) &= \lim_{h\rightarrow 0} \E \left[\frac{ Z^i_{t+dt} - Z^i_{t-dt}}{2 \, dt} \, \Big| \, X_t \right]\nonumber\\
	&= \frac{1}{2} \Big[ w_+^i(X_t,t) + w_-^i(X_t,t)\Big]
\end{align}
and a second order velocity field
\begin{equation}
	w_2^{ij}(X_t,t) 
	= \lim_{h\rightarrow 0} \E \left[\frac{ [Z^i_{t+dt} - Z^i_t] [Z^j_{t+dt} - Z^j_t]}{dt} \, \Big| \, X_t \right] ,
\end{equation}
which can be decomposed as
\begin{equation}
	w_2^{ij}(X_t,t) = v_2^{ij}(X_t,t) + \ri \, u_2^{ij}(X_t,t) \, ,
\end{equation}
and is fixed by the structure relation \eqref{eq:ComplexStructRelation}, such that
\begin{equation}\label{eq:w2fixed}
	w_2^{ij}(X_t,t) = \frac{\alpha}{m} \, \delta^{ij}\, .
\end{equation}

\subsection{Stochastic Action}
As in previous chapter, we must specify the phase space in order to define a Lagrangian. This phase space can be obtained by complexifying the tangent bundles from previous chapter. The Stratonovich bundle becomes $T_\circ^\mathbb{C}\R^d \cong \R^d \times \mathbb{C}^d$ and the It\^o bundles become $T_\pm^\mathbb{C}\R^d \cong \R^d \times \mathbb{C}^{\frac{d(d+3)}{2}}$. This allows to study processes  $(X_t,W_{\circ,t})$ on $T_\circ^\mathbb{C}\R^d$ and $(X_t,W_{\pm,t},\pm W_{2,t})$ on $T_\pm^\mathbb{C}\R^d$, using the relations\footnote{As in previous chapter, these differential expressions are defined by an integral expression similar to the one given in eq.~\eqref{eq:VelProcess}.}
\begin{align}
	W^i_{\circ,t} \, dt &= d_\circ Z^i_t \, ,\\
	W^i_{+,t} \, dt &= d_+ Z^i_t \, ,\\
	W^i_{-,t} \, dt &= d_- Z^i_t \, ,\\
	W^{ij}_{2,t} \, dt &= d[Z^i,Z^j]_t \,.
\end{align}
\par

We construct Lagrangians on these complexified tangent bundles by replacing the real velocity fields with complex fields. The Stratonovich Lagrangian is thus given by
\begin{equation}\label{SLagcomplex}
	L^\circ(x,w_\circ,t) = \frac{m}{2} \, \delta_{ij} \, w_\circ^i w_\circ^j + q \, A_i(x,t) \, w_\circ^i - \mathfrak{U}(x,t)\, ,
\end{equation}
and the It\^o Lagrangians by
\begin{equation}\label{ILagcomplex}
	L^\pm(x,w_\pm,w_2,t) = L_0^\pm(x,w_\pm,w_2,t) \pm L_\infty (x,w_\circ)
\end{equation}
with finite part
\begin{equation}
	L_0^\pm(x,w_\pm,w_2,t) = \frac{m}{2} \, \delta_{ij} \, w_\pm^i w_\pm^j + q \, A_i \, w_\pm^i \pm \frac{q}{2} \, \p_j A_i \, w_2^{ij} - \mathfrak{U}
\end{equation}
and a divergent part that satisfies the integral condition
\begin{equation}
	\E\left[ \int L_\infty (x,w_\circ) \, dt \right]
	=
	\E\left[ \int \frac{m}{2} \, \delta_{ij} \, d[z^i,w_\circ^j] \right] .
\end{equation}

\subsection{Equations of Motion}
All results from the real diffusion theory can be generalized to the complex case.  The Stratonovich-Euler-Lagrange equations \eqref{eq:SELeq} for the Lagrangian \eqref{SLagcomplex} are given by
\begin{align}
	d_\circ Z^i &= W^i_{\circ} \, dt\, , \nonumber\\
	m \, \delta_{ij} \, d_\circ W^j_{\circ}
	&= q \, F_{ij} \, W^j_{\circ} \, dt - q \, \p_t A_i \, dt - \p_i \mathfrak{U} \, dt \, .
\end{align}
Moreover, the It\^o-Euler-Lagrange equations \eqref{eq:IELeq} for the Lagrangian \eqref{ILagcomplex} become
\begin{align}
	d_\pm Z^i &= W^i_{\pm} \, dt \, ,\nonumber\\
	d[Z^i,Z^j] &= W^{ij}_{2} \, dt \, ,\nonumber\\
	m \, \delta_{ij} \, d_\pm W^j_{\pm} 
	&= 
	q \, F_{ij} \, W^j_{\pm} \, dt
	\pm \frac{q}{2} \, \p_k F_{ij} \, W^{jk}_2 \, dt
	- q \, \p_t A_i \, dt 
	- \p_i \mathfrak{U} \, dt \, . \label{IELComplex}
\end{align}
Hamilton's principal functions are complex valued functions given by
\begin{align}\label{eq:HamPrincC}
	S^+(x,t) 
	&= 
	-\E\left[\int_{t}^{t_f} L^+(X_s,W_{+,s},W_{2,s},s)\, ds \, \Big| \, X_t = x, X_{t_f} = x_f\right] ,
	\nonumber\\
	S^-(x,t)
	&= \E\left[\int_{t_0}^{t} L^-(X_s,W_{-,s},W_{2,s},s)\, ds \, \Big| \, X_t = x, X_{t_0} = x_0 \right],
\end{align}
and can be used to derive the Hamilton-Jacobi equations. These are given by equation \eqref{eq:HamJacStoch} with complex momenta and velocities. For the Lagrangian \eqref{ILagcomplex}, we then find
\begin{equation}
	\left[
	m \, \delta_{ij} \left(
	\p_t
	+ w_\pm^k \p_k 
	\pm \frac{1}{2} w_2^{kl} \p_l \p_k
	\right)
	- q \, F_{ij} 
	\right] w_\pm^j  
	= 
	\pm \frac{q}{2} \, w_2^{jk} \p_k F_{ij}
	- q \, \p_t A_i 
	- \p_i \mathfrak{U} \, ,
\end{equation}
which is equivalent to eq.~\eqref{IELComplex} with $W_t$ replaced by $w(X_t,t)$. In addition, the Hamilton-Jacobi equations provide an integral constraint for the velocity field $w_\pm(X_t,t)$, given by
\begin{equation}\label{eq:ComplexIntegralCond}
	\oint\left(  p^\pm_i \, w_\pm^i \pm \frac{1}{2} \, w_2^{ij} \p_j p^\pm_i \right) dt
	=
	\pm \, \alpha \, \pi \, \ri \, k^i_i \, ,
\end{equation}
where $k^i_j\in\mathbb{Z}^{d\times d}$ is a matrix of winding numbers.
\par 

Taking into account this integral constraint, the Hamilton-Jacobi equations can be solved for the velocity fields $w_\pm$, and the solution can be plugged into the stochastic differential equation
\begin{equation}
	\begin{cases}
	d_\pm Z^i_t &= w_\pm^i(X_t,t) \, dt + \delta^i_a \, dM^a_t \, ,\\
	d[M^a,M^b]_t &= \frac{\alpha}{m} \, \delta^{ab} \, dt \, .
	\end{cases}
\end{equation}
This It\^o equation can be solved for $Z$ yielding solutions $X_t={\rm Re}[Z_t]$ and $Y_t={\rm Im}[Z_t]$, which describe respectively the particle and the auxiliary process. 

\subsection{Boundary Conditions}\label{sec:Boundaries}
The complex stochastic Euler-Lagrange equations that were derived in the previous section admit solutions, and these are unique, when appropriate boundary conditions are provided. However, in the complex theory, there are new complex degrees of freedom, whose initial conditions must also be specified.
\par 

In the deterministic theory, which can be obtained by taking the limit $\alpha\rightarrow 0$, one must provide initial or terminal conditions for the positions $X,Y$ and velocities $V,U$. As discussed in section \ref{sec:Boundary}, such conditions for $X,V$ can be obtained by measurement, but there is no obvious measurement for the complex position $Y$ and velocity $U$. However, the equations of motion are independent of the complex position $Y$. Therefore, one can obtain unique solutions for $(X,V,U)$ without specifying boundary conditions for $Y$. Despite this simplification, one must still provide initial or terminal conditions for the complex velocity $U$ or equivalently for the field $u(x,t)$. We will return to this point and provide a physical interpretation of the complex velocity field $u(x,t)$ in section \ref{sec:Qfoam}.
\par 

In the stochastic theory, one must provide the initial or terminal probability measures $\mu_X$ and $\mu_Y$, but, by a similar reasoning, one only requires the initial or terminal measure $\mu_{X}$ supplemented with initial or terminal conditions for the velocity fields $u_\pm(x,t)$. As in the deterministic case, the fields, $u_\pm$ do not have a physical interpretation yet, but an interpretation will be provided in section \ref{sec:Qfoam} for the field $u_\circ=\frac{1}{2}(u_++u_-)$. This interpretation fixes both $u_+$ and $u_-$, since the four velocity fields $v_+$, $v_-$, $u_+$ and $u_-$ are not independent, due to the fact that the martingale $M$ is restricted to $e^{\frac{i \, \phi}{2}}\times \R^d \subset \mathbb{C}^d$. As a consequence, the velocity process $W$ is non-differentiable along this direction, but differentiable perpendicular to this hyperplane, which imposes the constraint
\begin{equation}\label{eq:ConstraintCVelFields}
	\cos \left( \frac{\phi}{2} \right) u_+^i - \sin \left( \frac{\phi}{2} \right) v_+^i
	=
	\cos \left( \frac{\phi}{2} \right) u_-^i - \sin \left( \frac{\phi}{2} \right) v_-^i \, .
\end{equation}
Therefore, there are only three independent velocity fields and, if $\phi\neq\pi$, one can take these fields to be $v_+$, $v_-$ and $u_\circ$.
\par 

We point out that $\phi=0$ for a Brownian motion, which implies that eq.~\eqref{eq:ConstraintCVelFields} reduces to
\begin{equation}
	u_+(x,t) = u_-(x,t) \, .
\end{equation}
The real Brownian motion from chapter \ref{sec:StochDynR} can then be recovered by adding the initial condition $u_\circ(x,t_0)=0$, as this ensures that $u_\pm(x,t)=0$ for all $t\in\mathcal{T}$.

\subsection{Diffusion Equations}
Combining the stochastic Hamilton-Jacobi equations \eqref{eq:StochHamJac}, with the real velocities $v$ replaced by complex velocity fields $w$, and plugging in eq.~\eqref{eq:w2fixed} for $w_2$ yields a complex partial differential equation
\begin{align}
	- 2 \, m \, \frac{\p S^\pm}{\p t}
	&=
	\p_i S^\pm \, \p^i S^\pm
	\pm \alpha \, \p_i\p^i S^\pm
	- 2 \, q \, A^i \, \p_i S^\pm
	\mp \alpha \, q \, \p_i A^i
	+ q^2 \, A_i A^i
	+ 2 \, m \, \mathfrak{U}\, .
\end{align}
This equation implies that the wave functions
\begin{equation*}
	\Psi_\pm(x,t) 
	= \exp\left( \pm \frac{ S^\pm(x,t)}{\alpha} \right)
\end{equation*}
satisfy complex diffusion equations. In particular, $\Psi_+(x,t)$ is a solution of the complex diffusion equation
\begin{equation}\label{eq:diffusionBwardComp}
	- \alpha \, \frac{\p}{\p t} \Psi
	=
	\left[
	\frac{\delta^{ij}}{2 \, m} 
	\left(\alpha \, \frac{\p}{\p x^i} - q \, A_i \right)
	\left(\alpha \, \frac{\p}{\p x^j} - q \, A_j \right)
	+ \mathfrak{U}
	\right] \Psi \, ,
\end{equation}
which reduces to the heat equation for $\alpha=-1$ and to the Schr\"odinger equation for $\alpha=-\ri$.
\par

Similarly, $\Psi_-(x,t)$ is a solution of the complex diffusion equation
\begin{equation}\label{eq:diffusionFwardComp}
	\alpha \, \frac{\p}{\p t} \Psi
	=
	\left[
	\frac{\delta^{ij}}{2 \, m} 
	\left(\alpha \, \frac{\p}{\p x^i} + q \, A_i \right)
	\left(\alpha \, \frac{\p}{\p x^j} + q \, A_j \right)
	+ \mathfrak{U}
	\right] \Psi \, ,
\end{equation}
which reduces to the heat equation for $\alpha=1$ and to the Schr\"odinger equation for $\alpha=\ri$.
\par

As in previous chapter, this result implies a correspondence between solutions of the diffusion equations \eqref{eq:diffusionBwardComp} and \eqref{eq:diffusionFwardComp} modulo the equivalence relation \eqref{eq:EquivRel},
\begin{equation*}
	\begin{cases}
		\tilde{\Psi}_+ \sim \Psi_+ \qquad {\rm if} \qquad \tilde{\Psi}_+ = \pm \Psi_+ \,\\
		\tilde{\Psi}_- \sim \Psi_- \qquad {\rm if} \qquad \tilde{\Psi}_- = \pm \Psi_- \, ,
	\end{cases}
\end{equation*}
and semi-martingale processes $Z$ that solve the It\^o equation
\begin{align}
	d_\pm Z^i_t &= w_\pm^i(X_t,t) \, dt + \delta^i_a \, dM^a_t \, \nonumber\\
	d[M^a,M^b]_t &= \frac{\alpha}{m} \, \delta^{ab} \, dt \, 
\end{align}
with velocity
\begin{equation}\label{eq:VelFromWave}
	w_\pm^i = \frac{\delta^{ij}}{m} \, \Big( \pm \alpha \, \p_j \ln\Psi_\pm - q \, A_j \Big)\, .
\end{equation}

\clearpage
\section{Relativistic Stochastic Dynamics on $\R^{d,1}$}\label{sec:RLT}
In previous chapters, we have discussed non-relativistic stochastic dynamics. In this chapter, we will extend this discussion to relativistic theories.
In a relativistic theory, the configuration space $\R^d$ is extended to the Minkowski space $\R^{d,1}$ with coordinates ${\mathbf{x}=(x^0,\vec{x})=(c \, t,\vec{x})}$ and $\vec{x}\in\R^d$, and one studies trajectories $\{X_\lambda:\lambda\in\mathcal{T}\}$ parameterized by an affine parameter $\lambda$ on this Minkowski space.
\par

The dynamics of a relativistic theory is governed by a Lagrangian $L:T\R^{d,1}\times \R^+\rightarrow \R$ defined on the tangent bundle $T\R^{d,1}\cong\R^{2d,2}$. The relativistic Lagrangian corresponding to the non-relativistic Lagrangian \eqref{eq:ClassLagrangian} is given by
\begin{equation}\label{eq:RelLag}
	L(x,v,\varepsilon) = \frac{1}{2\, \varepsilon} \, \eta_{\mu\nu} \, v^\mu v^\nu - \frac{\varepsilon \, m^2}{2} + q \, A_\mu(x) \, v^\mu \, ,
\end{equation}
where $\eta_{\mu\nu}={\rm diag}(-1,+1,...,+1)$ is the Minkowski metric, and $\varepsilon \in  (0,\infty)$ is an auxiliary variable that ensures that the action is invariant under affine reparameterizations
\begin{align}\label{eq:reparInv}
	\lambda &\rightarrow \tilde{\lambda} = a\, \lambda + b \, ,\nonumber\\
	\varepsilon &\rightarrow \tilde{\varepsilon} = \frac{\varepsilon}{a} \, .
\end{align}
\par

The relativistic Euler-Lagrange equations can be obtained by extremizing the action. For the Lagrangian \eqref{eq:RelLag}, this yields
\begin{align}
	\frac{dX^\mu_\lambda}{d\lambda} &= V_\lambda^\mu \, , \\
	\frac{1}{\varepsilon} \, \eta_{\mu\nu} \, \frac{dV_\lambda^\nu}{d\lambda} 
	&=
	q \, F_{\mu\nu}(X_\lambda) \, V_\lambda^\nu \, ,\\
	\eta_{\mu\nu} V_\lambda^\mu V_\lambda^\nu &= - \varepsilon^2 \, m^2 \, c^2 \, .
\end{align}
In addition, one must gauge fix the Lagrange multiplier $\varepsilon$ with a gauge condition that depends on the mass:
\begin{itemize}
	\item If $m^2>0$, we fix $\varepsilon=|m|^{-1}$, which fixes the affine parameter to be the proper time, i.e. $\lambda=\tau$.
	\item If $m=0$, we fix\footnote{For $m=0$, there is no canonical choice that fixes the affine parameter to a physical parameter, as the strict equality $m=0$ has already reduced the dimension of the phase space. Another common choice in the literature is $\varepsilon=1$ for which $[\lambda]=T/M$.} $\varepsilon=c^2 \, E(\lambda)^{-1}$, where $E$ is the energy of the particle and $\lambda$ is measured in units of time, i.e. $[\lambda]=T$.
	\item If $m^2<0$, we fix $\varepsilon= c^{-1} \, |m|^{-1}$, which fixes the affine parameter to be the proper length, i.e. $\lambda=s$.
\end{itemize}
\par

The classical relativistic theory can be generalized to a stochastic theory by considering semi-martingale processes
\begin{equation}\label{eq:DMDecomp}
	X^\mu_\lambda = C^\mu_{x,\lambda} + \delta^\mu_\alpha \, {\rm Re}[ M^\alpha_\lambda] \, ,
\end{equation}
where $X$ and $C_{x}$ are $\R^{d,1}$-valued and $M$ is $\mathbb{C}^{d+1}$-valued.
As in previous chapters, we can then fix the stochastic law of $X$ by imposing the structure relation\footnote{For $m=0$ with gauge choice $\varepsilon=c^2 \, E(\lambda)^{-1}$, $M^\alpha_\lambda$ is no longer a L\'evy process, as it does not have stationary increments. This does not affect the results, as our analysis requires continuity in probability and independent increments, but does not require stationarity of the increments. Moreover, $M$ can be made into a L\'evy process by changing the gauge to $\varepsilon=1$.}
\begin{equation}\label{eq:QuadVarRel}
	d[M^\alpha,M^\beta]_\lambda = \alpha \, \varepsilon  \, \eta^{\alpha\beta} \, d\lambda
\end{equation}
with $\alpha\in \mathbb{C}$, where we use natural units in which $\hbar,c=1$. Hence, we obtain
\begin{equation}\label{eq:QuadVarRelX}
	d[Z^\mu,Z^\nu]_\lambda 
	= \alpha \, \varepsilon \, \eta^{\mu\nu} \, d\lambda + o(d\lambda)\, ,
\end{equation}
where $Z=X + \ri \, Y$ and $Y^\mu= C^\mu_{y} + \delta^\mu_\alpha \, {\rm Im}[ M^\alpha]$ defines an auxiliary process.

\subsection{Equations of Motion}
Starting from the relativistic classical Lagrangian \eqref{eq:RelLag}, all results from chapter \ref{sec:StochDynC} can be generalized to relativistic theories.
The Stratonovich Lagrangian is given by
\begin{equation}\label{SLagcomplexRLT}
	L^\circ(x,w_\circ,\varepsilon) = \frac{1}{2 \, \varepsilon} \, \eta_{\mu\nu} \, w_\circ^\mu w_\circ^\nu - \frac{\varepsilon \, m^2}{2} + q \, A_\mu(x) \, w_\circ^\mu
\end{equation}
and the It\^o Lagrangian by
\begin{equation}\label{ILagcomplexRLT}
	L^\pm(x,w_\pm,w_2,\varepsilon) = L_0^\pm(x,w_\pm,w_2,\varepsilon) \pm L_\infty(x,w_\circ,\varepsilon)
\end{equation}
with finite part
\begin{equation}
	L_0^\pm(x,w_\pm,w_2,\varepsilon) = \frac{1}{2 \, \varepsilon} \, \eta_{\mu\nu} \, w_\pm^\mu w_\pm^\nu - \frac{\varepsilon \, m^2}{2} + q \, A_\mu \, w_\pm^\mu \pm \frac{q}{2} \, \p_\nu A_\mu \, w_2^{\mu\nu}
\end{equation}
and a divergent part that  is defined by the integral condition
\begin{equation}
	\E\left[ \int L_\infty^\pm (x,w_\circ,\varepsilon) \, d\lambda \right]
	=
	\E\left[\int \frac{1}{2 \, \varepsilon} \, \eta_{\mu\nu} \, d[z^\mu,w_\circ^\nu] \right] .
\end{equation}
\par 

The Stratonovich-Euler-Lagrange equations \eqref{eq:SELeq} for the Lagrangian \eqref{SLagcomplexRLT} are
\begin{align}
	d_\circ Z^\mu
	&= 
	W^\mu_{\circ} \, d\lambda  \, ,\\
	\eta_{\mu\nu} \, d_\circ W^\nu_{\circ}
	&= 
	\varepsilon \, q \, F_{\mu\nu} \, W^\nu_{\circ} \, d\lambda \, , \\
	\E\left[ \eta_{\mu\nu} \, W^\mu_{\circ} W^\nu_{\circ}\right] 
	&= - \varepsilon^2 \, m^2 \, .
\end{align}
Moreover, the It\^o-Euler-Lagrange equations \eqref{eq:IELeq} for the Lagrangian \eqref{ILagcomplexRLT} are
\begin{align}
	d_\pm Z^\mu	
	&= 
	W^\mu_{\pm} \, d\lambda  \, ,\\
	d[Z^\mu,Z^\nu]
	&=
	W^{\mu\nu}_{2} \, d\lambda \, , \\
	\eta_{\mu\nu} \, d_\pm W^\nu_{\pm}
	&= 
	\varepsilon \, q \, F_{\mu\nu} \, W^\nu_{\pm} \, d\lambda
	\pm \frac{\varepsilon \, q}{2} \, \p_\rho F_{\mu\nu} \, W^{\nu\rho}_{2} \, d\lambda \, ,\label{eq:IELRLT}\\
	\E\left[ \eta_{\mu\nu} \, W^\mu_{\pm} W^\nu_{\pm}\right] 
	&= - \varepsilon^2 \, m^2 \, . \label{eq:IELRLTConst}
\end{align}
\par 

Hamilton's principal functions are complex valued functions given by
\begin{align}
	S^+(x,\varepsilon,\lambda) 
	&= 
	- \E\left[\int_{\lambda}^{\lambda_f} L^+(X_\tau,W_{+,\tau},W_{2,\tau},\mathcal{E}_\tau)\, d\tau \, \Big| X_\lambda = x, X_{\lambda_f} = x_f, \mathcal{E}_{\lambda} = \varepsilon, \mathcal{E}_{\lambda_f} = \varepsilon_f \right]
	\nonumber\\
	S^-(x,\varepsilon,\lambda)
	&= \E\left[\int_{\lambda_0}^{\lambda} L^-(X_\tau,W_{-,\tau},W_{2,\tau},\mathcal{E}_\tau)\, d\tau \, \Big| X_\lambda = x, X_{\lambda_0} = x_0, \mathcal{E}_{\lambda} = \varepsilon, \mathcal{E}_{\lambda_0} = \varepsilon_0  \right]
\end{align}
and can be used to derive the Hamilton-Jacobi equations. These are given by a relativistic version of \eqref{eq:HamJacStoch}:
\begin{equation}\label{eq:HamJacStochRLT}
	\begin{cases}
		\frac{\p}{\p x^\mu} S^\pm(x,\varepsilon,\lambda) &= p^\pm_{\mu}\, ,\\
		\left( \frac{\p}{\p \lambda} + \frac{d\varepsilon(\lambda)}{d\lambda}  \, \frac{\p}{\p \varepsilon} \right) S^\pm(x,\varepsilon,\lambda) &= - p^\pm_{\mu} w_\pm^\mu \mp \frac{1}{2} \p_\nu p^\pm_{\mu} \, w_2^{\mu\nu} + L^\pm_0(x,w_\pm,w_2,\varepsilon)\, ,
	\end{cases}
\end{equation}
where the momenta $p^\pm$ are complex valued, and $w_2$ is determined by eq.~\eqref{eq:QuadVarRelX}, such that
\begin{equation}\label{eq:w2fixedRLT}
	w_2^{\mu\nu}(X_\lambda) = \alpha \, \varepsilon \, \eta^{\mu\nu}\, .
\end{equation}
Furthermore, the reparameterization invariance of the theory, given in eq.~\eqref{eq:reparInv}, implies the constraint
\begin{equation}
	\frac{\p S^\pm}{\p \lambda} + \frac{d \varepsilon}{d\lambda} \frac{\p S^\pm}{\p \varepsilon}
	= 0 \, .
\end{equation}
In addition, the velocity field $w_\pm(X_\lambda)$ satisfies the integral constraint
\begin{equation}\label{eq:IntegralconstRLT}
	\oint\left(  p^\pm_\mu \, w_\pm^\mu \pm \frac{1}{2} \, w_2^{\mu\nu} \, \p_\nu p^\pm_\mu \right) d\lambda
	=
	\pm \, \alpha \, \pi \, \ri \, k^\mu_\mu \, ,
\end{equation}
where $k^\mu_\nu\in\mathbb{Z}^{n\times n}$ is a matrix of winding numbers.
\par 

By taking a covariant derivative of the second Hamilton-Jacobi equation, and plugging in the Lagrangian \eqref{ILagcomplexRLT}, one finds
\begin{equation}
	\left[
	\eta_{\mu\nu} \left(
	w_\pm^\rho \p_\rho 
	\pm \frac{1}{2} w_2^{\rho\sigma} \p_\sigma \p_\rho
	\right)
	- \varepsilon \, q \, F_{\mu\nu}
	\right] w_\pm^\nu  
	= 
	\pm \frac{\varepsilon \, q}{2} \, w_2^{\nu\rho} \p_\rho F_{\mu\nu} \, ,
\end{equation}
which is equivalent to eq.~\eqref{eq:IELRLT} with $W$ replaced by $w(X)$. Taking into account the integral constraint \eqref{eq:IntegralconstRLT}, this equation can be solved for the velocity fields $w_\pm$. The solution can be plugged into the stochastic differential equation
\begin{equation}\label{eq:ItoRLT}
	\begin{cases}
		d_\pm Z^\mu_\lambda &= w_\pm^\mu(Z_\lambda) \, d\lambda + \delta^\mu_\alpha \, dM^\alpha_\lambda \, ,\\
		d[M^\alpha,M^\beta]_\lambda &= \alpha \, \varepsilon \, \eta^{\alpha\beta} \, d\lambda \, ,
	\end{cases}
\end{equation}
which can be solved for the process $X={\rm Re}[Z]$.
\par

We note that the velocity fields $w_\pm$ are also subjected to a relativistic generalization of the constraint \eqref{eq:ConstraintCVelFields}, which is given by
\begin{align}\label{eq:ConstraintCVelFieldsRLT}
	\sin \left( \frac{\phi}{2} \right) u_+^0 + \cos \left( \frac{\phi}{2} \right) v_+^0
	&=
	\sin \left( \frac{\phi}{2} \right) u_-^0 + \cos \left( \frac{\phi}{2} \right) v_-^0 \, , \nonumber\\
	\cos \left( \frac{\phi}{2} \right) u_+^i - \sin \left( \frac{\phi}{2} \right) v_+^i
	&=
	\cos \left( \frac{\phi}{2} \right) u_-^i - \sin \left( \frac{\phi}{2} \right) v_-^i \, .
\end{align}

\subsection{Diffusion Equations}
The stochastic relativistic Hamilton-Jacobi equations \eqref{eq:HamJacStochRLT} can be combined. Using the expression \eqref{eq:w2fixedRLT} for $w_2$, this yields the complex partial differential equation
\begin{align}
	- \frac{2}{\varepsilon} \, \frac{\p S^\pm}{\p \lambda}
	&=
	\p_\mu S^\pm \, \p^\mu S^\pm
	\pm \alpha \, \p_\mu\p^\mu S^\pm
	- 2 \, q \, A^\mu \, \p_\mu S^\pm
	\mp \alpha \, q \, \p_\mu A^\mu
	+ q^2 \, A_\mu A^\mu
	+ m^2 \, .
\end{align}
If we define the wave functions
\begin{align}\label{WaveFunctionRLT}
	\Psi_\pm(x,\varepsilon,\lambda) 
	&:= 
	\Phi_\pm(x) 
	\exp\left( \pm \frac{\varepsilon \, m^2}{2 \, \alpha} \, \lambda \right)
	,\nonumber\\
	\Phi_\pm(x) 
	&:= \exp\left( \pm \frac{ S^\pm(x)}{\alpha} \right),
\end{align}
we find that $\Psi_\pm$ satisfy the complex diffusion equations 
\begin{align}
	- \alpha \, \frac{\p}{\p \lambda} \Psi_+
	&=
	\frac{\varepsilon}{2} \, \eta^{\mu\nu}
	\left(\alpha \, \frac{\p}{\p x^\mu} - q \, A_\mu \right)
	\left(\alpha \, \frac{\p}{\p x^\nu} - q \, A_\nu \right)
	\Psi_+ \, ,
	\\
	\alpha \, \frac{\p}{\p \lambda} \Psi_-
	&=
	\frac{\varepsilon}{2}
	\, \eta^{\mu\nu}
	\left(\alpha \, \frac{\p}{\p x^\mu} + q \, A_\mu \right)
	\left(\alpha \, \frac{\p}{\p x^\nu} + q \, A_\nu \right)
	\Psi_- \, ,
\end{align}
while $\Phi_\pm$ satisfy the complex wave equations
\begin{align}
	\left[
	\eta^{\mu\nu}
	\left(\alpha \, \frac{\p}{\p x^\mu} - q \, A_\mu \right)
	\left(\alpha \, \frac{\p}{\p x^\nu} - q \, A_\nu \right)
	+ m^2
	\right] \Phi_+ 
	&=
	0\, ,
	\\
	\left[
	\eta^{\mu\nu}
	\left(\alpha \, \frac{\p}{\p x^\mu} + q \, A_\mu \right)
	\left(\alpha \, \frac{\p}{\p x^\nu} + q \, A_\nu \right)
	+ m^2
	\right] \Phi_- \, 
	&=
	0.
\end{align}
For $\alpha=\pm \ri$, these wave equations reduce to the Klein-Gordon equation. As in previous chapters, there is a correspondence between the solutions of these equations and semi-martingale processes $Z$ that solve the It\^o equation \eqref{eq:ItoRLT}
with velocity
\begin{equation}
	w_\pm^\mu = \varepsilon \, \eta^{\mu\nu}\, \Big( \pm \alpha \, \p_\nu \ln\Psi_\pm - q \, A_\nu \Big)\, .
\end{equation}

\clearpage

\section{Stochastic Dynamics on pseudo-Riemannian Manifolds}\label{sec:Manifolds}
In the previous chapters, we have discussed stochastic dynamics on the Euclidean space $\R^d$ and the Minkowski space $\R^{d,1}$. In this chapter, we will extend the stochastic theory to the case, where the underlying space is no longer flat, but given by a pseudo-Riemannian manifold $(\M,g)$. Here, we will focus on relativistic theories on $n$-dimensional Lorentzian manifolds $(\M,g)$ with $n=d+1$, but all the results in this chapter can be generalized in a straightforward manner to non-relativistic theories on $d$-dimensional Riemannian manifolds $(\M,g)$.
\par 

The main difference between a relativistic theory on $\R^{d,1}$ and a relativistic theory on the $n$-dimensional Lorentzian manifold  $(\M,g)$ with $n=d+1$ is that the Minkowski metric $\eta_{\mu\nu}$ is promoted to a smooth symmetric bilinear map $g_{\mu\nu}:T\M \times T\M \rightarrow \R $.
\par

As a consequence, the classical Euler-Lagrange equations are modified to incorporate the curvature of spacetime and are given by
\begin{align}
	\frac{dX^\mu_\lambda}{d\lambda} &= V_\lambda^\mu \, , \\
	g_{\mu\nu}(X_\lambda) \left( \frac{dV_\lambda^\nu}{d\lambda} 
	+ \Gamma^\nu_{\rho\sigma}(X_\lambda) \, V^\rho_\lambda V^\sigma_\lambda 
	\right)
	&=
	\varepsilon \, q \, F_{\mu\nu}(X_\lambda) \, V_\lambda^\nu \, ,\\
	g_{\mu\nu}(X_\lambda)\, V_\lambda^\mu V_\lambda^\nu &= - \varepsilon^2 \, m^2 \,,
\end{align}
where $\Gamma$ is the Christoffel symbol associated to the Levi-Civita connection and where we use natural units $\hbar=c=1$.
\par

In the stochastic theory, we promote the trajectories $\{X_\lambda:\lambda\in\mathcal{T}\}$ to $\M$-valued semi-martingale processes. These are processes that can, in any  local coordinate chart, be decomposed as
\begin{equation}
	X^\mu_\lambda = C^\mu_\lambda + e^\mu_\alpha(X_\lambda) \, {\rm {Re}}[M^\alpha_\lambda] \, ,
\end{equation}
where $X_\lambda$ and $C_\lambda$ are $\M$-valued and $M_\lambda$ is $\mathbb{C}^n$-valued. Moreover, the polyads $e^\mu_\alpha$ are now defined by the relation
\begin{equation}
	g_{\mu\nu} e^\mu_\alpha e^\nu_\beta = \eta_{\alpha\beta}\, .
\end{equation}
We will again impose the structure relation \eqref{eq:QuadVarRel}, i.e.
\begin{equation*}
	d[M^\alpha,M^\beta]_\lambda 
	= 
	\alpha \, \varepsilon \, \eta^{\alpha\beta} \, d\lambda
\end{equation*}
with $\alpha\in\mathbb{C}$, which implies
\begin{align}\label{eq:QVarRiem}
	d[Z^\mu,Z^\nu]_\lambda 
	= \alpha \, \varepsilon \, g^{\mu\nu}(X_\lambda) \, d\lambda + o(d\lambda)
\end{align}
with $Z_\lambda=X_\lambda+ \ri \, Y_\lambda$ and $Y_\lambda$ an auxiliary $\M$-valued process.

\subsection{Second Order Phase Space}
The major difficulty in extending the stochastic theory to a manifold resides in the fact that the velocity fields
\begin{align}
	w_+^\mu(X_\lambda) &= \lim_{d\lambda\rightarrow 0} \E \left[\frac{ Z^\mu_{\lambda+d\lambda} - Z^\mu_\lambda}{d\lambda} \, \Big| \, X_\lambda \right]\\
	w_-^\mu(X_\lambda) &= \lim_{d\lambda\rightarrow 0} \E \left[\frac{ Z^\mu_\lambda - Z^\mu_{\lambda-d\lambda}}{d\lambda} \, \Big| \, X_\lambda \right]\\
	w_2^{\mu\nu}(X_\lambda) &= \lim_{d\lambda\rightarrow 0} \E \left[\frac{ \left[ Z^\mu_{\lambda+d\lambda} - Z^\mu_\lambda\right] \left[  Z^\nu_{\lambda+d\lambda} - Z^\nu_\lambda \right] }{h} \, \Big| \, X_\lambda \right]
\end{align}
are well defined in every coordinate chart containing $X_\lambda$, but the fields $w_\pm$ are not covariant.
This issue can be resolved using second order geometry, cf. appendix \ref{Ap:2Geometry}. In second order geometry, the $n$-dimensional tangent spaces $T_x\M$ at every point $x$ are extended to $\frac{1}{2}n(n+3)$-dimensional tangent spaces $T_{\pm,x}\M$. Moreover, given a coordinate chart, any vector $w_\pm\in T_{\pm,x}\M$ can be represented with respect to the canonical basis of $T_{\pm,x}\M$ as
\begin{equation}
	w_\pm = w_\pm^\mu \, \p_\mu \pm \frac{1}{2} \, w_2^{\mu\nu} \, \p_{\mu\nu}\,.
\end{equation}
This allows to construct covariant objects
\begin{equation}\label{eq:Covvector}
	\hat{w}_\pm^\mu = w_\pm^\mu \pm \frac{1}{2} \, \Gamma^\mu_{\nu\rho} w_2^{\nu\rho} \, ,
\end{equation}
such that the fields $\hat{w}_\pm^\mu$ transform in a covariant manner.
We note that $\hat{w}_\circ=w_\circ$, which reflects the fact that the Stratonovich field is a first order vector field and thus transforms covariantly. Therefore, at any point $x\in\M$, one can define a Stratonovich tangent space $T_{\circ,x}\M$, and, given a coordinate chart, vectors $w_\circ \in T_{\circ,x}\M$ can be represented with respect to the canonical basis of $T_{\circ,x}\M$ as
\begin{equation}
	w_\circ = w_\circ^\mu \, \p_\mu \, .
\end{equation} 
\par

The Stratonovich and It\^o tangent bundles can be constructed from the respective tangent spaces, such that
\begin{align*}
	T_\circ\M &= \bigsqcup_{x\in\M} T_{\circ,x}\M \, ,\\
	T_\pm\M &= \bigsqcup_{x\in\M} T_{\pm,x}\M \, .
\end{align*}
This allows to study processes $(X_\lambda,W_{\circ,\lambda})$ on $T^\mathbb{C}_\circ\M$ and $(X_\lambda,W_{\pm,\lambda},\pm W_{2,\lambda})$ on $T^\mathbb{C}_\pm\M$ that satisfy\footnote{These differential expressions are defined by an integral expression as in eq.~\eqref{eq:VelProcess}.}
\begin{align*}
	W^\mu_{\circ,\lambda} \, d\lambda &= d_\circ Z^\mu_\lambda \, ,\\
	W^\mu_{\pm,\lambda} \, d\lambda &= d_\pm Z^\mu_\lambda \, ,\\
	W^{\mu\nu}_{2,\lambda} \, d\lambda &= d[Z^\mu,Z^\nu]_\lambda \,.
\end{align*}
Alternatively, one can express the processes on $T^\mathbb{C}_\pm\M$ in an explicitly covariant form $(X_\lambda,\hat{W}_{\pm,\lambda},\pm W_{2,\lambda})$, where
\begin{equation}
	\hat{W}^\mu_{\pm,\lambda} = W^{\mu}_{\pm,\lambda} \pm \frac{1}{2} \Gamma^\mu_{\nu\rho}(X_\lambda) \, W_{2,\lambda}^{\nu\rho}
\end{equation}
and
\begin{equation*}
	\hat{W}^\mu_{\pm,\lambda} \, d\lambda = d_\pm Z^\mu_\lambda \pm \frac{1}{2} \Gamma^\mu_{\nu\rho}(X_\lambda) \, d[Z^\nu,Z^\rho]_\lambda \, .
\end{equation*}

\subsection{Equations of Motion}\label{sec:EQMRiemann}
Using second order geometry, all results from chapters \ref{sec:StochDynR}, \ref{sec:StochDynC} and \ref{sec:RLT} can be generalized to pseudo-Riemannian manifolds. In this section, we state the results for the generalization of the complex relativistic theory discussed in chapter \ref{sec:RLT}.
\par

The Stratonovich Lagrangian is given by
\begin{equation}\label{eq:LagStratMan}
	L^\circ(x,w_\circ,\varepsilon) = \frac{1}{2 \, \varepsilon} \, g_{\mu\nu}(x) \, w_\circ^\mu w_\circ^\nu - \frac{\varepsilon \, m^2}{2} + q \, A_\mu(x) \, w_\circ^\mu \,
\end{equation}
and obeys the Stratonovich-Euler-Lagrange equations
\begin{align}
	d_\circ \frac{\p L^\circ}{\p w_\circ^\mu} &= \frac{\p L^\circ}{\p x^\mu} \,  d\lambda \, ,\\
	\E\left[ \frac{\p L^\circ}{\p \varepsilon}\right] &= 0 \, .
\end{align}
For the Lagrangian \eqref{eq:LagStratMan}, this yields
\begin{align}
	d_\circ Z^\mu
	&= 
	W^\mu_{\circ} \, d\lambda  \, , \\
	g_{\mu\nu} \, \Big( 
		d_\circ W^\nu_{\circ}
		+ \Gamma^\nu_{\rho\sigma} \, W^\rho_{\circ} W^\sigma_{\circ} \, d\lambda
	\Big)
	&= 
	\varepsilon \, q \, F_{\mu\nu} \, W^\nu_{\circ} \, d\lambda \, ,\\
	\E\left[ g_{\mu\nu} \, W^\mu_{\circ} W^\nu_{\circ} \right] 
	&= - \varepsilon^2 \, m^2 \, ,
\end{align}
where the last equation is the energy-momentum relation, which can be rewritten as the stochastic line element
\begin{equation}
	\E\left[ g_{\mu\nu} \, d_{\circ}X^\mu \, d_{\circ}X^\nu \right] 
	= - \varepsilon^2 \, m^2 \, d\lambda^2 \, .
\end{equation}
\par

The generalization of the It\^o formulation to manifolds is less straightforward. The It\^o Lagrangian is derived in appendix \ref{sec:ItoLagrangian}, where it is found that
\begin{equation}
	L^\pm(x,w_\pm,w_2,\varepsilon) = L_0^\pm(x,w_\pm,w_2,\varepsilon) \pm L_\infty(x,w_\circ,\varepsilon) \, 
\end{equation}
with finite part
\begin{align}\label{eq:LagItoMan}
	L_0^\pm(x,w_\pm,w_2,\varepsilon) 
	&= 
	\frac{1}{2 \, \varepsilon} \, g_{\mu\nu}(x) \, 
	\left(w_\pm^\mu \pm \frac{1}{2} \, \Gamma^\mu_{\rho\sigma}(x) \, w_2^{\rho\sigma}  \right)
	\left(w_\pm^\nu \pm \frac{1}{2} \, \Gamma^\nu_{\kappa\lambda}(x) \, w_2^{\kappa\lambda}  \right)
	\nonumber\\
	&\quad
	+ \frac{1}{12 \, \varepsilon} \, \mathcal{R}_{\mu\nu\rho\sigma}(x) \, w_{2}^{\mu\rho} w_{2}^{\nu\sigma} - \frac{\varepsilon \, m^2}{2} 
	\nonumber\\
	&\quad
	+ q \, A_\mu \left( w_\pm^\mu \pm \frac{1}{2} \, \Gamma^\mu_{\nu\rho}(x) \, w_2^{\nu\rho} \right)
	\pm \frac{q}{2} \, \nabla_\nu A_\mu \, w_2^{\mu\nu}
\end{align}
and a divergent part such that
\begin{equation}
	\E \left[ \int L_\infty^\pm (x,w_\circ,\varepsilon) \, d\lambda \right]
	=
	\pm \,\E \left[ \int \frac{1}{2 \, \varepsilon} \, g_{\mu\nu}(x) \, d[z^\mu,w_\circ^\nu] \right] .
\end{equation}
The corresponding It\^o-Euler-Lagrange equations are given by, cf.  appendix \ref{Ap:ItoELEqs},
\begin{align}
	d_\pm \frac{\p L_0^\pm}{\p w_\pm^\mu} 
	&= 
	\frac{\p L_0^\pm}{\p x^\mu} \, d\lambda 
	- \Gamma^\rho_{\mu\nu} \left( \frac{\p L_0^\pm}{\p w_2^{\rho\sigma}} + \frac{\p L_0^\pm}{\p w_2^{\sigma\rho}} \right) \, d[Z^\nu,Z^\sigma]
	\pm \Gamma^\rho_{\mu \nu} \, \frac{\p^2 L_0^\pm}{\p x^\sigma \p w_\pm^\rho} \, d[Z^\nu,Z^\sigma] 
	\nonumber\\
	&\quad
	\pm \Gamma^\rho_{\mu \nu} \, \frac{\p^2 L_0^\pm}{\p w_\pm^\sigma \p w_\pm^\rho} \, d[Z^\nu,W_\pm^\sigma]
	\pm \Gamma^\rho_{\mu \nu} \, \frac{\p^2 L_0^\pm}{\p w_2^{\sigma\kappa} \p w_\pm^\rho} \, d[Z^\nu,W_2^{\sigma\kappa}] \, ,\\
	\E\left[ \frac{\p L_0^\pm}{\p \varepsilon}\right] 
	&= 0 \, .
\end{align}
For the Lagrangian \eqref{eq:LagItoMan}, this yields
\begin{align}
	d_\pm Z^\mu 
	\pm \Gamma^\mu_{\nu\rho} \, d[Z^\nu,Z^\rho] 
	&= 
	\hat{W}^\mu_{\pm} \, d\lambda \, , \\
	d[Z^\mu,Z^\nu]
	&=
	W^{\mu\nu}_{2} \, d\lambda \, ,\\
	{\rm LHS} 
	&= 
	{\rm RHS}  \, \label{eq:IELMan},\\
	\E\left[ g_{\mu\nu} \, \hat{W}^\mu_{\pm} \hat{W}^\nu_{\pm}
	+ \frac{1}{6} \, \mathcal{R}_{\mu\nu\rho\sigma} \, W_2^{\mu\rho} W_2^{\nu\sigma} \right]
	&= - \varepsilon^2 \, m^2 \, \label{eq:IELManConst}
\end{align}
with
\begin{align}
	{\rm LHS} 
	&=
	g_{\mu\nu} \, d_\pm \hat{W}_\pm^\nu
	+ g_{\mu\nu} \Gamma^{\nu}_{\rho\sigma} \hat{W}_\pm^{\rho} \left( \hat{W}_\pm^{\sigma} - \frac{1}{2} \, \Gamma^\sigma_{\kappa\lambda} W_2^{\kappa\lambda} \right) d\lambda
	\pm g_{\mu\nu} \Gamma^{\nu}_{\rho\sigma} \, d[Z^\rho,\hat{W}_\pm^\sigma]
	\nonumber\\
	&\quad
		\pm \frac{1}{2} \left[ 
		\mathcal{R}_{\mu\rho\sigma\nu}  
		+ g_{\mu\kappa} \left(
			\p_\rho \Gamma^\kappa_{\sigma\nu} 
			+ \Gamma^\kappa_{\rho\lambda} \Gamma^\lambda_{\sigma\nu} 
		\right)
	\right]
	\hat{W}_2^\nu \, W_2^{\rho\sigma}  \, d\lambda
	\, , \\
	{\rm RHS} 
	&=
	\varepsilon \, q \, F_{\mu\nu} \, \hat{W}_\pm^{\nu} \, d\lambda
	\pm \frac{\varepsilon \, q}{2} \, \nabla_\rho F_{\mu\nu} \, W_2^{\nu\rho} \, d\lambda
	+ \frac{1}{12} \, \nabla_\mu \mathcal{R}_{\nu\rho\sigma\kappa} \, W_2^{\nu\sigma} W_2^{\rho\kappa} \, d\lambda \, .
\end{align}
Moreover, the energy-momentum relation \eqref{eq:IELManConst} can be rewritten as the stochastic line element
\begin{equation}
	\E\left[ g_{\mu\nu} \left( d_\pm\hat{X}^\mu \, d_\pm\hat{X}^\nu \mp d[X^\mu,X^\nu] \right)
	+ \frac{1}{6} \, \mathcal{R}_{\mu\nu\rho\sigma} \, d[X^\mu,X^\rho] \, d[X^\nu,X^\sigma] \right]
	= - \varepsilon^2 \, m^2 \, d\lambda^2\, .
\end{equation}

\subsection{Hamilton-Jacobi Equations}
Hamilton's principal functions are complex valued functions given by
\begin{align}\label{eq:HamPrincCMan}
	S^+(x,\varepsilon,\lambda) 
	&= 
	- \E\left[\int_{\lambda}^{\lambda_f} L^+(X_\tau,W_{+,\tau},W_{2,\tau},\mathcal{E}_\tau)\, d\tau \, \Big| X_\lambda = x, X_{\lambda_f} = x_f, \mathcal{E}_{\lambda} = \varepsilon, \mathcal{E}_{\lambda_f} = \varepsilon_f \right]
	\nonumber\\
	S^-(x,\varepsilon,\lambda)
	&= \E\left[\int_{\lambda_0}^{\lambda} L^-(X_\tau,W_{-,\tau},W_{2,\tau},\mathcal{E}_\tau)\, d\tau \, \Big| X_\lambda = x, X_{\lambda_0} = x_0, \mathcal{E}_{\lambda} = \varepsilon, \mathcal{E}_{\lambda_0} = \varepsilon_0  \right]
\end{align}
and can be used to derive the Hamilton-Jacobi equations. These are derived in appendix \ref{ap:HamJac} and given by 
\begin{equation}\label{eq:HamJacRiem}
	\begin{cases}
		\frac{\p}{\p x^\mu} S^\pm(x,\varepsilon,\lambda) &= p^\pm_{\mu}\, ,\\
		\left( \frac{\p}{\p \lambda} + \frac{d\varepsilon(\lambda)}{d\lambda} \frac{\p}{\p \varepsilon} \right) S^\pm(x,\varepsilon,\lambda) &= - p^\pm_{\mu} \hat{w}_\pm^\mu \mp \frac{1}{2} \nabla_\nu p^\pm_{\mu} \, w_2^{\mu\nu} + L^\pm_0(x,w_\pm,w_2,\varepsilon)\, ,
	\end{cases}
\end{equation}
where the momentum $p^\pm$ is complex valued, and $w_2$ is determined by eq.~\eqref{eq:QVarRiem}, i.e.
\begin{equation}\label{eq:w2fixedRiem}
	w_2^{\mu\nu}(X_\lambda) = \alpha \, \varepsilon \, g^{\mu\nu}(X_\lambda)\, .
\end{equation}
Furthermore, the reparameterization invariance of the theory, given in eq.~\eqref{eq:reparInv}, implies the constraint
\begin{equation}
	\frac{\p S^\pm}{\p \lambda} + \frac{d \varepsilon}{d\lambda} \frac{\p S^\pm}{\p \varepsilon}
	= 0 \, .
\end{equation}
In addition, the velocity field $\hat{w}_\pm(X_\lambda)$ satisfies an integral constraint given by
\begin{equation}\label{eq:IntConstRiem}
	\oint\left(  p^\pm_\mu \, \hat{w}_\pm^\mu \pm \frac{1}{2} \, w_2^{\mu\nu} \, \nabla_\nu p^\pm_\mu \right) d\lambda
	=
	\pm \, \alpha \, \pi \, \ri \, k^\mu_\mu \, ,
\end{equation}
where $k^\mu_\nu\in\mathbb{Z}^{n\times n}$ is a matrix of winding numbers.
\par

If we take a covariant derivative of the second Hamilton-Jacobi equation, we find
\begin{equation}
	\, g_{\nu\rho} \hat{w}_\pm^\rho \nabla_\mu \hat{w}_\pm^\nu
	\pm \frac{1}{2} \, g_{\nu\rho} w_2^{\rho\sigma}   \nabla_\mu \nabla_\sigma \hat{w}_\pm^\nu
	=
	\frac{1}{12} \, w_2^{\nu\sigma} w_2^{\rho\kappa} \nabla_\mu \mathcal{R}_{\nu\rho\sigma\kappa} \, .
\end{equation}
Then, by applying eq.~\eqref{eq:pivjpjvi} in the form
\begin{align}
	g_{\nu\rho} \, \nabla_\mu \hat{w}_\pm^\nu 
	= g_{\mu\nu}\, \nabla_\rho \hat{w}_\pm^\nu - \varepsilon \, q \, F_{\mu\rho} \, ,
\end{align}
we obtain
\begin{align}\label{eq:HJMan}
	\left[
	g_{\mu\nu} \hat{w}_\pm^\rho \nabla_\rho 
	\pm \frac{1}{2} \, w_2^{\rho\sigma} \, \Big(g_{\mu\nu} \nabla_\rho\nabla_\sigma + \mathcal{R}_{\mu\rho\sigma\nu} \Big)
	- \varepsilon \, q \, F_{\mu\nu} 
	\right] \hat{w}_\pm^\nu  
	&= 
	\frac{1}{12}  w_2^{\nu\sigma} w_2^{\rho\kappa} \nabla_\mu \mathcal{R}_{\nu\rho\sigma\kappa}
	\nonumber\\
	&\quad
	\pm \frac{\varepsilon \, q}{2} \, w_2^{\nu\rho} \, \nabla_\rho F_{\mu\nu} \, ,
\end{align}
which is equivalent to eq.~\eqref{eq:IELMan} with $W_\lambda$ replaced by $w(X_\lambda)$.
\par 

Taking into account the integral constraint \eqref{eq:IntConstRiem}, eq.~\eqref{eq:HJMan} can be solved for the velocity fields $\hat{w}_\pm$. The solution can be plugged into the stochastic differential equation
\begin{equation}
	\begin{cases}
		d_\pm Z^\mu_\lambda &= \left( \hat{w}_\pm^\mu \mp \frac{1}{2}\, \Gamma^\mu_{\nu\rho}  w_2^{\nu\rho} \right) d\lambda + e^\mu_\alpha \, dM^\alpha_\lambda \, ,\\
		d[M^\alpha,M^\beta]_\lambda &= \alpha \, \varepsilon \, \delta^{\alpha\beta} \, d\lambda \, ,
	\end{cases}
\end{equation}
which can be solved for $Z$. The fields $w_\pm$ are subjected to the constraint \eqref{eq:ConstraintCVelFieldsRLT}, which reflects that the boundary conditions for $w_+$ and $w_-$ cannot be chosen completely independently.

\subsection{Diffusion Equations}\label{sec:CovDiffussion}
Combining the stochastic Hamilton-Jacobi equations \eqref{eq:HamJacRiem} and plugging in expression \eqref{eq:w2fixedRiem} for $w_2$ yields a complex partial differential equation
\begin{align}
	- \frac{2}{\varepsilon} \, \frac{\p S^\pm}{\p \lambda}
	&=
	\nabla_\mu S^\pm \, \nabla^\mu S^\pm
	\pm \alpha \, \Box S^\pm
	- 2 \, q \, A^\mu \, \nabla_\mu S^\pm
	\mp \alpha \, q \, \nabla_\mu A^\mu
	+ q^2 \, A_\mu A^\mu
	- \frac{\alpha^2}{6} \, \mathcal{R}
	+ m^2\, .
\end{align}
If we define the wave functions \eqref{WaveFunctionRLT}:
\begin{align*}
	\Psi_\pm(x,\varepsilon,\lambda) 
	&:= 
	\Phi(x) 
	\exp\left( \pm \frac{\varepsilon \, m^2}{2 \, \alpha} \, \lambda \right)
	,\nonumber\\
	\Phi_\pm(x) 
	&:= \exp\left( \pm \frac{ S^\pm(x)}{\alpha} \right),
\end{align*}
we find that $\Psi_\pm$ satisfy the complex diffusion equations 
\begin{align}
	- \alpha \, \frac{\p}{\p \lambda} \Psi_+
	&=
	\frac{\varepsilon}{2} 
	\left[
	\Big(\alpha \, \nabla_\mu - q \, A_\mu \Big)
	\Big(\alpha \, \nabla^\mu - q \, A^\mu \Big)
	- \frac{\alpha^2}{6} \, \mathcal{R}
	\right]
	\Psi_+ \, ,
	\\
	\alpha \, \frac{\p}{\p \lambda} \Psi_-
	&=
	\frac{\varepsilon}{2} 
	\left[
	\Big(\alpha \, \nabla_\mu + q \, A_\mu \Big)
	\Big(\alpha \, \nabla^\mu + q \, A^\mu \Big)
	- \frac{\alpha^2}{6} \, \mathcal{R}
	\right]
	\Psi_- \, ,
\end{align}
while $\Phi_\pm$ satisfy the complex wave equations
\begin{align}
	\left[
	\Big(\alpha \, \nabla_\mu - q \, A_\mu \Big)
	\Big(\alpha \, \nabla^\mu - q \, A^\mu \Big)
	- \frac{\alpha^2}{6} \, \mathcal{R}
	+ m^2
	\right] \Phi_+ 
	&=
	0 \, ,
	\\
	\left[
	\Big(\alpha \, \nabla_\mu + q \, A_\mu \Big)
	\Big(\alpha \, \nabla^\mu + q \, A^\mu \Big)
	- \frac{\alpha^2}{6} \, \mathcal{R}
	+ m^2
	\right]\Phi_- \, 
	&=
	0 \, .
\end{align}
For $\alpha=\pm \ri$ these equations reduce to the Klein-Gordon equation including the Pauli-DeWitt term \cite{Pauli:1973,DeWitt:1957}, which represents a non-minimal coupling $\xi=\frac{1}{6}$ to the Ricci scalar. For $n=4$, this coupling coincides with the conformal coupling $\xi=\frac{n-2}{4(n-1)}$, but this is not true for any other dimension.

\subsection{Spacetime Symmetries}
Any non-relativistic theory is locally invariant under the action of the Galilean group ${\rm Gal}(d)$, which can be written as a Cartesian product
\begin{equation}
	{\rm Gal}(d) = O(d) \times \R^d \times \R^d \times \R
\end{equation}
with group multiplication
\begin{equation}
	(R',v',a',s') \, (R,v,a,s)
	=
	(R'\, R, \, R' \, v + v', \, R' \, a + a' + v' \, s,\, s'+s)
\end{equation}
for all $R,R'\in O(d)$, $v,v'\in(\R^d,+)$, $a,a'\in(\R^d,+)$ and $s,s'\in(\R,+)$. The left action of this group on the space $\R^{d+1}$ is given by
\begin{equation}
	(R,v,a,s) \, (x,t) = (R \, x + v \, t + a, \, t+s)
\end{equation}
for all $(R,v,a,s)\in {\rm Gal}(d)$ and $(x,t)\in\R^{d+1}$.
\par 

Relativistic theories, on the other hand, are locally invariant under the action of the Poincar\'e group ${\rm IO}(d,1)$, which can be written as a semi-direct product
\begin{equation}
	{\rm IO}(d,1) = O(d,1) \ltimes \R^{d,1}.
\end{equation}
with group multiplication
\begin{equation}
	(\Lambda', b') \, (\Lambda, b)
	=
	(\Lambda'\, \Lambda,\, \Lambda' \, b + b') 
\end{equation}
for all $\Lambda,\Lambda'\in O(d,1)$ and $b,b'\in(\R^{d,1},+)$. The left action of this group on $\R^{d,1}$ is given by
\begin{equation}
	(\Lambda,b) \, x = \Lambda \, x + b
\end{equation}
for all $(\Lambda,b)\in {\rm IO}(d,1)$ and $x\in\R^{d,1}$.
\par 

In stochastic theories on curved spacetime the Galiliean symmetry and Poincar\'e symmetry are no longer the correct symmetries of spacetime, since the structure group of the (co)tangent bundle is deformed, cf. Appendix \ref{sec:Structuregroup}.
\par 

In a non-relativistic theory, the structure group of the tangent bundle is ${\rm GL}(d,\R)$, which has the orthogonal group $O(d)$ as a subgroup. In a stochastic theory, the structure group of the It\^o tangent bundle is the It\^o group $G^d_I$, which can be written as a Cartesian product
\begin{equation}
	G^d_I = {\rm GL}(d,\R) \times {\rm Lin}(\R^d\otimes \R^d, \R^d)
\end{equation}
with group multiplication 
\begin{equation}\label{eq:groupMultiIto}
	(g',\kappa') \, (g,\kappa) = (g'\, g,\, g'\circ\kappa + \kappa'\circ(g\otimes g))
\end{equation}
for all $g,g'\in{\rm GL}(d,\R)$ and $\kappa,\kappa'\in{\rm Lin}(\R^d\otimes\R^d,\R^d)$.
Furthermore, its left action on $\R^d\times {\rm Sym}(T\R^d\otimes T \R^d)$ is given by
\begin{equation}\label{eq:LActionIto}
	(g,\kappa) \, (x,x_2) = (g \, x + \kappa \, x_2, \, (g\otimes g) \, x_2)
\end{equation}
for all $(g,\kappa)\in G^d_I$, $x\in\R^d$ and $x_2 \in {\rm Sym}(T\R^d\otimes T \R^d)$.
\par 

This It\^o deformation of the ${\rm GL}(d,\R)$ is inherited by the orthogonal group, which induces a deformation of the Galilean group to ${\rm Gal}_I(d)$. We can define this group as a Cartesian product
\begin{equation}
	{\rm Gal}_I(d) = {\rm Gal}(d) \times {\rm Lin}(\R^d\otimes \R^d, \R^d)
\end{equation}
with group multiplication
\begin{equation}
	(R',v',a',s',\kappa') \, (R,v,a,s,\kappa)
	=
	(R'\, R, \, R' v + v', \, R' a + a' + v' s,\, s'+s, \, R' \circ \kappa + \kappa' \circ (R\otimes R))
\end{equation}
for all $R,R'\in O(d)$, $v,v'\in(\R^d,+)$, $a,a'\in(\R^d,+)$, $s,s'\in(\R,+)$ and ${\kappa,\kappa'\in{\rm Lin}(\R^d\otimes\R^d,\R^d)}$. The left action of this group on the space $\R^{d+1}\times {\rm Sym}(T\R^d\otimes T \R^d)$ is given by
\begin{equation}
	(R,v,a,s,\kappa) \, (x,t,x_2) = (R \, x + v \, t + a + \kappa \, x_2, \, t+s, \, (R\otimes R) \, x_2)
\end{equation}
for all $(R,v,a,s,\kappa)\in {\rm Gal}_I(d)$, $(x,t)\in\R^{d+1}$ and $x_2\in{\rm Sym}(T\R^d\otimes T \R^d)$.
\par 

A similar analysis can be applied to relativistic theories. For this, we first construct a relativistic version of the It\^o group given by the Cartesian product 
\begin{equation}
	G^{d,1}_I = {\rm GL}(d+1,\R) \times {\rm Lin}(\R^{d,1}\otimes \R^{d,1}, \R^{d,1})
\end{equation}
with group multiplication as given in eq.~\eqref{eq:groupMultiIto} and a left action on ${\R^{d,1}\times {\rm Sym}(T\R^{d,1}\otimes T \R^{d,1})}$ as given in eq.~\eqref{eq:LActionIto}. This It\^o deformation of ${\rm GL}(d+1,\R)$ is inherited by the Lorentz group $O(d,1)$, which induces a deformation of the Poincar\'e group to ${\rm IO}_I(d,1)$. We define this group as the Cartesian product
\begin{equation}
	{\rm IO}_I(d,1) = {\rm IO}(d,1) \times {\rm Lin}(\R^{d,1}\otimes \R^{d,1}, \R^{d,1})
\end{equation}
with group multiplication
\begin{equation}
	(\Lambda',b',\kappa') \, (\Lambda,b,\kappa)
	=
	(\Lambda'\, \Lambda, \, \Lambda' \, b + b', \, \Lambda' \circ \kappa + \kappa' \circ (\Lambda\otimes \Lambda))
\end{equation}
for all $\Lambda,\Lambda'\in O(d,1)$, $b,b'\in(\R^{d,1},+)$ and $\kappa,\kappa'\in{\rm Lin}(\R^{d,1}\otimes\R^{d,1},\R^{d,1})$. The left action of this group on the space $\R^{d,1}\times {\rm Sym}(T\R^{d,1}\otimes T \R^{d,1})$ is then given by
\begin{equation}
	(\Lambda,b,\kappa) \, (x,x_2) = (\Lambda \, x + b + \kappa \, x_2, \, (\Lambda\otimes \Lambda) \, x_2)
\end{equation}
for all $(\Lambda,b,\kappa)\in {\rm IO}_I(d,1)$, $(x)\in\R^{d,1}$ and $x_2\in{\rm Sym}(T\R^{d,1}\otimes T \R^{d,1})$.
\par 

We conclude this section with two remarks. First, the It\^o deformations of the spacetime symmetries only apply to the It\^o formulation of the theory. In the Stratonovich framework the symmetries remain undeformed.
\par

Secondly, the It\^o deformations of the spacetime symmetries vanish in the limits $\kappa\rightarrow 0$ and $x_2\rightarrow 0$. In this book, these limits have a clear physical interpretation, as we have fixed the quadratic variation to be proportional to the inverse metric. Due to this choice, we have
\begin{itemize}
	\item $x_2 = \frac{\alpha \, \hbar}{m} \, g \, t$ and $\kappa=\Gamma$ for non-relativistic theories,
	\item $x_2 = \alpha \, \hbar \, \varepsilon \,  g \, \lambda $ and $\kappa=\Gamma$ for relativistic theories,
\end{itemize}
where $\Gamma$ is the Levi-Civita connection associated to the metric $g$. Consequently, the It\^o deformations vanish in the classical limit $\hbar\rightarrow 0$, as this implies $x_2\rightarrow0$, and in the limit of vanishing gravity $G\rightarrow 0$, as this implies $\Gamma\rightarrow0$.

\clearpage

\section{Stochastic Interpretation}\label{sec:HiddenVariables}
The theory that we have presented in this book is inherently stochastic and does not contain any deterministic hidden variables. Instead, the theory should be regarded as a mathematically rigorous implementation of the statement `God plays dice'. Here, the dice game is played in the probability space $(\Omega,\Sigma)$, which determines the possible outcomes of the dice rolls, and the probability for each of these outcomes is determined by the probability measure $\mathbb{P}$. 
\par 

This stochastic theory describes a generalization of both quantum mechanics, which is obtained for $\phi=\frac{\pi}{2}$ and the theory of Brownian motion, which is obtained for $\phi=0$. In physics, this mathematical theory of Brownian motion is only used as an effective theory to replace the more fundamental physical theory of Brownian motion. In this fundamental theory, the Brownian motion is not induced by a dice roll, but by the interaction of the macroscopic particle under consideration with a large number of microscopic particles. These interactions are fully deterministic, but their average effect can be modeled more efficiently by the stochastic theory. Therefore, in this physical picture, the probability measure $\mathbb{P}$ can in principle be derived from the underlying deterministic physics.
\par 

The above discussion suggests that stochastic mechanics may be reinterpreted as a hidden variable theory. In such a hidden variable interpretation, the stochastic law models the effective interactions of the particle with an omnipresent background field. One might wonder whether the dynamics of this background can itself be described by some (non-)deterministic physical laws. In this chapter, we explore this possibility by discussing various aspects of the stochastic theory in more detail. Moreover, we show that the Bell inequalities do not provide any objection to such a hidden variable theory.

\subsection{Locality}\label{sec:locality}
The stochastic theory is a local theory, since it is governed by local physical laws of motion. The drift of the stochastic theory obeys a stochastic generalization of the principle of least action, which is a local principle. Moreover, the stochastic fluctuations of the theory are introduced by a local structure relation. The locality of the theory is reflected by the fact that the equations of motion that govern the motion of a stochastic particle are stochastic differential equations, and these are intrinsically local.
\par 

We emphasize that this notion of locality only applies to the stochastic theory of a single particle discussed in this book. It is expected that, in order to explain entanglement in multi-particle systems, the notion of locality in the stochastic theory must be relaxed. For example, for two particles, described by semi-martingales $X^\mu_{1,2}=C^\mu_{1,2}+e^\mu_\alpha \, M^\alpha_{1,2}$, the entanglement will introduce a stochastic coupling such that $d[M_1,M_2]\neq0$. If this coupling is preserved, when $X_1$ and $X_2$ are separated, this coupling will introduce a non-locality in the stochastic law that governs this two particle state.
\par

Furthermore, we note that, even though the single particle theory is local, we have defined objects that are not local. These are the wave functions $\Psi_\pm$ and the velocity fields $w_\pm$. The wave functions are solutions of the diffusion equations and therefore depend on the global properties of the manifold and the scalar and vector potential. A similar reasoning applies to the velocity fields, which are solutions of another set of partial differential equations and closely related to the wave functions.
\par 

These non-local objects do not render the theory non-local, as they do not represent physical properties of the particle. Only $w_\pm(X_t,t)$ along the trajectory $X_t$ has a physical meaning, as it represents the expected velocity of the particle along its trajectory. The extension of $w(X_t,t)$ to $w(x,t)$ for all $x\in\M$ is introduced for mathematical convenience, but does not represent physical properties.
\par 

Since $w_\pm(x,t)$ is related to $\Psi_\pm(x,t)$ the same reasoning can be applied to the wave functions. This implies that the extension of the wave function $\Psi(x,t)$ to $x\in\M$ does not represent any intrinsic properties of the particle. Instead, it provides the best possible prediction that an observer can make about the particle. In making this prediction, observers use their knowledge about an initial condition $\Psi_-(X_0,t_0)$ or terminal condition $\Psi_+(X_f,t_f)$ and about the global properties of the manifold and the scalar and vector potential.
\par

We conclude that the wave function is not intrinsic to the particle, but only describes how an observer sees the particle.\footnote{This interpretation is standard in the study of Brownian motion. The complex stochastic theory shows that this interpretation of the wave function must also be adopted in quantum mechanics.} Therefore, the wave function is not a universal object. Due to this fact, wave function collapse does not introduce any peculiarities in the stochastic formulation. In the stochastic theory, a wave function collapse occurs when an observer updates its own knowledge about the particle after making a measurement. This leads to a collapse in the wave function that the observer uses to describe the particle, but does not affect the particle under consideration.
\par 

Another consequence of this non-universal character is that the wave function is an observer dependent object: different observers can describe the same particle using a different wave function, as their knowledge about the state of the particle might differ.

\subsection{Causality}\label{sec:Causality}
The stochastic theory is by construction causal with respect to any affine parameter, since both the deterministic and stochastic laws are defined with respect to an affine parameter. In a non-relativistic theory, the notion of time is an affine parameter. Therefore, the non-relativistic stochastic theory is causal with respect to time.
\par 

In a relativistic theory, time is no longer an affine parameter, as it is promoted to a coordinate in the spacetime. Therefore, one must impose further conditions to obtain causality with respect to time. In relativistic theories, causality with respect to time is obtained by imposing that all trajectories are causal. In a stochastic theory, this can be defined as follows:\footnote{Note that this definition depends on the choice for the $(-+..+)$ signature.}
\begin{itemize}
	\item a trajectory $X_\lambda$ is causal, if $\E\left[ g(v_\circ,v_\circ)\right] \leq 0$ along $X_\lambda$ for all $\lambda\in\mathcal{T}$.
\end{itemize}
\par 

In practice, causality is ensured by imposing that all particles are time-like or null-like. In a stochastic theory, these notions can be defined by\footnote{Cf. appendix \ref{sec:L2StochProc}.}
\begin{itemize}
	\item a particle is time-like, if $\E\left[ g(v_\circ,v_\circ)\right] < 0$ along its trajectory $X_\lambda$ for all $\lambda\in\mathcal{T}$;
	\item a particle is null-like, if $\E\left[ g(v_\circ,v_\circ)\right] = 0$ along its trajectory $X_\lambda$ for all $\lambda\in\mathcal{T}$;
	\item a particle is space-like, if $\E\left[ g(v_\circ,v_\circ)\right] > 0$ along its trajectory $X_\lambda$ for all $\lambda\in\mathcal{T}$.
\end{itemize}
As in the deterministic theory, we can ensure that all particles fall into one of these categories by imposing an energy momentum relation or on-shell condition. In the stochastic theory, this condition is given by
\begin{equation}
	\E\left[g(v_\circ,v_\circ)\right] 
	=
	- \varepsilon^2 \, m^2 
	=
	\begin{cases}
		-1 \quad &{\rm if}\; m^2>0 \, ,\\
		0 \quad &{\rm if} \; m^2=0 \, ,\\
		+1 \quad &{\rm if} \; m^2<0\, ,
	\end{cases} \,
\end{equation}
where we gauge fixed $\varepsilon$ in the second equality. This condition can also be written in the It\^o formulation using that
\begin{equation}
	\E\left[g(v_\circ,v_\circ)\right] = \E\left[g(\hat{v}_\pm,\hat{v}_\pm) + \frac{1}{6} \, \mathcal{R}_{\mu\nu\rho\sigma} v_2^{\mu\rho} v_2^{\nu\sigma} \right] ,
\end{equation} 
which follows from section \ref{sec:EQMRiemann} and appendix \ref{sec:ItoLagrangian}.
If an on-shell condition is imposed, one finds that
\begin{itemize}
	\item a particle is time-like, if $m^2>0$;
	\item a particle is null-like, if $m^2=0$;
	\item a particle is space-like, if $m^2<0$.
\end{itemize}
Therefore, the theory is causal, if the rest mass of all particles is fixed to be $m^2\geq0$.
\par 

We point out that the stochastic notion of causality is slightly weaker than the classical notion of causality. In a classical theory, causality is defined without taking an expectation value, while in a stochastic theory the expectation value is necessary to make the causality condition well defined. As a consequence, stochastic causality ensures that the expectation value $\E[X]=\int_\Omega X(\omega) \, d\mathbb{P}(\omega)$ satisfies the classical notion of causality, but the sample paths $X(\omega)$ can violate this stricter classical notion of causality, as they can go off-shell.
\par 

For massless particles, deviations from the on-shell condition ${g(v_\circ,v_\circ)=0}$, immediately imply that the particle is no longer null-like, and, therefore, that the classical causality condition is violated. For massive particles, on the other hand, deviations from the on-shell condition $g(v,v)=-1$, do not imply violation of the classical causality condition, as they may retain their time-like character. Nevertheless, for large deviations from the on-shell condition a violation of classical causality may occur.
\par

We can provide an estimate for the likelihood of such violations by considering a free massive particle in a flat spacetime. Its trajectory is a solution of the It\^o equation
\begin{equation}
	\begin{cases}
		d_+ X^\mu_\tau &= v_+^\mu(X_\tau,\tau) \, d\tau + {\rm Re}[dM^\mu_\tau] \, ,\\
		d[M^\mu,M^\nu]_\tau &= \frac{\hbar}{m} \, \eta^{\mu\nu} \, d\tau \, ,
	\end{cases}
\end{equation}
where we fixed $|\alpha|=1$ and gauge-fixed $\varepsilon=m^{-1}$ such that $\tau$ is the proper time. Classical causality is violated, when the particle moves faster than light. In the rest frame of the particle, where $v_+^i=0$, we can estimate this by the violation condition
\begin{equation}\label{eq:Causalityviolation}
	\sum_{i=1}^d\big|\delta_{ij}\Delta X^i \, \Delta X^j\big|^2 = \sum_{i=1}^d \Big|\delta_{ij}{\rm Re}(\Delta M^i) \,{\rm Re}(\Delta M^j)\Big|^2 \geq c^2 \, \Delta\tau^2\, ,
\end{equation}
where $\Delta X$ represents the distance traveled over a time interval $\Delta \tau$. Then, using that
\begin{equation}
	{\rm Re}(\Delta M^i) \sim\mathcal{N}\left( 0 ,\frac{\hbar \, (1+ \cos \phi) }{2 \, m}  \, \Delta \tau\right),
\end{equation}
we find the characteristic scale for the causality violations. Causality violations are likely to occur for length scales
\begin{equation}
	c \, \Delta \tau \lesssim \frac{\hbar \, d}{2 \, m \, c} \, (1 + \cos\phi) \, ,
\end{equation}
where $d$ is the dimension of space, $\phi=0$ for Brownian motion and $\phi=\frac{\pi}{2}$ for quantum mechanics. For larger scales, the probability of such violations quickly decays to $0$.
\par 

We conclude that a stochastic trajectory satisfies the stochastic notion of causality, but the sample paths may violate the stricter classical notion of causality on segments shorter than the Compton wavelength. This provides a stochastic explanation why a quantum particle cannot be localized within its own Compton wavelength.

\subsection{Bell's Theorem}\label{sec:Bell}
The stochastic theory presented in this book does not exclude the existence of an underlying (non-)deterministic hidden variable interpretation. However, since the stochastic theory reduces to quantum mechanics in the limit $\alpha=\ri$, Bell's theorem could provide an argument against such a hidden variable interpretation. As the Bell inequalities apply to multi-particle systems and typically involve the notion of spin, a proper analysis of Bell's theorem cannot be performed within the stochastic framework presented in this book. The stochastic theory can, however, be extended to multi-particle systems with spin. Thus, it is worth investigating whether the Bell inequalities apply to the stochastic theory.
\par 

Bell's theorem relies on the derivation of Bell inequalities for certain systems. The system considered by Bell \cite{Bell:1964kc,Bell:1964fg} is a system of two particles with spin-$\frac{1}{2}$. These particles are prepared in the singlet state
\begin{equation}\label{eq:singlet}
	| s_{1,x} \, s_{2,x} \rangle = \frac{1}{2} \, \Big( |\uparrow \downarrow \rangle - |\uparrow \downarrow \rangle \Big)
\end{equation}
with respect to the unit vector $x$. The particles are then separated, in such a way that the entanglement is preserved, and the spin of the individual particles is measured by two observers. These observers, Alice and Bob, measure the normalized spin of the particle along the unit vectors $a$ and $b$, which they may choose freely. We will denote the measurement performed by Alice by $s_1$ and the measurement performed by Bob by $s_2$. Both measurements may depend on both orientations $a$ and $b$, but, if we assume that the choice of orientation $a$ cannot affect the measurement outcome at $s_2$ and vice versa,\footnote{In refs.~\cite{Nelson:1986,Schulz}, this is called the assumption of active locality.} we find
\begin{equation}\label{eq:localasmpt}
	s_1(a,b) = s_1(a) \qquad {\rm and} \qquad s_2(a,b)= s_2(b) \, .
\end{equation}
The outcomes of these measurements are given by
\begin{equation}\label{eq:BellResults}
	s_1(a) = \pm 1  \qquad {\rm and} \qquad
	s_2(b) = \pm 1 \, . 
\end{equation}
Moreover, since the system is in a singlet state, we have 
\begin{equation}\label{eq:BellResults2}
	s_2(c) = - s_1(c)\, \qquad \forall\, c\, . 
\end{equation}
\par 

One can calculate the expectation value of the product of two states. Quantum mechanics predicts that this expectation value is given by
\begin{equation}
	\E\left[ s_1(a) \, s_2(b) \right] = - a \cdot b \, .
\end{equation}
On the other hand, one can derive that in a locally real theory the expectation satisfies the Bell inequality
\begin{equation}\label{eq:BellInequality}
	\Big| \E [ s_1(a) \, s_2(b) ]
	- \E [ s_1(a) \, s_2(c) ] \Big| \leq 1 + \E[ s_1(c) \, s_2(b) ] \, ,
\end{equation}
where $c$ is a third direction along which the spin can be measured. 
It is easy to find configurations for the unit vectors such that the quantum mechanics prediction violates the Bell inequality. One possible choice is the configuration that satisfies
\begin{equation}
	a \cdot b = 0\, \qquad a\cdot c = \frac{\sqrt{2}}{2} \qquad b\cdot c = \frac{\sqrt{2}}{2} \, .
\end{equation}
\par 

The Bell inequality \eqref{eq:BellInequality} is derived assuming that there is a hidden variable $\Lambda$ that determines both the spins, such that their product can be calculated using the tower property for conditional expectations:
\begin{align}\label{eq:BellHypothesis}
	\E [ s_1(a) \, s_2(b) ] 
	&= \E [ \, \E[ s_1(a) \, s_2(b)  \, | \, \Lambda]] \, \nonumber\\
	&= \int \E[ s_1(a) \, s_2(b)  \, | \, \Lambda=\lambda] \, d\mu_\Lambda(\lambda) \nonumber\\
	&= \int s_1(a,\lambda) \, s_2(b,\lambda)  \, d\mu_\Lambda(\lambda) \, ,
\end{align}
where $d\mu_\Lambda(\lambda)$ is the probability distribution of the hidden variable $\Lambda$. The Bell inequality \eqref{eq:BellInequality} then follows from this assumption and eqs.~\eqref{eq:BellResults} and \eqref{eq:BellResults2}. 
\par 

The fundamental assumption made in eq.~\eqref{eq:BellHypothesis} is that both $s_1$ and $s_2$ are $\Lambda$-measurable. However, this assumption may fail for any theory that (i) allows for the creation of entangled states and (ii) predicts an uncertainty principle. This can be seen by treating the singlet state \eqref{eq:singlet} as a composite particle. Then, if there exists an uncertainty principle between the spin projections $s_x$, $s_y$ and $s_z$, it immediately follows that the product $s_1(a)\, s_2(b)$ is only defined, if $a$ and $b$ are parallel. Therefore, the presence of an uncertainty relation implies that there does not exist a measure $\mu_\Lambda$ that determines $s_1(a)\, s_2(b)$ for any choice of orientations $a$ and $b$. 
\par 

Consequently, the failure of eq.~\eqref{eq:BellHypothesis} can be regarded as a failure of the assumption of realism, i.e. the assumption that the product $s_1(a) \, s_2(b)$ is well-defined for any $a$ and $b$. Alternatively, one can argue that the treatment of the entangled state as a composite object is only applicable before the separation of the two particles. Following this line of reasoning, the failure of eq.~\eqref{eq:BellHypothesis} is due to the interplay of an uncertainty principle for the singlet state, when it is created, and a violation of locality that is implied by the existence of spatially separated entangled systems.\footnote{A similar argument is given in refs.~\cite{Nelson:1986,Schulz}. There, it is shown that stochastic theories can violate a principle of passive locality, meaning that the observations of Alice and Bob are not conditionally independent given the prior preparation \eqref{eq:singlet}.}
\par

For the stochastic theories presented in this book, (i) and (ii) hold, such that eq.~\eqref{eq:BellHypothesis} fails. Indeed, in this framework, the position $x^i$ and momentum operator $p_j=-\alpha\,\p_j$ satisfy a commutation relation $[x^i,p_j]=\alpha \, \delta^i_j$. For $d=3$, this implies a commutation relation on the angular momenta of the form $[L_i,L_j]=\alpha \, \epsilon_{ij}^{\;\;k}L_k$. Hence, once spin is included in the theory, one expects a spin-commutation relation of the form $[s_i,s_j]=\alpha \, \epsilon_{ij}^{\;\;k}s_k$. Moreover, as discussed in section \ref{sec:locality}, one can construct entangled states by allowing for a weak violation of locality. Then, for any $\alpha\in\mathbb{C}\setminus\{0\}$, one can derive a Bell inequality, and find a detector configuration for which this inequality is violated. This statement applies to the stochastic theory irrespective of whether a hidden variable interpretation is given. In particular, for $\alpha=1$ the stochastic theory reduces to a Brownian motion, for which the existence of a hidden variable interpretation is expected.

\subsection{The Quantum Foam}\label{sec:Qfoam}
As discussed in section \ref{sec:Boundaries}, the complex diffusion theory requires the existence of four velocity fields $v_+,v_-,u_+,u_-$ that satisfy the constraint \eqref{eq:ConstraintCVelFields} in a non-relativistic theory and the constraint \eqref{eq:ConstraintCVelFieldsRLT} in a relativistic theory. We can introduce the averages of these velocities and deviations thereof given by
\begin{alignat}{3}
	v_\circ &= \frac{1}{2} (v_+ + v_-) \, , \qquad\qquad  && u_\circ &&= \frac{1}{2} (u_+ + u_-) \, , \nonumber\\
	v_\perp &= \frac{1}{2} (v_+ - v_-) \, , \qquad\qquad  && u_\perp &&= \frac{1}{2} (u_+ - u_-) \, .
\end{alignat}
Since $v_+(X_t,t)$ and $u_+(X_t,t)$ describe the velocity fields of $X_t$ shortly after time $t$, while $v_-(X_t,t)$ and $u_-(X_t,t)$ describe the velocity fields of $X_t$ shortly before time $t$, these new fields may be preferred to describe the velocity field of $X$ at time $t$.
\par 

Since the fields $v_+$ and $v_-$ are associated to the probability measure $\mu_{X}$, they can be measured in experiment, which implies that $v_\circ$ and $v_\perp$ have a clear physical interpretation. This is not true for $u_\circ$ and $u_\perp$, but, using the constraint \eqref{eq:ConstraintCVelFields}, $u_\perp$ can be expressed in terms of $v_\circ,v_\perp,u_\circ$, if $\phi\in(-\pi,\pi)$. Therefore, only the field $u_\circ$ lacks a physical interpretation. 
\par 

If one supplements the stochastic theory with a hidden variable in the sense of Brownian motion, one obtains a physical interpretation of this velocity field $u_\circ$. In this physical picture, the Brownian particle is continuously interacting with many microscopic particles that induce the Brownian motion. These particles together make up a fluid and the average motion of these particles defines a velocity field for this fluid. In this interpretation, we can associate $u_\circ$ to the velocity of this fluid.
\par

One can speculate on the nature of this background fluid with which all matter interacts. Since the fluid acts as a medium through which all matter propagates, it has a strong resemblance with the aether theory, which has been discarded as it does not respect Lorentz invariance. The aether was then replaced by the notion of spacetime in general relativity and the notion of a quantum vacuum in quantum field theory. Both spacetime and the quantum vacuum can be interpreted as a medium through which matter propagates, but, contrary to the aether theory, they are compatible with general covariance, Lorentz invariance and the gauge symmetries. 
\par 

In a yet to be discovered theory of quantum gravity, it is expected that the notions of spacetime and the quantum vacuum will be replaced by the notion of a quantum foam, which is expected to be compatible with general covariance, gauge symmetries and a possibly deformed Lorentz invariance.
\par 

The fluid arising in the hidden variable interpretation of the stochastic theory is compatible with general covariance, gauge symmetries and with an It\^o deformed Lorentz symmetry. Moreover, the It\^o deformations vanish in the limit $\hbar\rightarrow0$, where the theory reduces to general relativity and the limit $G\rightarrow0$, where the theory reduces to quantum theory on flat spacetime. Therefore, this fluid can be interpreted as the quantum foam arising in various approaches to quantum gravity.\footnote{A similar observation was made by Calogero \cite{Calogero:1997,Calogero:2004}.}
\par 

We emphasize that the stochastic theory does not provide any physical laws for the microscopic behavior of the quantum foam. Instead, it describes the effective interactions between the quantum foam and all matter by means of a stochastic law. In this book, we have assumed that this law is given by eq.~\eqref{eq:QuadVarRel}, as this particular assumption allows to recover ordinary quantum theory on curved spacetime.
\par 

Theories of quantum gravity often predict Planck scale deviations from quantum theory on curved spacetime. Such deviations may be incorporated in the stochastic theory by considering more general structure relations as given in eq.~\eqref{eq:StructRelationGen}. In particular, one can consider $\beta\neq0$, which allows the particle to make jumps at randomly distributed times. The characteristic length and time scales for these jumps are proportional to $\kappa^{-1}$ and $(c\,\kappa)^{-1}$, which can be associated to the Planck length and Planck time.

\clearpage
\section{Discussion}\label{sec:Conclusion}
\subsection{Conclusion}\label{sec:ConclusionI}
In this book, we have reviewed and extended the theory of stochastic mechanics. In particular, we have used a generalization of the stochastic quantization procedure to describe a large class of stochastic theories. This stochastic quantization procedure can be formulated by the following postulates:
\begin{itemize}
	\item a stochastic particle is described by a two-sided semi-martingale $X=C_\pm+{\rm Re}(M)$;
	\item the finite variation processes $C_\pm$ are such that $X$ minimizes a stochastic action $S(X)$;
	\item the complex martingale $M=M_x + \ri \, M_y$ is such that $M_x$ and $M_y$ are correlated two-sided L\'evy processes;
	\item the quadratic variation of these L\'evy processes is fixed by the structure relation
	\begin{equation}\label{eq:StructConc}
		d[M^a,M^b] = \frac{\alpha \, \hbar}{m} \, A^{ab} \, dt + \frac{\beta}{\kappa} \, B^{ab}_c \, dM^c_t
	\end{equation} 
	with $\alpha,\beta\in\mathbb{C}$.
\end{itemize}
We have studied this theory for $\beta=0$ and $A^{ab}=\delta^{ab}$, and shown that the resulting theory describes a general complex diffusion theory with diffusion parameter $\alpha=|\alpha|\, e^{\ri \, \phi}$. This theory reduces to a Brownian theory for ${\alpha\in\R}$, to a quantum theory for $\alpha\in \ri \times \R$ and to a classical theory for $\alpha=0$. Moreover, the time reversal operation acts as $T(\alpha)=-\alpha$.
\par 

The equivalence between the stochastic theory and a complex diffusion theory is established by the derivation of a complex diffusion equation for non-relativistic theories and a complex wave equation for relativistic theories. This derivation implies that all predictions following from these diffusion and wave equations are also predictions of the stochastic theory. 
\par 

In particular, one can easily verify that the stochastic theory satisfies an energy-momentum uncertainty principle given by
\begin{equation}
	\sigma_{x^i} \, \sigma_{p_{j}} \geq \frac{|\alpha| \, \hbar}{2} \, \Big[1+\cos (\phi) \Big] \, \delta^i_j \, .
\end{equation}
In the relativistic theory, this is supplemented by an energy momentum relation for $x^0=c\,t$ and $E=c\,p_0$, which is given by
\begin{equation}
	\sigma_{t} \, \sigma_{E} \geq \frac{|\alpha| \, \hbar}{2} \, \Big[1-\cos (\phi) \Big] .
\end{equation}
For $\phi=\frac{\pi}{2}$, these relations reduce to the well-known uncertainty relations from quantum mechanics. The time-energy relation can be rewritten in terms of a temperature-energy relation, by defining a temperature as a Wick rotated time:
\begin{equation}
	T^{-1} = \pm \frac{\ri \, k_B \, t}{\hbar} \, .
\end{equation}
This yields a temperature-energy uncertainty relation of the form
\begin{equation}
	\sigma_{T^{-1}} \sigma_{E} \geq \frac{|\alpha| \, k_B}{2} \, \Big[1+\cos (\phi) \Big] \, ,
\end{equation}
which, for $\phi=0$, reduces to the uncertainty relation encountered in statistical physics \cite{Mandelbrot,Roupas:2021soz}.
\par 

Starting from the diffusion equation, one finds that, for any $\alpha\in\mathbb{C}\setminus\{0\}$, the theory can be described using an operator formalism with operators given by 
\begin{equation}
	\hat{x}^\mu = x^\mu \qquad  {\rm and} \qquad \hat{p}_\mu = - \alpha \, \hbar \, \frac{\p}{\p x^\mu} 
\end{equation}
in the position representation. This immediately yields the canonical quantization condition given by the commutation relation
\begin{equation}
	[\hat{x}^\mu,\hat{p}_\nu] = \alpha \, \hbar \, \delta^\mu_\nu \, .
\end{equation}
\par

The theory can also be related to the path integral formulation by considering the characteristic functional\footnote{Alternatively, one can consider the moment generating functional $M_{X}(J)=\varphi_{X}(- \ri \, J)$.} 
\begin{equation}
	\varphi_{X}(J) = \E\left[ e^{\ri \int_\mathcal{T} J_{\mu}(\lambda)\, X^\mu(\lambda) \, d\lambda}\right], 
\end{equation}
where the process $X$ is a solution of the stochastic equations of motion. If we interpret this process as a single random variable on the path space $\M^\mathcal{T}$, this expression can formally be related to the partition function $Z_X(J)$, defined by a path integral, such that
\begin{equation}
	\varphi_{X}(J) = \frac{Z_X(J)}{Z_X(0)} := \frac{\int DX \, \exp\left[-\frac{S(X)}{\alpha} + \ri \, \langle J,X\rangle\right]}{\int DX \, \exp\left[-\frac{S(X)}{\alpha}\right]}\, ,
\end{equation}
where $\langle .,.\rangle$ denotes the inner product on the path space.
\par

The stochastic theory contains two limits that are of particular interest. The first is given by $\alpha\in[0,\infty)$, which represents a Brownian motion. This limit is interesting, as it is the only diffusion theory that can be described using real martingales. In this real theory, which was discussed in chapter \ref{sec:StochDynR}, the probability density is related to the solution of the heat equation and is given by
\begin{equation}
	\rho_{X_t}(x,t)\Big|_{\alpha\in(0,\infty)} = \frac{|\Psi(x,t)|}{\int|\Psi(x,t)| \, d^dx} \, ,
\end{equation}
where $\Psi$ is a real integrable function.
\par 

For all other values of $\alpha$, one requires the complex description, introduced in chapter \ref{sec:StochDynC}, using complex martingales and complex square integrable wave functions. For these complex theories the probability density is given by
\begin{equation}
	\rho(x,t) = \frac{|\Psi(x,t)|^2}{\int|\Psi(x,t)|^2 \, d^dx} \, .
\end{equation}
\par 

The second interesting limit is the quantum limit, given by $\alpha \in \ri \times \R$. In this limit, the theory is unitary. This unitarity is reflected by the fact
\begin{equation}
	\frac{d}{dt} \int|\Psi(x,t)|^2 \, d^dx\Big|_{\alpha\in \ri \times \R} = 0 \, ,
\end{equation}
which implies the Born rule
\begin{equation}
	\rho(x,t)\big|_{\alpha\in \ri \times \R} = |\Psi(x,t)|^2 \, .
\end{equation}
\par 

As shown in chapter \ref{sec:RLT}, the complex processes introduced in chapter \ref{sec:StochDynC} also provide a natural framework for the study of relativistic stochastic processes. This relativistic theory provides a consistent framework for the study of relativistic Brownian motion, but also for the study of a single relativistic quantum particle and anything in between. 
\par 

The fact that the relativistic stochastic theory is well defined for the single particle is particularly interesting, as the standard description of relativistic quantum particles requires an extension to quantum field theory. In the stochastic theory, such an extension is not necessary, since the norm is defined through an expectation value. As discussed in section \ref{sec:Causality}, the sample paths of the process may go off-shell and may even violate the classical notion of causality for short time intervals, but the stochastic notion of causality is obeyed at all times, which ensures that the sign of the norm is preserved in the theory.\footnote{When compared to a field theoretic formulation of relativistic quantum theories, the violation of classical causality is associated to the presence of eigenstates with negative eigenenergy, while the condition of stochastic causality is associated to the condition that commutators vanish outside the lightcone.}
\par

In chapter \ref{sec:Manifolds}, we extended the complex stochastic theory to Riemannian and Lorentzian geometry. This extension was performed using the framework of second order geometry, which provides an elegant way to construct covariant It\^o integrals. Using this framework, we found that the usual spacetime symmetries are deformed to the It\^o deformed Galilean symmetry in non-relativistic theories and the It\^o deformed Poincar\'e symmetry in relativistic theories. Moreover, we noticed that these deformations vanish in the classical limit $\hbar\rightarrow 0$ and the no-gravity limit $G\rightarrow 0$.
\par 

The It\^o deformations only occur when the theory is evaluated in the It\^o framework. In a Stratonovich framework, the Galilean group and Poincar\'e group remain to be the correct symmetries of the stochastic theory in curved space(time). Therefore, there may exist equivalent formulations of a quantum theory in curved spacetime: one that respects the classical spacetime symmetries and one that It\^o deforms the classical spacetime symmetries.
\par 

Canonical formulations of quantum theories respect the classical spacetime symmetries, as they are typically constructed in a first order formalism. However, this is not true for path integrals. As these are second order objects, the path integral measure is not covariant. Thus, for path integrals, covariance can be regained by It\^o deforming the Galilean and Poincar\'e symmetries.
\par 

We emphasize that these It\^o deformations of the spacetime symmetries follow from imposing that ordinary quantum theories on curved space(time) must respect general covariance. Other deformations of spacetime symmetries have been suggested in the literature, cf. e.g.~\cite{Arzano:2021scz}, but such deformations are different form the It\^o deformations, as they typically require the introduction of a minimal length scale, which is inspired by quantum gravity.

\subsection{Outlook}
The complex stochastic theory unifies quantum mechanics with Brownian motion in a single mathematical framework. The presented theory is, however, limited to the dynamics of a single (non-)relativistic particle subjected to a scalar and vector potential on a pseudo-Riemannian manifold. An important next step is the extension of this framework to multi-particle systems, field theories and to particles with spin. Various examples of such extensions have been studied in the literature, cf. e.g. \cite{Nelson,Guerra:1981ie,Dankel,Faris:1982,DeAngelis:1985hp,Dohrn:1978gd,Guerra:1973ck,Guerra:1980sa,Kodama:2014dba,Morato:1995ty,Garbaczewski:1995fr,Koide:2015,NelsonPath,Nelson:2014exa,Guerra:1973ck,Yasue:1978JMP,Davidson:1980,Koide:2014zkj,DeSiena:1983bx,DeSiena:1986nc,Davidson:1980df}, but no general framework exists yet. The complex stochastic theory, as presented in this book, allows for a systematic study of such extensions.
\par 

Of particular interest is the analysis of the It\^o deformations in a field theoretic framework. Given the Klein-Gordon equation that was derived for the single particle in section \ref{sec:CovDiffussion}, it is expected that the Pauli-DeWitt term must be present in the quantum action of any scalar field. This implies that a minimally coupled classical scalar theory in $3+1$ dimensions will become conformally coupled, when the theory is quantized. As we have not discussed spin, it is unclear whether further corrections can be expected for fermions or vector fields. In general, any such correction to the Lagrangian will predict deviations from the standard models of particle physics and cosmology in the regime where neither quantum mechanics nor gravity can be neglected.
\par 

Further corrections are to be expected from the complexification of the theory. In the complexified theory, the Lagrangian depends on the real position $X$ and the complex velocity $W$. In the Stratonovich framework, this Lagrangian is given by
\begin{equation}
	L^\circ(x,w_\circ,\varepsilon) = \frac{1}{2 \, \varepsilon} \, g_{\mu\nu}(x) \, w_\circ^\mu w_\circ^\nu - \frac{\varepsilon \, m^2}{2} + q \, A_\mu(x) \, w_\circ^\mu \, .
\end{equation}
Hence, the complexification introduces new terms, related to the velocity $u_\circ$, given by
\begin{equation}
	L_{q}^\circ(x,v_\circ,u_\circ,\varepsilon) 
	= 
	- \frac{1}{2 \, \varepsilon} \, g_{\mu\nu}(x) \, u_\circ^\mu u_\circ^\nu 
	+ \frac{\ri}{\varepsilon} \, g_{\mu\nu}(x) \, v_\circ^\mu u_\circ^\nu 
	+ \ri \, q \, A_\mu(x) \, u_\circ^\mu  \, .
\end{equation}
Since these corrections involve the complex velocity $u_\circ$, they can be associated to the velocity of the background field or quantum foam, cf. section \ref{sec:Qfoam}. Therefore, it is expected that, in a field theoretic framework, the stochastic theory predicts the existence of a background field that contributes to the energy balance of the universe.
\par 

As discussed in section \ref{sec:Qfoam}, the stochastic theory is a natural framework for the study of a fluctuating spacetime, which appears in various theories of quantum gravity. However, the interpretation slightly differs from most approaches to quantum gravity. According to the stochastic theory, all quantum effects are induced by the stochastic fluctuations of the spacetime foam. The stochastic theory does  not predict any Planck scale corrections to ordinary quantum physics. However, various classes of such corrections can be incorporated by studying the general structure relation \eqref{eq:StructConc} with $\beta\neq0$. Stochastic theories defined by such a general structure relation can be studied as a model for quantum gravity and may be related to other approaches to quantum gravity.
\par 

In the stochastic theory, gravity has itself a rather special status, due to the fact that the variation of the action with respect to a process $X$ induces a variation with respect to both the velocity $V_\pm$ and the second order velocity $V_2$.
This second order velocity is proportional to the metric $g$. Therefore, a variation with respect to the metric is implied by the variation with respect to any other field. This verifies that gravity couples to all matter, but also suggests that gravity is an emergent force. In this picture, gravity emerges as the field must minimize its action with respect to both its classical trajectory and its quantum fluctuations. The latter may be achieved by deforming spacetime, thus by inducing gravity.
\par

Various of these extensions of the stochastic theory are subject of current investigation. More generally, the stochastic theory, as presented in this book, deepens the connection between quantum mechanics and stochastic theories. This might be used in various ways: it allows to further exploit the tools from stochastic analysis to study quantum systems, while, on the other hand, it allows for further exploitation of the tools from quantum theory to study stochastic processes in applications ranging from soft condensed matter theory to finance.

\section*{Acknowledgements}
I am grateful to Michele Arzano, Can G\"okler, Qiao Huang and Zacharias Roupas for valuable discussions. This research was carried out in the frame of Programme STAR Plus, financially supported  by UniNA and Compagnia di San Paolo. 
\vspace{3cm}

\noindent
\fbox{%
	\parbox{\textwidth}{ 
		This is a preprint of the following work:\\
		Folkert Kuipers, ``Stochastic Mechanics: the Unification of Quantum Mechanics with Brownian Motion'', SpringerBriefs in Physics, Springer (2023),\\[0.1cm]
		reproduced with permission of Springer Nature Switzerland AG. \\[0.2cm]
		The final authenticated version is available online at:\\
		\href{https://link.springer.com/book/10.1007/978-3-031-31448-3}{http://dx.doi.org/10.1007/978-3-031-31448-3}
	}
}
\clearpage
\appendix

\section{Review of Probability Theory}\label{ap:ReviewStochProb}
In this appendix, we list various standard definitions and results from probability theory. For a more detailed discussion of these definitions and results, we refer to textbooks on probability theory and stochastic processes, e.g. Refs.~\cite{Karatzas,Cinlar,SchreveI,SchreveII,Emery}.
\subsection{Probability Spaces}
The most elementary object in probability theory is a probability space. This is a sample space $\Omega$, i.e. a collection of the possible outcomes in a probability experiment, together with the probability of occurrence for each outcome $\omega\in\Omega$. Crucially, these probabilities cannot be defined on the outcomes $\omega\in\Omega$ themselves. Instead, they must be defined on events in a sigma algebra over $\Omega$.
\begin{mydef}{\rm (Sigma algebra)}\\
Given a set $\Omega$, a sigma algebra over $\Omega$ is a set $\Sigma(\Omega)=\{A\,|\,A\subseteq \Omega\}$ satisfying the following properties
	\begin{itemize}
		\item $\Omega\in\Sigma\,$; 
		\item $\forall \, A\in\Sigma \, , \quad \Omega\setminus A\in \Sigma\,$; 
		\item $\forall\,A_1,A_2,...\in\Omega\, , \quad A_1\cup A_2 \cup ...\in \Sigma\,$.
	\end{itemize}
\end{mydef}
\begin{mydef}{\rm (Borel sigma algebra)}\\
	Given a topological space $(\M,\mathfrak{T})$, a Borel set is a set that can be obtained by taking countable unions, countable intersections and complements of the sets in the topology $\mathfrak{T}$. The collection of all Borel sets is the Borel sigma algebra $\mathcal{B}(\M)$.
\end{mydef}
\begin{mydef}{\rm (Measurable space)}\\
	Given a set $\Omega$, a measurable space is a tuple $(\Omega,\Sigma)$, where $\Sigma=\Sigma(\Omega)$ is a sigma algebra over $\Omega$.
\end{mydef}
\begin{mydef}{\rm (Probability measure)}\\
	Given a measurable space $(\Omega,\Sigma)$, a probability measure is a function $\mathbb{P}:\Sigma\rightarrow[0,1]$, such that
	\begin{itemize}
		\item $\mathbb{P}(\Omega)=1\,$,
		\item $\mathbb{P}(\bigcup_{i}A_i)=\sum_i\mathbb{P}(A_i)$ for any countable collection $\{A_i\in\Sigma \,| \, i\in \{1,2,...\}\}$ of pairwise disjoint sets.
	\end{itemize}
\end{mydef}
\begin{mydef}{\rm (Probability space)}\\
	A probability space is a tuple $(\Omega,\Sigma,\mathbb{P})$, where $(\Omega,\Sigma)$ is a measurable space and ${\mathbb{P}:\Sigma\rightarrow [0,1]}$ is a probability measure.
\end{mydef}
\begin{mydef}{\rm (Almost surely)}\\
	Given a probability space $(\Omega,\Sigma,\mathbb{P})$, we say that an event $A\in\Sigma$ occurs almost surely (abbreviated to a.s.), if $\mathbb{P}(A)=1$.
\end{mydef}
\subsection{Random Variables}
Random variables are the main object of study in probability theory. Essentially, these are functions that translate the outcomes of a probability experiment into probabilistic events in the real world. 
\begin{mydef}{\rm (Measurable function)}\\
	Given two measurable spaces $(\Omega_1,\Sigma_1)$ and $(\Omega_2,\Sigma_2)$, a function $X:(\Omega_1,\Sigma_1)\rightarrow(\Omega_2,\Sigma_2)$ is a measurable function, if
	\begin{itemize}
		\item $\forall \; U\in\Sigma_2\,, \quad X^{-1}(U):=\{\omega\in\Omega\,|\, X(\omega)\in U\}\in\Sigma_1\,$.
	\end{itemize}
\end{mydef}
\begin{mydef}{\rm (Borel measurable function)}\\
	Given two topological spaces $\M$ and $\mathcal{N}$, a function $f:\M\rightarrow\mathcal{N}$ is called Borel measurable, if
	\begin{itemize}
		\item $\forall \; U\in\mathcal{B}(\mathcal{N})\,, \quad f^{-1}(U):=\{x\in\M\,|\, f(x)\in U\}\in\mathcal{B}(\M)$ .
	\end{itemize}
\end{mydef}
\begin{mydef}{\rm (Random variable)}\\
	Given a probability space $(\Omega,\Sigma,\mathbb{P})$ and a measurable space $(\M,\mathcal{B}(\M))$, an $\M$-valued random variable is a measurable function $X:(\Omega,\Sigma,\mathbb{P})\rightarrow(\M,\mathcal{B}(\M))$.	
\end{mydef}
\begin{mydef}{\rm (Real random variable)}\\
	Given a probability space $(\Omega,\Sigma,\mathbb{P})$ and the real space $\R^n$, an $n$-dimensional real random variable is a random variable $X:(\Omega,\Sigma,\mathbb{P})\rightarrow(\R^n,\mathcal{B}(\R^n))$.
\end{mydef}
\begin{mydef}{\rm (Complex random variable)}\\
	Given a probability space $(\Omega,\Sigma,\mathbb{P})$ and the complex space $\mathbb{C}^n$, an $n$-dimensional complex random variable is a random variable $Z:(\Omega,\Sigma,\mathbb{P})\rightarrow(\mathbb{C}^n,\mathcal{B}(\mathbb{C}^n))$, such that $Z= X + \ri \, Y$, where ${X,Y:(\Omega,\Sigma,\mathbb{P})\rightarrow(\R^n,\mathcal{B}(\R^n))}$ are real random variables.
\end{mydef}
\begin{mydef}{\rm (Distribution of a random variable)}\\
	A random variable $X:(\Omega,\Sigma,\mathbb{P})\rightarrow(\M,\mathcal{B}(\M))$ induces a probability measure ${\mu_X=\mathbb{P} \circ X^{-1}}$ on $(\M,\mathcal{B}(\M))$. The induced function $d\mu_X:\M\rightarrow[0,1]$ is called the distribution of $X$.
\end{mydef}
\begin{theorem}{\rm (Functions of random variables)}\\
	Given a random variable $X:(\Omega,\Sigma,\mathbb{P})\rightarrow(\M,\mathcal{B}(\M))$ and a Borel measurable function $f$ defined on $\M$, the composition $f\circ X$ is a random variable.
\end{theorem}

\subsection{Expectation Value}
In the study of random variables, one is often interested in their expectation values. The expectation value of a random variable is defined by a Lebesgue integral.
\begin{mydef}{\rm (Lebesgue integrability)}\\
	A real random variable $X:(\Omega,\Sigma,\mathbb{P})\rightarrow(\R,\mathcal{B}(\R))$ is integrable, if and only if the Lebesgue integral $ \int_\Omega \big|X(\omega)]\big| \, d\mathbb{P}(\omega)$ converges.\\
	More generally, given a smooth manifold $\M$, a random variable $X:(\Omega,\Sigma,\mathbb{P})\rightarrow(\M,\mathcal{B}(\M))$ is integrable, if and only if the Lebesgue integral $ \int_\Omega \big|f[X(\omega)]\big| \, d\mathbb{P}(\omega)$ converges for all $f\in C^\infty(\M)$.
\end{mydef}
\begin{mydef}{\rm (Square integrability)}\\
	A real random variable $X:(\Omega,\Sigma,\mathbb{P})\rightarrow(\R,\mathcal{B}(\R))$ is square integrable, if and only if the Lebesgue integral $ \int_\Omega \big|X(\omega)]\big|^2 \, d\mathbb{P}(\omega)$ converges.\\	
	More generally, given a smooth manifold $\M$, a random variable $X:(\Omega,\Sigma,\mathbb{P})\rightarrow(\M,\mathcal{B}(\M))$ is square integrable, if and only if the Lebesgue integral $ \int_\Omega \big|f[X(\omega)]\big|^2 \, d\mathbb{P}(\omega)$ converges for all $f\in C^\infty(\M)$.
\end{mydef}
\begin{mydef}{\rm (Expectation value)}\\
	Given an integrable random variable $X:(\Omega,\Sigma,\mathbb{P})\rightarrow(\R^n,\mathcal{B}(\R^n))$, the expectation value of the $i$-th component $X^i$ is given by the Lebesgue integral
	\begin{equation}
		\E[X^i]= \int_\Omega X^i(\omega)\, d\mathbb{P}(\omega) \, .
	\end{equation}
	More generally, given an integrable random variable $X:(\Omega,\Sigma,\mathbb{P})\rightarrow(\M,\mathcal{B}(\M))$ and a function $f\in C^\infty(\M)$, the expectation value of $f(X)$ is given by the Lebesgue integral
	\begin{equation}
		\E[f(X)]= \int_\Omega f[X(\omega)]\, d\mathbb{P}(\omega) \, .
	\end{equation}
\end{mydef}
\begin{theorem}{\rm (Expectation value I)}\\
	Given a probability space $(\Omega,\Sigma,\mathbb{P})$, a measurable space $(\M,\mathcal{B}(\M))$, a random variable $X:\Omega\rightarrow\M$ and a Borel measurable function $f:\mathcal{M}\rightarrow \R$ such that $f\circ X$ is integrable, then 
	\begin{equation}
		\E[f(X)] = \int_\M f(x) \, d\mu_X(x) \, ,
	\end{equation}
	where $d\mu_X$ is the distribution of $X$ on $\M$.
\end{theorem}
\begin{mydef}{\rm (Probability density)}\\
	Given a probability space $(\Omega,\Sigma,\mathbb{P})$, an $n$-dimensional pseudo-Riemannian manifold $(\M,g)$ and an integrable random variable $X:\Omega\rightarrow\M$, a Borel measurable function ${\rho_X:\M \rightarrow [0,\infty)}$ is a probability density for $X$, if
	\begin{equation}
		\mu_X(B) = \int_B \rho_X(x) \, dV_R(x) = \int_B \sqrt{|g(x)|} \, \rho_X(x) \, d^nx \qquad \forall \; B\in \mathcal{B}(\M) \, ,
	\end{equation}
	where $V_R$ is the Riemann measure on $(\M,g)$.
\end{mydef}
\begin{theorem}{\rm (Expectation value II)}\\
	Given a probability space $(\Omega,\Sigma,\mathbb{P})$, an $n$-dimensional pseudo-Riemannian manifold $(\M,g)$, a random variable $X:\Omega\rightarrow\M$ with probability density $\rho_X$ and a Borel measurable function $f:\M\rightarrow\R$, such that $f\circ X$ is integrable, then
	\begin{equation}
		\E[f(X)] = \int_\M \sqrt{|g(x)|} \, f(x) \, \rho_X(x) \, d^nx \, .
	\end{equation}
\end{theorem}

\subsection{Conditional Expectation}
In practice, one is often interested in properties of a random variable, while one is already given some information. In order to study random variables, when such partial information is given, we require the notion of conditional expectation.
\begin{mydef}{\rm (Sigma algebra generated by a random variable)}\\
	Given a random variable $X:(\Omega,\Sigma,\mathbb{P})\rightarrow(\M,\mathcal{B}(\M))$, the sigma algebra generated by $X$ is given by
	\begin{equation}
		\sigma(X) = \big\{\{\omega\in \Omega \, | \, X(\omega)\in B\} \, \big|\, B\in\mathcal{B}(\M)\big\}\, .
	\end{equation}
\end{mydef}
\begin{mydef}{\rm ($\mathcal{G}$-measurability)}\\
	Given a random variable $X:(\Omega,\Sigma,\mathbb{P})\rightarrow(\M,\mathcal{B}(\M))$ and a sigma algebra $\mathcal{G}(\Omega)$, $X$ is $\mathcal{G}$-measurable, if $\sigma(X)\subseteq\mathcal{G}$.
\end{mydef}
\begin{mydef}{\rm (Stochastic independence)}\\
	Given a probability space $(\Omega,\Sigma,\mathbb{P})$ and two sub sigma algebras $\mathcal{G}_{1},\mathcal{G}_{2}\subseteq\Sigma$, we call $\mathcal{G}_1$ and $\mathcal{G}_2$ independent, denoted by $\mathcal{G}_1\indep\mathcal{G}_2$, if
	\begin{equation}
		\mathbb{P}(A\cap B) = \mathbb{P}(A) \, \mathbb{P}(B) \quad \forall\; A\in\mathcal{G}_1, \, B\in\mathcal{G}_2 \, .
	\end{equation}
\end{mydef}
\begin{mydef}{\rm (Independent random variables)}\\
	Given a probability space $(\Omega,\Sigma,\mathbb{P})$ and two random variables $X,Y$ defined on this space, we say that $X$ and $Y$ are independent, denoted by $X\indep Y$, if $\sigma(X)$ and $\sigma(Y)$ are independent sigma algebras.
\end{mydef}
\begin{theorem}{\rm (Functions of independent random variables)}\\
	Given two independent random variables $X,Y:(\Omega,\Sigma,\mathbb{P})\rightarrow(\M,\mathcal{B}(\M))$, and two Borel measurable functions $f,g$ defined on $\M$, then $f(X)\indep g(Y)$.
\end{theorem}

\begin{mydef}{\rm (Conditional expectation)}\\
	Given an integrable random variable $X:(\Omega,\Sigma,\mathbb{P})\rightarrow(\R,\mathcal{B}(\R))$ and a sub sigma algebra $\mathcal{G}\subseteq\Sigma$, the conditional expectation of $X$ given $\mathcal{G}$, denoted by $\E[X\,|\,\mathcal{G}]$, is any real random variable that is $\mathcal{G}$-measurable and satisfies 
	\begin{equation}
		\int_A \E[X\,|\,\mathcal{G}](\omega) \, d\mathbb{P}(\omega)
		=
		\int_A X(\omega) \, d\mathbb{P}(\omega)
		\qquad \forall \; A\in\mathcal{G}\, .
	\end{equation}
	If $\mathcal{G}=\sigma(Y)$ for some other random variable $Y$, we write $\E[X\,|\,Y]:=\E[X\,|\,\sigma(Y)]$.
\end{mydef}
\begin{theorem}{\rm (Properties of conditional expectation)}\\
	Given two integrable random variables $X,Y:(\Omega,\Sigma,\mathbb{P})\rightarrow(\R,\mathcal{B}(\R))$ and a sub sigma algebra $\mathcal{G}\subseteq\Sigma$, the conditional expectation satisfies the following properties:
	\begin{itemize}
		\item (Linearity): $\E[a \, X + b \, Y \, |\,\mathcal{G}] = a \, \E[X\,|\,\mathcal{G}] + b \, \E[Y\,|\,\mathcal{G}]$ for any $a,b\in \R$;
		\item (Taking out what is known): If $X$ is $\mathcal{G}$-measurable, then $\E[X\,Y\,|\,\mathcal{G}]=X\, \E[Y\,|\,\mathcal{G}]$;
		\item (Tower property): $\E[\,\E[X\,|\,\mathcal{G}]\,|\,\mathcal{H}] = \E[X\,|\,\mathcal{H}]$ for any sub sigma algebra ${\mathcal{H}\subset\mathcal{G}}$;
		\item (Independence): If $X\indep\mathcal{G}$, then $\E[X\,|\,\mathcal{G}]=\E[X]$;
		\item (Jensen's inequality): $\phi(\E[X|\mathcal{G}]) \leq \E[\phi(X)|\mathcal{G}]$ for any convex function $\phi$.
	\end{itemize}
\end{theorem}

\subsection{Change of Measure}
The measure on a measurable space is not uniquely defined, but there exist relations between different measures on the same measurable space. These relations are expressed by the Radon-Nykod\'ym theorem.
\begin{theorem}{\rm (Change of measure)}\\
	Given a probability space $(\M,\mathcal{B}(\M),\mu)$ and a Borel measurable function ${h:\M\rightarrow[0,\infty)}$, such that $\int_\M h(x) \, d\mu(x) =1$, the object defined by
	\begin{equation}
		\tilde{\mu}(A) = \int_B h(x) \, d\mu(x) \qquad \forall\; B\in\mathcal{B}(\M)
	\end{equation}
	is a probability measure on $(\M,\mathcal{B}(\M))$. Furthermore, for every Borel measurable function $f:\M\rightarrow[0,\infty)$, 
	\begin{equation}
		\int_\M f(x) \, d\tilde{\mu}(x) = \int_\M f(x) \, h(x) \, d\mu(x).
	\end{equation}
	Moreover, if $h:\M\rightarrow(0,\infty)$, then
	\begin{equation}
		\int_\M g(x) \, d\mu(x) = \int_\M \frac{g(x)}{h(x)} \, d\tilde{\mu}(x)
	\end{equation}
	for every Borel measurable function $g:\M\rightarrow[0,\infty)$.
\end{theorem}
\begin{mydef}{\rm (Absolutely continuous)}\\
	Given a measurable space $(\M,\mathcal{B}(\M))$ and two measures $\mathbb{\mu}$ and $\tilde{\mu}$ defined on this space, the measure $\tilde{\mu}$ is said to be absolutely continuous with respect to $\mu$, if
	\begin{equation}
		\tilde{\mu}(B) = 0 \qquad \forall \, B\in\{A\in\mathcal{B}(\M) \, | \, \mu(A) =0\} \,.
	\end{equation}
\end{mydef}
\begin{theorem}{\rm (Radon-Nykod\'ym)}\\
	Given a measurable space $(\M,\mathcal{B}(\M))$ and two probability measures $\mu$ and $\tilde{\mu}$ defined on this space, such that $\tilde{\mu}$ is absolutely continuous with respect to $\mu$, then there exists a Borel measurable function $h:\M\rightarrow[0,\infty)$, such that
	\begin{itemize}
		\item $\int_\M h(x) \, d\mu(x) =1$;
		\item $\tilde{\mu}(B) = \int_B h(x) \, d\mu(x) \quad \forall \; B\in\mathcal{B}(\M)$.
	\end{itemize}
	This function is denoted by
	\begin{equation}
		h = \frac{d\tilde{\mu}}{d\mu}
	\end{equation}
	and called the Radon-Nykod\'ym derivative of $\tilde{\mu}$ with respect to $\mu$.
\end{theorem}
\begin{corollary}{(Probability density)}\\
	Given a pseudo-Riemannian manifold $\M$, a random variable $X:(\Omega,\Sigma,\mathbb{P})\rightarrow(\M,\mathcal{B}(\M))$, such that $\mu_X=\mathbb{P}\circ X^{-1}$ is absolutely continuous with respect to the Riemann measure $V_R$ on $(\M,\mathcal{B}(\M))$, then $X$ has a probability density $\rho_X$, which is given by the  Radon-Nykod\'ym derivative
	\begin{equation}
		\rho_X = \frac{d\mu_X}{dV_R}\, .
	\end{equation}
\end{corollary}

\subsection{$L^2$-spaces}\label{ap:L2spaces}
In order to do analysis and to study convergence of random variables, one requires a notion of distance. This is provided by the $L^p$-norm. Here, we focus on the case $p=2$.
\begin{mydef}{\rm ($L^2$-norm)}\\
	Given a random variable $X:(\Omega,\Sigma,\mathbb{P})\rightarrow(\mathbb{K}^n,\mathcal{B}(\mathbb{K}^n))$ with $\mathbb{K}\in\{\R,\mathbb{C}\}$, the $L^2$-norm for $X$ is defined by 
	\begin{equation}
		||X|| = \sqrt{\E\big[\delta_{ij}\overline{X}{}^i X^j\big]} \, .
	\end{equation}
\end{mydef}
\begin{mydef}{\rm ($L^2$-space)}\\
	Given a probability space $(\Omega,\Sigma,\mathbb{P})$, the spaces
	\begin{equation}
		L^2(\Omega,\Sigma,\mathbb{P}) = \left\{ X:(\Omega,\Sigma,\mathbb{P})\rightarrow(\mathbb{K}^n,\mathcal{B}(\mathbb{K}^n)) \, \Big| \, ||X||<\infty \right\}
	\end{equation}
	with $\mathbb{K}\in\{\R,\mathbb{C}\}$ are the real and complex $L^2$-space over $(\Omega,\Sigma,\mathbb{P})$.
\end{mydef}
\begin{mydef}(Stochastic Riemannian distance)\\
	Given a Riemannian manifold $(\M,g)$ and random variables ${X,Y:(\Omega,\Sigma,\mathbb{P})\rightarrow(\M,\mathcal{B}(\M))}$, for any $\omega\in\Omega$ we denote the set of piecewise smooth curves between $X(\omega)$ and $Y(\omega)$ by
	\begin{equation}
		\mathcal{C}_{X,Y}(\omega) = \left\{ \gamma(\omega): [0,T]\rightarrow \M \, \Big| \, \gamma_0(\omega)=X(\omega), \, \gamma_T(\omega)=Y(\omega) \right\} .
	\end{equation}
	We define a stochastic Riemannian distance function on the space of all square integrable random variables $X:\Omega\rightarrow\M$ by
	\begin{equation}
		d(X,Y)(\omega) =  \begin{cases} \inf \left\{ \int_0^T \sqrt{ g(\dot{\gamma}_t,\dot{\gamma}_t)(\gamma_t) }\, dt \, \Big| \, \gamma\in \mathcal{C}_{X,Y}(\omega)\right\} \quad &{\rm if } \; C_{X,Y}(\omega) \neq \emptyset \, , \\
			\infty \quad & {\rm if } \; C_{X,Y}(\omega) = \emptyset\, .
		\end{cases}
	\end{equation}
\end{mydef}

\begin{mydef}(Stochastic Lorentzian distance)\\
	Given a Lorentzian manifold $(\M,g)$ with $(-+...+)$ signature and random variables ${X,Y:(\Omega,\Sigma,\mathbb{P})\rightarrow(\M,\mathcal{B}(\M))}$, for any $\omega\in\Omega$ we denote the set of causal and anti-causal piecewise smooth curves between $X(\omega)$ and $Y(\omega)$ by respectively
	\begin{align}
		\mathcal{C}^{ca}_{X,Y}(\omega) &= \left\{ \gamma(\omega): [0,T]\rightarrow \M \, \Big| \, \gamma_0(\omega)=X(\omega), \, \gamma_T(\omega)=Y(\omega), \, g(\dot{\gamma}_t,\dot{\gamma}_t)(\gamma_t)(\omega)\leq0 \; \forall\, t\in[0,T] \right\}  \nonumber\\
		\mathcal{C}^{ac}_{X,Y}(\omega) &= \left\{ \gamma(\omega): [0,T]\rightarrow \M \, \Big| \, \gamma_0(\omega)=X(\omega), \, \gamma_T(\omega)=Y(\omega), \, g(\dot{\gamma}_t,\dot{\gamma}_t)(\gamma_t)(\omega) \geq0 \; \forall\, t\in[0,T] \right\} 
	\end{align}
	We define stochastic Lorentzian distance functions on the space of all square integrable random variables $X:\Omega\rightarrow\M$ by
	\begin{align}
		\Delta\tau(X,Y)(\omega) &=  
		\begin{cases} 
			\sup \left\{ \int_0^T \sqrt{ -g(\dot{\gamma}_t,\dot{\gamma}_t)(\gamma_t) }\, dt \, \Big| \, \gamma\in \mathcal{C}^{ca}_{X,Y}(\omega)\right\} \quad &{\rm if } \; C^{ca}_{X,Y}(\omega) \neq \emptyset \, , \\
			-\infty \quad & {\rm if } \; C^{ca}_{X,Y}(\omega) = \emptyset\, .
		\end{cases}\\
		\Delta s(X,Y)(\omega) &=  
		\begin{cases} 
			\sup \left\{ \int_0^T \sqrt{g(\dot{\gamma}_t,\dot{\gamma}_t)(\gamma_t) }\, dt \, 	\Big| \, \gamma\in \mathcal{C}^{ac}_{X,Y}(\omega)\right\} \quad &{\rm if } \; C^{ac}_{X,Y}(\omega) \neq \emptyset \, , \\
			-\infty \quad & {\rm if } \; C^{ac}_{X,Y}(\omega) = \emptyset\, ,
		\end{cases}
	\end{align}
	which we call the proper time and proper distance between $X$ and $Y$.
\end{mydef}
\begin{mydef}{\rm (Induced $L^2$-norm)}\\
	Given a random variable $X:(\Omega,\Sigma,\mathbb{P})\rightarrow(\M,\mathcal{B}(\M))$ and a Borel measurable function $f:\M\rightarrow \mathbb{K}$ with $\mathbb{K}\in\{\R,\mathbb{C}\}$, the $L^2$-norm for $f\circ X$ is defined by 
	\begin{equation}
		||f(X)|| = \sqrt{\E\big[|f(X)|^2\big]} \, .
	\end{equation}
\end{mydef}
\begin{mydef}{\rm (Induced $L^2$-space)}\\
	Given a random variable $X:(\Omega,\Sigma,\mathbb{P})\rightarrow(\M,\mathcal{B}(\M))$, the spaces
	\begin{equation}
		L^2(\M,\mathcal{B}(\M),\mu_X) = \left\{ f:(\M,\mathcal{B}(\M)) \rightarrow (\mathbb{K},\mathcal{B}(\mathbb{K})) \, \Big| \, ||f(X)||<\infty \right\} 
	\end{equation}
	with $\mathbb{K}\in\{\R,\mathbb{C}\}$ are the real and complex $L^2$-space over $(\M,\mathcal{B}(\M),\mu_X)$.
\end{mydef}
\begin{theorem}{\rm (Properties of $L^2$-spaces)}\\
	$L^2$-spaces have the following properties:
	\begin{itemize}
		\item $L^2$ is a Hilbert space with inner product 
		\begin{itemize}
			\item $X \cdot Y = \E[\delta_{ij} \overline{X}{}^i Y^j]$ on $L^2(\Omega,\Sigma,\mathbb{P})\,,$
			\item $\langle f,g\rangle_{\mu_X} = \int_\M \overline{f(x)} \ g(x) \, d\mu_X(x)$ on $L^2(\M,\mathcal{B}(\M),\mu_X)\,;$
		\end{itemize}
		\item $L^2$ is self-dual, i.e. $L^2\cong(L^2)^\ast$.
	\end{itemize}
\end{theorem}

\subsection{Generating Functions}\label{ap:momgenfnct}
\begin{mydef}{\rm (Dual norm)}\\
	Given the space $L^2(\Omega,\Sigma,\mathbb{P})$, the norm on the dual space $[L^2(\Omega,\Sigma,\mathbb{P})]^\ast$ is given by
	\begin{align}
		||a|| &= \sup\left\{ \E[|a(X)|] \, \Big| \, X\in L^2(\Omega,\Sigma,\mathbb{P}), \, ||X|| \leq 1  \right\} .
	\end{align}
\end{mydef}
\begin{mydef}{\rm (Moment generating function)}\\
	Given a real random variable $X:(\Omega,\Sigma,\mathbb{P})\rightarrow(\R^n,\mathcal{B}(\R^n))$ and a linear form ${a\in [L^2(\Omega,\Sigma,\mathbb{P})]^\ast}$, the moment generating function of $X$ is defined by
	\begin{equation}
		M_{X}(a) := \E\left[ e^{a(X)}\right] ,
	\end{equation}
	where $||a||<\rho$ with $\rho$ the radius of convergence of $M_{X}$.\\
	Given a complex random variable $Z:(\Omega,\Sigma,\mathbb{P})\rightarrow(\mathbb{C}^n,\mathcal{B}(\mathbb{C}^n))$ and a linear form ${a\in [L^2(\Omega,\Sigma,\mathbb{P})]^\ast}$, the moment generating function of $Z$ is defined by
	\begin{equation}
		M_{Z}(a) := \E\left[ e^{\overline{a}(X)}\right] ,
	\end{equation}
	where $||a||<\rho$ with $\rho$ the radius of convergence of $M_{Z}$.
\end{mydef}

These functions are called the moment generating functions, since they generate the moments of random variables. Indeed, it can easily be verified that
\begin{align}
	\E\left[\prod_{i=1}^n (X^i)^{k_i} \right] = \prod_{i=1}^n \frac{d^{k_i}}{d a_i^{k_i}} \, \E\left[ e^{a_j X^j}\right]\Big|_{a=0} \qquad \forall \, k\in\hat{\mathbb{N}}^n \, ,\nonumber\\
	\E\left[\prod_{i=1}^n (Z^i)^{k_i} \right] = \prod_{i=1}^n \frac{d^{k_i}}{d \overline{a}{}_i^{k_i}} \, \E\left[ e^{\overline{a}{}_j Z^j}\right]\Big|_{a=0} \qquad \forall \, k\in\hat{\mathbb{N}}^n \, .
\end{align}

\begin{mydef}{\rm (Characteristic function)}\\
	Given a real random variable $X:(\Omega,\Sigma,\mathbb{P})\rightarrow(\R^n,\mathcal{B}(\R^n))$ and a linear form ${a\in [L^2(\Omega,\Sigma,\mathbb{P})]^\ast}$, the characteristic function of $X$ is defined by
	\begin{equation}
		\varphi_{X}(a) := \E\left[ e^{\ri \, a(X)}\right] .
	\end{equation}
	Given a complex random variable $Z:(\Omega,\Sigma,\mathbb{P})\rightarrow(\mathbb{C}^n,\mathcal{B}(\mathbb{C}^n))$ and a linear form ${a\in [L^2(\Omega,\Sigma,\mathbb{P})]^\ast}$, the characteristic function of $Z$ is defined by
	\begin{equation}
		\varphi_{Z}(a) := \E\left[ e^{\ri \, {\rm Re}[\overline{a}(Z)]}\right] .
	\end{equation}
\end{mydef}

If $X$ admits a probability density, then the characteristic function $\varphi_X$ and probability density $\rho_X$ are each others Fourier transform. Similarly, if $Z=X+ \ri \, Y$ admits a probability density, then the characteristic function $\varphi_Z$ and probability density ${\rho_Z=\rho_{(X,Y)}}$ are each others Fourier transform.

\clearpage
\section{Review of Stochastic Processes}\label{ap:ReviewStochProc}
In this appendix, we list various standard definitions and results from the theory of stochastic processes. For a more detailed discussion of these definitions and results, we refer to textbooks on probability theory and stochastic processes, e.g. Refs.~\cite{Karatzas,Cinlar,SchreveI,SchreveII,Emery}.
\subsection{Stochastic Processes}
\begin{mydef}{\rm (Stochastic process)}\\
	Given a probability space $(\Omega,\Sigma,\mathbb{P})$, a measurable space $(\M,\mathcal{B}(\M))$ and a set ${\mathcal{T}\subseteq\R}$, an $\M$-valued stochastic process is a family of random variables $\{X_t \, | \, t\in\mathcal{T}\}$ with \\ ${X_t:(\Omega,\Sigma,\mathbb{P})\rightarrow(\M,\mathcal{B}(\M)) \quad \forall\, t\in\mathcal{T}} \,$.
\end{mydef}
The possible outcomes of the stochastic process are given by the sample paths
\begin{mydef}{\rm (Sample path)}\\
	Given a stochastic process $X:\mathcal{T}\times (\Omega,\Sigma,\mathbb{P}) \rightarrow (\M,\mathcal{B}(\M))$ and an event $\omega\in\Omega$, a sample path is a function $X(\cdot,\omega) : \mathcal{T} \rightarrow \M$.
\end{mydef}

\subsection{Conditioning}\label{ap:filtration}
One is often interested in properties of a stochastic process at a time $t$, given that certain events occur in the past or future. In order to study such properties, we require an object that collects information about past or future events. This object is called a filtration.
\begin{mydef}{\rm (Filtration)}\\
	Given a  measurable space $(\Omega,\Sigma,\mathbb{P})$ and a set $\mathcal{T}\subseteq\R$, a filtration (or past filtration) is a family $\vec{\mathcal{F}}=\{\vec{\mathcal{F}}_t \, | \, t\in\mathcal{T}\}$ such that $\{\emptyset,\Omega\}\subseteq\vec{\mathcal{F}}_s \subseteq \vec{\mathcal{F}}_t \subseteq \Sigma \quad \forall s<t\in\mathcal{T}\,$.
\end{mydef}
\begin{mydef}{\rm (Time reversed filtration)}\\
	Given a  measurable space $(\Omega,\Sigma,\mathbb{P})$ and a set $\mathcal{T}\subseteq\R$, a time reversed filtration (or future filtration) is a family $\cev{\mathcal{F}}=\{\cev{\mathcal{F}}_t \, | \, t\in\mathcal{T}\}$ such that $\{\emptyset,\Omega\}\subseteq\cev{\mathcal{F}}_s \subseteq \cev{\mathcal{F}}_t \subseteq \Sigma \quad \forall s>t\in\mathcal{T}\,$.
\end{mydef}
There exist many different filtrations on a probability space $(\Omega,\Sigma,\mathbb{P})$, but not all of them are equally useful for the study of a process $X$. If we can determine the state $X_t$ from the information given by $\mathcal{F}_t$, we say that $X$ is adapted to the filtration $\mathcal{F}$.
\begin{mydef}{\rm (Adaptedness)}\\
	Given a measurable space $(\Omega,\Sigma,\mathbb{P})$ equipped with a filtration $\mathcal{F}$ and a stochastic process $X:\mathcal{T}\times (\Omega,\Sigma,\mathbb{P}) \rightarrow (\M,\mathcal{B}(\M))$, $X$ is said to be adapted to the filtration $\mathcal{F}$, if $X_t$ is $\mathcal{F}_t$-measurable for every $t\in\mathcal{T}\, $.
\end{mydef}
For any process $X$, we can also define a natural filtration, which is the minimal filtration that contains all the information about the process $X$.
\begin{mydef}{\rm (Natural filtration)}\\
	Given a stochastic process $X:\mathcal{T}\times (\Omega,\Sigma,\mathbb{P}) \rightarrow (\M,\mathcal{B}(\M))$, the natural (past) filtration of $\Sigma$ with respect to $X$ is given by $\vec{\mathcal{F}}^X=\{\vec{\mathcal{F}}^X_t\,|\,t\in\mathcal{T}\}$ with
	\begin{equation}
		\vec{\mathcal{F}}_t^X = \sigma\big(\{X_s \, | \, s\leq t\in\mathcal{T} \}\big)\,
	\end{equation}
	and the natural future filtration of $\Sigma$ with respect to $X$ is given by ${\cev{\mathcal{F}}{}^X=\{\cev{\mathcal{F}}{}^X_t \,|\,t\in\mathcal{T}\}}$ with
	\begin{equation}
		\cev{\mathcal{F}}{}_t^X = \sigma\big(\{X_s \, | \, s\geq t\in\mathcal{T}\}\big)\,.
	\end{equation}
\end{mydef}

\noindent It immediately follows that any stochastic process $X$ is adapted to its natural filtration.
\begin{mydef}{\rm (Filtered probability space)}\\
	A filtered probability space is a tuple $(\Omega,\Sigma,\mathcal{F},\mathbb{P})$, where $(\Omega,\Sigma,\mathbb{P})$ is a probability space and $\mathcal{F}$ is a filtration of $\Sigma\,$.
\end{mydef}
\noindent
If a process $X:(\Omega,\Sigma,\mathbb{P})\rightarrow(\M,\mathcal{B}(\M))$ is adapted to a filtration $\mathcal{F}$, it can be defined on the filtered probability, such that $X:(\Omega,\Sigma,\mathcal{F},\mathbb{P})\rightarrow(\M,\mathcal{B}(\M))$.
\subsection{Stopping Times}
Stopping times are random variables that define the time at which a stochastic process exhibits a certain type of behavior. A typical example is the first time at which the stochastic process starting at $x\in U\subset\M$ hits the boundary $\p U$ of $U\subset\M$.
\begin{mydef}{\rm (Stopping time)}\\
	Given a set $\mathcal{T}\subseteq\R$ and a filtered probability space $(\Omega,\Sigma,\vec{\mathcal{F}},\mathbb{P})$, a random variable ${\tau: (\Omega,\Sigma,\vec{\mathcal{F}},\mathbb{P}) \rightarrow (\mathcal{T},\mathcal{B}(\mathcal{T}))}$ is a stopping time with respect to the past $\vec{\mathcal{F}}$, if $\{\tau\leq t\}\in\vec{\mathcal{F}}_t$ for all $t\in\mathcal{T}$.
\end{mydef}
\begin{mydef}{\rm (Time reversed stopping time)}\\
	Given a set $\mathcal{T}\subseteq\R$ and a filtered probability space $(\Omega,\Sigma,\cev{\mathcal{F}},\mathbb{P})$, a random variable ${\tau: (\Omega,\Sigma,\cev{\mathcal{F}},\mathbb{P}) \rightarrow (\mathcal{T},\mathcal{B}(\mathcal{T}))}$ is a stopping time with respect to the future $\cev{\mathcal{F}}$, if ${\{\tau\geq t\}\in\cev{\mathcal{F}}_t}$ for all $t\in\mathcal{T}$.
\end{mydef}
\noindent If we stop a process at a stopping time, we obtain a stopped process.
\begin{mydef}{\rm (Stopped process)}\\
	Given a stochastic process $X:\mathcal{T}\times(\Omega,\Sigma,\vec{\mathcal{F}},\mathbb{P})\rightarrow(\M,\mathcal{B}(\M))$, and a stopping time $\tau: (\Omega,\Sigma,\vec{\mathcal{F}},\mathbb{P}) \rightarrow (\mathcal{T},\mathcal{B}(\mathcal{T}))$, the process $\{X^\tau_t\,|\,t\in\mathcal{T}\}$ such that $X^\tau_t=X_{\min\{t,\tau\}}$ is called a stopped process.
	\\
	Similarly, given a stochastic process $X:\mathcal{T}\times(\Omega,\Sigma,\cev{\mathcal{F}},\mathbb{P})\rightarrow(\M,\mathcal{B}(\M))$, and a time reversed stopping time $\tau: (\Omega,\Sigma,\cev{\mathcal{F}},\mathbb{P}) \rightarrow (\mathcal{T},\mathcal{B}(\mathcal{T}))$, the process $\{X^\tau_t\,|\,t\in\mathcal{T}\}$ such that $X^\tau_t=X_{\max\{t,\tau\}}$ is called a stopped process.
\end{mydef}
\subsection{Semi-Martingales}
Semi-martingales are processes that can be decomposed into a deterministic drift process and a driftless stochastic process. Semi-martingales are particularly useful, as they form the largest class of stochastic processes for which there exists a stochastic calculus. 
\begin{mydef}{\rm (C\`adl\`ag process)}\\
	A process $X:\mathcal{T}\times(\Omega,\Sigma,\mathbb{P})\rightarrow (\M,\mathcal{B}(\M))$ is  c\`adl\`ag, if
	\begin{itemize}
		\item $\lim_{s\uparrow t} X(s)$ exists a.s. for all $t\in\mathcal{T}$;
		\item $\lim_{s\downarrow t} X(s)= X(t)$ a.s. for all $t\in\mathcal{T}$.
	\end{itemize}
\end{mydef}
\begin{mydef}{\rm (C\`agl\`ad process)}\\
	A process $X:\mathcal{T}\times(\Omega,\Sigma,\mathbb{P})\rightarrow (\M,\mathcal{B}(\M))$ is  c\`agl\`ad, if
	\begin{itemize}
		\item $\lim_{s\uparrow t} X(s) = X(t)$ a.s. for all $t\in\mathcal{T}$;
		\item $\lim_{s\downarrow t} X(s)$ exists a.s. for all $t\in\mathcal{T}$.
	\end{itemize}
\end{mydef}
\begin{mydef}{\rm (Continuous process)}\\
A process $X:\mathcal{T}\times(\Omega,\Sigma,\mathbb{P})\rightarrow (\M,\mathcal{B}(\M))$ is continuous, if
	\begin{itemize}
		\item $\lim_{s\uparrow t} X(s) = X(t)$ a.s. for all $t\in\mathcal{T}$;
		\item $\lim_{s\downarrow t} X(s)= X(t)$ a.s. for all $t\in\mathcal{T}$.
	\end{itemize}
\end{mydef}
\begin{mydef}{\rm (Finite variation)}\\
	A process $X:\mathcal{T}\times(\Omega,\Sigma,\mathbb{P})\rightarrow(\R^n,\mathcal{B}(\R^n))$ has finite variation, if for any finite interval $I\subseteq\mathcal{T}$
	\begin{equation}
		\mathbb{P}\left[\sup_\Pi\left(\sum_{k} \left|X^i_{t_k} - X^i_{t_{k-1}}\right|\right) < \infty\right] = 1 \qquad \forall\, i\in\{1,...,n\}\,,
	\end{equation}
	where the supremum is taken over all partitions $\Pi$ of $I$.
\end{mydef}
\begin{mydef}{\rm (Martingale process)}\\
	A stochastic process $X:\mathcal{T}\times(\Omega,\Sigma,\vec{\mathcal{F}},\mathbb{P})\rightarrow(\R^n,\mathcal{B}(\R^n))$ is a martingale with respect to the past $\vec{\mathcal{F}}$, if $X$ is adapted to $\vec{\mathcal{F}}$, $X_t$ is integrable for every $t\in\mathcal{T}$ and
	\begin{equation}\label{eq:MartingaleProperty}
		\E\left[X_t \, \Big| \, \vec{\mathcal{F}}_s \right] = X_s \qquad \forall \, s<t\in\mathcal{T} \, .
	\end{equation} 
\end{mydef}
\begin{mydef}{\rm (Time reversed martingale)}\\
	A stochastic process $X:\mathcal{T}\times(\Omega,\Sigma,\cev{\mathcal{F}},\mathbb{P})\rightarrow(\R^n,\mathcal{B}(\R^n))$ is a martingale with respect to the future $\cev{\mathcal{F}}$, if $X$ is adapted to $\cev{\mathcal{F}}$, $X_t$ is integrable for every $t\in\mathcal{T}$ and
	\begin{equation}
		\E\left[X_t \, \Big| \, \cev{\mathcal{F}}_s \right] = X_s \qquad \forall \, s>t\in\mathcal{T} \,.
	\end{equation} 
\end{mydef}
The martingale property \eqref{eq:MartingaleProperty}, which ensure that the process is driftless, is a global property, as it must hold for all pairs $s,t\in\mathcal{T}$, but the property can be localized. All martingales are local martingales, but the opposite is not true.
\begin{mydef}{\rm (Local martingale)}\\
	A stochastic process $X:\mathcal{T}\times(\Omega,\Sigma,\vec{\mathcal{F}},\mathbb{P})\rightarrow(\R^n,\mathcal{B}(\R^n))$ is a local martingale with respect to the past $\vec{\mathcal{F}}$, if $X$ is adapted to $\vec{\mathcal{F}}$ and there exists a sequence of stopping times $\tau_k:(\Omega,\Sigma,\vec{\mathcal{F}},\mathbb{P})\rightarrow(\mathcal{T},\mathcal{B}(\mathcal{T}))$ such that
	\begin{itemize}
		\item $\mathbb{P}(\tau_{k+1}>\tau_k) = 1\,$;
		\item $\mathbb{P}[\lim_{k\rightarrow\infty} \tau_k = \sup(\mathcal{T}) ] = 1\,$;
		\item the stopped process $X_{{\min\{t,\tau_k\}}}$ is a martingale with respect to $\vec{\mathcal{F}}$ for every $k\in\mathbb{N}\,$.
	\end{itemize}
\end{mydef}
\begin{mydef}{\rm (Time reversed local martingale)}\\
	A stochastic process $X:\mathcal{T}\times(\Omega,\Sigma,\cev{\mathcal{F}},\mathbb{P})\rightarrow(\R^n,\mathcal{B}(\R^n))$ is a local martingale with respect to the future $\cev{\mathcal{F}}$, if $X$ is adapted to $\cev{\mathcal{F}}$ and there exists a sequence of time reversed stopping times $\tau_k:(\Omega,\Sigma,\cev{\mathcal{F}},\mathbb{P})\rightarrow(\mathcal{T},\mathcal{B}(\mathcal{T}))$ such that
	\begin{itemize}
		\item $\mathbb{P}(\tau_{k+1} < \tau_k) = 1\,$;
		\item $\mathbb{P}[\lim_{k\rightarrow\infty} \tau_k = \inf(\mathcal{T}) ] = 1\,$;
		\item the stopped process $X_{\max\{t,\tau_k\}}$ is a martingale with respect to $\cev{\mathcal{F}}$ for every $k\in\mathbb{N}\,$.
	\end{itemize}
\end{mydef}
\begin{mydef}{\rm (Semi-martingale)}\\
	A stochastic process $X:\mathcal{T}\times(\Omega,\Sigma,\vec{\mathcal{F}},\mathbb{P})\rightarrow(\R^n,\mathcal{B}(\R^n))$ is a semi-martingale with respect to the past $\vec{\mathcal{F}}$, if it can be decomposed as
	\begin{equation}
		X_t = C_t + M_t \, ,
	\end{equation}
	where $C$ is a c\`adl\`ag process of finite variation adapted to $\vec{\mathcal{F}}$ and $M$ is a local martingale with respect to $\vec{\mathcal{F}}$.
\end{mydef}
\begin{mydef}{\rm (Time reversed semi-martingale)}\\
	A stochastic process $X:\mathcal{T}\times(\Omega,\Sigma,\cev{\mathcal{F}},\mathbb{P})\rightarrow(\R^n,\mathcal{B}(\R^n))$ is a semi-martingale with respect to the future $\cev{\mathcal{F}}$, if it can be decomposed as
	\begin{equation}
		X_t = C_t + M_t \, ,
	\end{equation}
	where $C$ is a c\`agl\`ad process of finite variation adapted to $\cev{\mathcal{F}}$ and $M$ is a local martingale with respect to $\cev{\mathcal{F}}$.
\end{mydef}
\begin{mydef}{\rm (Two-sided semi-martingale)}\\
	A stochastic process $X:\mathcal{T}\times(\Omega,\Sigma,\vec{\mathcal{F}},\cev{\mathcal{F}},\mathbb{P})\rightarrow(\R^n,\mathcal{B}(\R^n))$ is a two-sided semi-martingale with respect to the past $\vec{\mathcal{F}}$ and the future $\cev{\mathcal{F}}$, if it can be decomposed as
	\begin{equation}
		X_t = C_{\pm,t} + M_t \, ,
	\end{equation}
	where $C_+$ is a c\`adl\`ag process of finite variation adapted to $\vec{\mathcal{F}}$, $C_-$ is a c\`agl\`ad process of finite variation adapted to $\cev{\mathcal{F}}$ and $M$ is a local martingale with respect to both $\vec{\mathcal{F}}$ and $\cev{\mathcal{F}}$.
\end{mydef}
\begin{mydef}{\rm (Complex valued semi-martingale)}\\
	A process ${Z:\mathcal{T}\times(\Omega,\Sigma,\mathcal{F},\mathbb{P})\rightarrow(\mathbb{C}^n,\mathcal{B}(\mathbb{C}^n)}$ is a complex semi-martingale with respect to $\mathcal{F}$, if it can be decomposed as $Z=X+\ri \, Y$, where $X,Y:\mathcal{T}\times(\Omega,\Sigma,\mathcal{F},\mathbb{P})\rightarrow(\R^n,\mathcal{B}(\R^n))$ are real semi-martingales with respect to $\mathcal{F}$.
\end{mydef}
\begin{theorem}{\rm (Functions of semi-martingales)}\\
	Given a semi-martingale $X:\mathcal{T}\times(\Omega,\Sigma,\mathcal{F},\mathbb{P})\rightarrow(\R^n,\mathcal{B}(\R^n))$ and a Borel measurable function $f\in C^2(\R^n)$, then $f\circ X$ is a semi-martingale.
\end{theorem}

\begin{mydef}{\rm (Manifold valued semi-martingale)}\\
	Given a smooth manifold $\M$, a stochastic process $X:\mathcal{T}\times(\Omega,\Sigma,\mathcal{F},\mathbb{P})\rightarrow(\M,\mathcal{B}(\M))$ is a semi-martingale with respect to $\mathcal{F}$, if $f\circ X$ is a semi-martingale for every $f\in C^\infty(\M)$.
\end{mydef}

\subsection{Markov Processes}
Markov processes form a class of stochastic processes that obey a memoryless property, which is known as the Markov property.
\begin{mydef}{\rm (Markov process)}\\
	A stochastic process $X:\mathcal{T}\times(\Omega,\Sigma,\vec{\mathcal{F}},\mathbb{P})\rightarrow(\R^n,\mathcal{B}(\R^n))$ is a Markov process with respect to the past $\vec{\mathcal{F}}$, if $X$ is adapted to $\vec{\mathcal{F}}$, $X_t$ is integrable for every $t\in\mathcal{T}$ and
	\begin{equation}
		\E[X_t \, | \, \vec{\mathcal{F}}_s] = \E[X_t \, | \, X_s] \qquad \forall \; s<t\in\mathcal{T} \, .
	\end{equation}
\end{mydef}
\begin{mydef}{\rm (Time reversed Markov process)}\\
	A stochastic process $X:\mathcal{T}\times(\Omega,\Sigma,\cev{\mathcal{F}},\mathbb{P})\rightarrow(\R^n,\mathcal{B}(\R^n))$ is a Markov process with respect to the future $\cev{\mathcal{F}}$, if $X$ is adapted to $\cev{\mathcal{F}}$, $X_t$ is integrable for every $t\in\mathcal{T}$ and 
	\begin{equation}
		\E[X_t \, | \, \cev{\mathcal{F}}_s] = \E[X_t \, | \, X_s] \qquad \forall \; s>t\in\mathcal{T} \, .
	\end{equation}
\end{mydef}
\begin{mydef}{\rm (Complex valued Markov process)}\\
	A complex stochastic process ${Z:\mathcal{T}\times(\Omega,\Sigma,\mathcal{F},\mathbb{P})\rightarrow(\mathbb{C}^n,\mathcal{B}(\mathbb{C}^n))}$ is a Markov process, if it can be decomposed as $Z=X+\ri\,Y$, where $X,Y:\mathcal{T}\times(\Omega,\Sigma,\mathcal{F},\mathbb{P})\rightarrow(\M,\mathcal{B}(\M))$ are real Markov processes.
\end{mydef}
\begin{mydef}{\rm (Manifold valued Markov process)}\\
	Given a smooth manifold $\M$, a stochastic process $X:\mathcal{T}\times(\Omega,\Sigma,\mathcal{F},\mathbb{P})\rightarrow(\M,\mathcal{B}(\M))$ is a Markov process, if $f\circ X$ is a Markov process for every $f\in C^\infty(\M)$.
\end{mydef}

\subsection{Quadratic Variation}\label{ap:QVar}
A characteristic feature of stochastic processes is their non-vanishing quadratic variation. On $\R^n$ this quadratic variation can be defined as a Riemann sum of squares. In the remaining sections of this appendix, we focus on the future directed process, i.e. adapted to the past $\vec{\mathcal{F}}$, but all results can be generalized to a past directed process, i.e. adapted to the future $\cev{\mathcal{F}}$.

\begin{mydef}{\rm (Quadratic variation)}\\
	Given two stochastic processes $X,Y:\mathcal{T}\times(\Omega,\Sigma,\vec{\mathcal{F}},\mathbb{P})\rightarrow(\R^n,\mathcal{B}(\R^n))$, the quadratic covariation of $X$ and $Y$ is defined by
	\begin{equation}\label{eq:defQVar}
		[X,Y]_t = \lim_{||\Pi||\rightarrow0} \sum_k \left( X_{t_k} - X_{t_{k-1}}\right) \otimes \left(Y_{t_k} - Y_{t_{k-1}} \right) ,
	\end{equation}
	where $\Pi$ is a partition of $[\inf(\mathcal{T}),t]$ and $||\Pi||$ is its mesh. Moreover, the processes $[X,X]$ and $[Y,Y]$ are called the quadratic variation of $X$ and $Y$ respectively.
\end{mydef}
\begin{theorem}{\rm (Quadratic variation)}\\
	Given two semi-martingales $X,Y:\mathcal{T}\times(\Omega,\Sigma,\vec{\mathcal{F}},\mathbb{P})\rightarrow(\R^n,\mathcal{B}(\R^n))$, the quadratic covariation $[X,Y]$ is a $(\R^{n\times n})$-valued c\`adl\`ag process of finite variation adapted to $\vec{\mathcal{F}}$.
\end{theorem}

\begin{theorem}{\rm (Properties of quadratic variation I)}\\
	Given semi-martingales $X,Y,Z:\mathcal{T}\times(\Omega,\Sigma,\vec{\mathcal{F}},\mathbb{P})\rightarrow(\R^n,\mathcal{B}(\R^n))$, the quadratic covariation satisfies the following properties
	\begin{itemize}
		\item (symmetry): $[X,Y]=[Y,X]\,$;
		\item (bilinearity): $[a \, X + b \, Y ,Z] = a \, [X,Z] + b \, [Y,Z]\;$ for all $a,b\in\R\,$;
		\item (positivity): $[X,X]_t-[X,X]_s$ is a.s. positive semi-definite for all $s\leq t\in\mathcal{T}$.
	\end{itemize}
\end{theorem}
A stochastic processes $X:\mathcal{T}\times(\Omega,\Sigma,\mathcal{F},\mathbb{P})\rightarrow(\R^n,\mathcal{B}(\R^n))$ is not necessarily continuous, but, if it is a semi-martingale, the number of discontinuities is countable, such that
\begin{align}
	X_t = X_t^c + \sum_{k} \Delta X_{t_k} \, ,
\end{align}
where $X^c$ denotes the continuous part of $X$, $\Delta X_{t_k}$ the size of the jumps and $k$ ranges over all jumps up to time $t$. Similarly, for the quadratic variation of two semi-martingales $X,Y$, we can write
\begin{align}
	[X,Y]_t &= [X,Y]^c_t + \sum_{k} \Delta [X,Y]_{t_k} ,
\end{align}
where
\begin{equation}
	\Delta [X,Y]_{t_k} = \Delta X_{t_k} \, \Delta Y_{t_k}\, .
\end{equation}
Using this decomposition, we can formulate a few more properties.
\begin{lemma}{\rm (Properties quadratic variation III)}\\
	Given semi-martingales $X,Y:\mathcal{T}\times(\Omega,\Sigma,\mathcal{F},\mathbb{P})\rightarrow(\R^n,\mathcal{B}(\R^n))$, the quadratic covariation satisfies the following properties
	\begin{itemize}
		\item $[X,Y]_t^c=0$, if $X_t$ or $Y_t$ has finite variation;
		\item $\Delta[X,Y]_{t_k}=0$, if $\Delta X_{t_k}=0$ or $\Delta Y_{t_k}=0$.
	\end{itemize}
\end{lemma}
\begin{corollary}{\rm (Properties quadratic variation IV)}\\
	Given semi-martingales $X,Y,Z:\mathcal{T}\times(\Omega,\Sigma,\mathcal{F},\mathbb{P})\rightarrow(\R^n,\mathcal{B}(\R^n))$, the quadratic covariation satisfies the following properties
	\begin{itemize}
		\item $[X,Y]$ is continuous, if at least one of the processes $X,Y$ is continuous;
		\item $[[X,Y],Z]=0$, if at least one of the processes $X,Y,Z$ is continuous.
	\end{itemize}
\end{corollary}
In appendix \ref{Ap:2Geometry}, we show that the sum \eqref{eq:defQVar} used in the definition of the quadratic variation is intrinsic on a pseudo-Riemannian manifold. This allows to construct a manifold valued quadratic variation process $[X,Y]:\mathcal{T}\times(\Omega,\Sigma,\mathcal{F},\mathbb{P})\rightarrow(\M^{\otimes2},\mathcal{B}(\M)^{\otimes2})$ on all pseudo-Riemannian manifolds $(\M,g)$, which satisfies the same properties.
\par

Furthermore, since all complex valued processes can be decomposed into real valued processes, one can straightforwardly generalize the quadratic variation and its properties to complex valued processes.
\par 

We emphasize that all properties of the quadratic variation rely on the fact that the semi-martingales are adapted to the same filtration $\vec{F}$. If $X$ and $Y$ cannot be adapted to a common filtration, the limit \eqref{eq:defQVar} may not exist. This fact is crucial for the discussion in section \ref{sec:MomentumProcess}.

\subsection{L\'evy Processes}\label{ap:Levy}
L\'evy processes form a class of Markov processes, and can be regarded as the continuous time analog of the random walk.
\begin{mydef}{\rm (L\'evy process)}\\
	A stochastic process $X:\mathcal{T}\times(\Omega,\Sigma,\mathbb{P})\rightarrow(\R^n,\mathcal{B}(\R^n))$ is a L\'evy process, if it is continuous in probability and has independent stationary increments, i.e.
	\begin{itemize}
		\item $\forall \, \epsilon>0,\, t\in\mathcal{T}, \; \lim_{s\rightarrow t} \mathbb{P}(||X_{s}-X_t||\geq\epsilon)=0\,;$
		\item $\forall \, t_1<t_2<t_3<t_4 \in \mathcal{T}, \; (X_{t_4} - X_{t_3}) \indep (X_{t_2} - X_{t_1}) \, ;$
		\item $\forall \, t_1,t_2,t_3,t_4 \in \mathcal{T},\; {\rm s.t.} \; |t_4-t_3|=|t_2-t_1|, \; d\mu_{(X_{t_4} - X_{t_3})} = d\mu_{(X_{t_2} - X_{t_1})} \, .$
	\end{itemize}
\end{mydef}
\begin{mydef}{\rm (Complex L\'evy process)}\\
A stochastic process $Z:\mathcal{T}\times(\Omega,\Sigma,\mathbb{P})\rightarrow(\mathbb{C}^n,\mathcal{B}(\mathbb{C}^n))$ is a complex valued L\'evy process, if it can be decomposed as $Z= X +\ri \, Y$, where $X,Y:\mathcal{T}\times(\Omega,\Sigma,\mathbb{P})\rightarrow(\R^n,\mathcal{B}(\R^n))$ are independent real valued L\'evy processes.
\end{mydef}
\begin{mydef}{\rm (L\'evy process w.r.t. a filtration)}\\
	Given a L\'evy process $X:\mathcal{T}\times(\Omega,\Sigma,\mathbb{P})\rightarrow(\R^n,\mathcal{B}(\R^n))$, we call ${X:\mathcal{T}\times(\Omega,\Sigma,\mathcal{F},\mathbb{P})\rightarrow(\R^n,\mathcal{B}(\R^n))}$ a L\'evy process with respect to the filtration $\mathcal{F}$, if $X$ is adapted to $\mathcal{F}$ and $(X_t-X_s)\indep \mathcal{F}_s$ for all $t>s\in\mathcal{T}$ ($s<t$, if $\mathcal{F}$ is a time reversed filtration).
\end{mydef}

\begin{theorem}{\rm (L\'evy process is Markov)}\\
	A L\'evy process $X:\mathcal{T}\times(\Omega,\Sigma,\mathcal{F},\mathbb{P})\rightarrow(\R^n,\mathcal{B}(\R^n))$ is a Markov process w.r.t. $\mathcal{F}$.
\end{theorem}
\begin{theorem}{\rm (L\'evy process is semi-martingale)}\\
	A L\'evy process $X:\mathcal{T}\times(\Omega,\Sigma,\mathcal{F},\mathbb{P})\rightarrow(\R^n,\mathcal{B}(\R^n))$ is a semi-martingale w.r.t. $\mathcal{F}$.
\end{theorem}

\subsection{Wiener Processes}\label{ap:Wiener}
The Wiener process is a continuous L\'evy process that is best known for its application in describing Brownian motion. For this reason, the terms Wiener process and Brownian motion are often used interchangeably.
\begin{mydef}{\rm (Wiener process)}\\
	A stochastic process $X:\mathcal{T}\times(\Omega,\Sigma,\mathbb{P})\rightarrow(\R^n,\mathcal{B}(\R^n))$ is a Wiener process (a.k.a. Brownian motion), if it is almost surely continuous and has independent normally distributed increments, i.e.
	\begin{itemize}
		\item $\mathbb{P}\big(\lim_{s\rightarrow t} ||X_s - X_t|| = 0\big)=1\,;$
		\item $\forall \, t_1<t_2<t_3<t_4 \in \mathcal{T}, \; (X_{t_4} - X_{t_3}) \indep (X_{t_2} - X_{t_1})\, ;$
		\item $\forall \ t_1,t_2\in\mathcal{T}, \; (X_{t_2}-X_{t_1})\sim \mathcal{N}(0,\alpha \, \delta^{ij} \, |t_2-t_1|)\,,$
	\end{itemize}
	where $\mathcal{N}(\mu,\sigma^2)$ denotes the normal distribution with mean $\mu\in\R^n$ and covariance matrix $\sigma^2\in\R^{n\times n}$. Furthermore, $\alpha\in(0,\infty)$ is a scaling parameter.
\end{mydef}
\begin{mydef}{\rm (Complex Wiener process)}\\
	A stochastic process $Z:\mathcal{T}\times(\Omega,\Sigma,\mathbb{P})\rightarrow(\mathbb{C}^n,\mathcal{B}(\mathbb{C}^n))$ is a complex valued Wiener process, if it can be decomposed as $Z= X +\ri \, Y$, where $X,Y:\mathcal{T}\times(\Omega,\Sigma,\mathbb{P})\rightarrow(\R^n,\mathcal{B}(\R^n))$ are independent real valued Wiener processes.
\end{mydef}
\begin{theorem}{\rm (Wiener process is L\'evy)}\\
	A Wiener process $X:\mathcal{T}\times(\Omega,\Sigma,\mathbb{P})\rightarrow(\R^n,\mathcal{B}(\R^n))$ is a continuous L\'evy process.
\end{theorem}
\begin{theorem}{\rm (Wiener process is martingale)}\\
	A Wiener process $X:\mathcal{T}\times(\Omega,\Sigma,\mathcal{F},\mathbb{P})\rightarrow(\R^n,\mathcal{B}(\R^n))$ is a martingale with respect to $\mathcal{F}$.
\end{theorem}
There exists an alternative definition of the Wiener process, known as the L\'evy characterization \cite{Levy}. In this equivalent definition, the requirement that the increments are normally distributed is replaced by a condition on the quadratic variation.
\begin{theorem}{(\rm L\'evy characterization of Brownian motion)}\\
	Given a continuous local martingale $X:\mathcal{T}\times(\Omega,\Sigma,\vec{\mathcal{F}},\mathbb{P})\rightarrow(\R^n,\mathcal{B}(\R^n))$ and a continuous finite variation process $\sigma^2:\mathcal{T}\rightarrow\R^{n\times n}$, such that $\sigma^2_t-\sigma^2_s$ is positive semi-definite for all $t\geq s\in\mathcal{T}$, then the following statements are equivalent
	\begin{enumerate}[label=(\roman*)]
		\item $(X_t-X_s)\indep \vec{\mathcal{F}}_s$ and $(X_t-X_s)\sim\mathcal{N}(0,\sigma^2_t-\sigma^2_s)$ for all $t\geq s\in\mathcal{T}\,$;
		\item $X_t\otimes X_t - \sigma^2_t$ is a local martingale;
		\item $X$ has a quadratic variation given by $[X,X]_t = \sigma^2_t\,$.
	\end{enumerate}
	In particular, if $(\sigma^2_t - \sigma^2_s)^{ij} =\alpha \, \delta^{ij} \, |t-s|$, (i) implies that $X$ is a Brownian motion (Wiener process) with respect to $\vec{\mathcal{F}}$ with scaling parameter $\alpha$.
\end{theorem}

\subsection{$L^2$-spaces}\label{sec:L2StochProc}
The definitions of the $L^2$-norm and the $L^2$-spaces encountered in appendix \ref{ap:momgenfnct} can be generalized to the context of stochastic processes.
\begin{mydef}{\rm ($L^2$-norm)}\\
	Given a stochastic process $X:\mathcal{T}\times(\Omega,\Sigma,\mathbb{P})\rightarrow(\mathbb{K}^n,\mathcal{B}(\mathbb{K}^n))$ with $\mathbb{K}\in\{\R,\mathbb{C}\}$, the \hbox{$L^2$-norm} for $X$ is defined by 
	\begin{equation}
		||X|| = \sqrt{\E\left[\int_\mathcal{T} \delta_{ij}\overline{X}{}_t^i X_t^j \, dt\right]} \, .
	\end{equation}
\end{mydef}
\begin{mydef}{\rm ($L^2$-space)}\\
	Given a probability space $(\Omega,\Sigma,\mathbb{P})$ and a set $\mathcal{T}\subseteq\R$, the spaces
	\begin{equation}
		L_{\mathcal{T}}^2(\Omega,\Sigma,\mathbb{P}) = \left\{ X:\mathcal{T}\times(\Omega,\Sigma,\mathbb{P})\rightarrow(\mathbb{K}^n,\mathcal{B}(\mathbb{K}^n)) \, \Big| \, ||X||<\infty \right\} \nonumber\\
	\end{equation}
	with $\mathbb{K}\in\{\R,\mathbb{C}\}$ are the real and complex $L^2$-space over $\mathcal{T}\times(\Omega,\Sigma,\mathbb{P})$.
\end{mydef}
\begin{mydef}{\rm (Velocity)}\\
	Given a stochastic process ${X:\mathcal{T}\times (\Omega,\Sigma,\mathbb{P}) \rightarrow (\M,\mathcal{B}(\M))}$, we define the forward It\^o velocity by
	\begin{equation}
		v_+(X_t,t) = \lim_{dt\rightarrow 0} \E\left[ \frac{X_{t+dt} - X_t}{dt} \, \Big| \, X_t \right] ,
	\end{equation}
	the backward It\^o velocity by
	\begin{equation}
		v_-(X_t,t) = \lim_{dt\rightarrow 0} \E\left[ \frac{X_{t} - X_{t-dt}}{dt} \, \Big| \, X_t \right],
	\end{equation}
	the Stratonovich velocity by
	\begin{equation}
		v_\circ(X_t,t) = \frac{1}{2} \, \Big[v_+(X_t,t) + v_-(X_t,t) \Big] ,
	\end{equation}
	and the second order velocity by
	\begin{equation}
		v_2(X_t,t) = \lim_{dt\rightarrow 0} \E\left[ \frac{(X_{t+dt} - X_t)\, (X_{t+dt} - X_t)}{dt} \, \Big| \, X_t \right] .
	\end{equation}
\end{mydef}

\begin{mydef}{\rm (Length)}\\
	Given a Riemannian manifold $(\M,g)$ and a stochastic process ${X:\mathcal{T}\times (\Omega,\Sigma,\mathbb{P}) \rightarrow (\M,\mathcal{B}(\M))}$, we define the length of $X$ by
	\begin{equation}
		\Delta X_\mathcal{T} = \E\left[ \int_\mathcal{T} \sqrt{g(v_\circ,v_\circ)(X_t,t)} \, dt \right] .
	\end{equation}
\end{mydef}
\vbox{
\begin{mydef}{\rm (Time-like, null-like and space-like)}\\
	Given a Lorentzian manifold $(\M,g)$ with $(-+...+)$ signature and a stochastic process $X:\mathcal{T}\times (\Omega,\Sigma,\mathbb{P}) \rightarrow (\M,\mathcal{B}(\M))$, we call the process $X$
	\begin{itemize}
		\item time-like, if $\E\left[ g(v_\circ,v_\circ)(X_t,t)\right] < 0 \quad \forall\, t\in\mathcal{T}$,
		\item null-like, if $\E\left[ g(v_\circ,v_\circ)(X_t,t)\right] = 0 \quad \forall\, t\in\mathcal{T}$,
		\item space-like, if $\E\left[ g(v_\circ,v_\circ)(X_t,t)\right] > 0 \quad \forall\, t\in\mathcal{T}$.
	\end{itemize}
\end{mydef}
}
\begin{mydef}{\rm (Proper time)}\\
	Given a Lorentzian manifold $(\M,g)$ with $(-+...+)$ signature and a time-like stochastic process $X:\mathcal{T}\times (\Omega,\Sigma,\mathbb{P}) \rightarrow (\M,\mathcal{B}(\M))$, we define the proper time of $X$ by
	\begin{equation}
		\Delta\tau(X_\mathcal{T}) = \E\left[ \int_\mathcal{T} \sqrt{-g(v_\circ,v_\circ)(X_t,t)} \, dt \right].
	\end{equation}
\end{mydef}
\begin{mydef}{\rm (Proper distance)}\\
	Given a Lorentzian manifold $(\M,g)$ with $(-+...+)$ signature and a space-like stochastic process $X:\mathcal{T}\times (\Omega,\Sigma,\mathbb{P}) \rightarrow (\M,\mathcal{B}(\M))$, we define the proper length of $X$ by
	\begin{equation}
		\Delta s(X_\mathcal{T}) = \E\left[ \int_I \sqrt{g(v_\circ,v_\circ)(X_t,t)} \, dt \right].
	\end{equation}
\end{mydef}
\begin{mydef}{\rm (Induced $L^2$-norm)}\\
	Given a stochastic process $X:\mathcal{T}\times(\Omega,\Sigma,\mathbb{P})\rightarrow(\M,\mathcal{B}(\M))$ and a Borel measurable function $f:\M\times\mathcal{T}\rightarrow\mathbb{K}$ with $\mathbb{K}\in\{\R,\mathbb{C}\}$, the $L^2$-norm for $f\circ X$ is defined by 
	\begin{equation}
		||f(X)|| = \sqrt{\int_\mathcal{T}\E\Big[|f(X_t,t)|^2\Big] \, dt} \, .
	\end{equation}
\end{mydef}
\begin{mydef}{\rm (Induced $L^2$-space)}\\
	Given a stochastic process $X:\mathcal{T}\times(\Omega,\Sigma,\mathbb{P})\rightarrow(\M,\mathcal{B}(\M))$, the spaces
	\begin{equation}
		L_{\mathcal{T}}^2(\M,\mathcal{B}(\M),\mu_X) = \left\{ f:(\M,\mathcal{B}(\M))\times\mathcal{T} \rightarrow (\mathbb{K},\mathcal{B}(\mathbb{K})) \, \Big| \, ||f(X)||<\infty \right\}
	\end{equation}
	with $\mathbb{K}\in\{\R,\mathbb{C}\}$ are the real and complex $L^2$-space over $(\M,\mathcal{B}(\M),\mu_X)\times \mathcal{T}$.
\end{mydef}
\begin{theorem}{\rm (Properties of $L^2$-spaces)}\\
	$L^2$-spaces have the following properties:
	\begin{itemize}
		\item $L^2$ is a Hilbert space with inner product 
		\begin{itemize}
			\item $X \cdot Y = \E[\int_\mathcal{T}\delta_{ij} \overline{X}{}^i Y^j\, dt]$ on $L_\mathcal{T}^2(\Omega,\Sigma,\mathbb{P})\,,$
			\item $\langle f,g\rangle_{\mu_X} = \int_\mathcal{T}\int_\M \overline{f(x,t)} \ g(x,t) \, d\mu_{X_t}(x,t)\,dt$ on $L_{\mathcal{T}}^2(\M,\mathcal{B}(\M),\mu_X)\,;$
		\end{itemize}
		\item $L^2$ is self-dual, i.e. $L^2\cong(L^2)^\ast$.
	\end{itemize}
\end{theorem}
\subsection{Generating Functionals}
The definitions of the characteristic function and moment generating function encountered in appendix \ref{ap:momgenfnct} can be generalized to the context of stochastic processes.
\begin{mydef}{\rm (Moment generating functional)}\\
	Given a real stochastic process $X:\mathcal{T}\times(\Omega,\Sigma,\mathbb{P})\rightarrow(\R^n,\mathcal{B}(\R^n))$ and a family of linear forms $a=\{a_t\in[L^2(\Omega,\Sigma,\mathbb{P})]^\ast\,|\,t\in\mathcal{T}\}$, the moment generating functional of $X$ is defined by
	\begin{equation}
		M_{X}(a) := \E\left[ e^{\int_\mathcal{T}a_t(X_t) \, dt}\right] ,
	\end{equation}
	where $||a_t||<\rho_t$ with $\rho_t$ the radius of convergence of $M_{X_t}$.\\
	Given a complex stochastic process $Z:\mathcal{T}\times(\Omega,\Sigma,\mathbb{P})\rightarrow(\mathbb{C}^n,\mathcal{B}(\mathbb{C}^n))$ and a family of linear forms $a=\{a_t\in[L^2(\Omega,\Sigma,\mathbb{P})]^\ast\,|\,t\in\mathcal{T}\}$, the moment generating functional of $Z$ is defined by
	\begin{equation}
		M_{Z}(a) := \E\left[ e^{\int_\mathcal{T}\overline{a}_t(Z_t)\, dt}\right],
	\end{equation}
	where $||a_t||<\rho_t$ with $\rho_t$ the radius of convergence of $M_{Z_t}$.
\end{mydef}
\begin{mydef}{\rm (Characteristic functional)}\\
	Given a real stochastic process $X:\mathcal{T}\times(\Omega,\Sigma,\mathbb{P})\rightarrow(\R^n,\mathcal{B}(\R^n))$ and a family of linear forms $a=\{a_t\in[L^2(\Omega,\Sigma,\mathbb{P})]^\ast\,|\,t\in\mathcal{T}\}$, the characteristic functional of $X$ is defined by
	\begin{equation}
		\varphi_{X}(a) := \E\left[ e^{\ri \int_\mathcal{T} a_t(X_t) \, dt}\right] .
	\end{equation}
	Given a complex stochastic process $Z:\mathcal{T}\times(\Omega,\Sigma,\mathbb{P})\rightarrow(\mathbb{C}^n,\mathcal{B}(\mathbb{C}^n))$ and a family of linear forms $a=\{a_t\in[L^2(\Omega,\Sigma,\mathbb{P})]^\ast\,|\,t\in\mathcal{T}\}$, the characteristic functional of $Z$ is defined by
	\begin{equation}
		\varphi_{Z}(a) := \E\left[ e^{\ri \int_\mathcal{T} {\rm Re}[\overline{a}_t(Z_t)]\, dt}\right]  .
	\end{equation}
\end{mydef}

\clearpage
\section{Review of Stochastic Calculus}\label{ap:ReviewStochCalc}
Ordinary analysis and its associated calculus allow to describe deterministic trajectories of finite variation, but it is no longer applicable when the trajectories are stochastic. In this case, one must resort to stochastic analysis and stochastic calculus.
\par 

A main difficulty in stochastic analysis is the fact that generic stochastic processes are almost surely not differentiable. As a consequence, one cannot unambiguously define a derivative of a stochastic process, which prevents the construction of a differential calculus. One can, however, construct an integral along any stochastic trajectory that is a semi-martingale, which induces an integral calculus. In this appendix, we review some elementary properties of this calculus.
\par 

Given a semi-martingale $X:\mathcal{T}\times(\Omega,\Sigma,\mathcal{F},\mathbb{P})\rightarrow(\R,\mathcal{B}(\R))$ and a Borel measurable function $f:\R\times \mathcal{T}\rightarrow \R$ that is twice continuously differentiable, one can construct an integral along $X$ in various ways. The first is called the \textit{Stratonovich integral} and is given by
\begin{equation}\label{eq:Stratonovich}
	\int_\mathcal{T} f(X_t,t) \circ dX_t
	:=
	\lim_{||\Pi||\rightarrow 0} \sum_k
	\frac{1}{2} \big[ f(X_{t_k},t_k) + f(X_{t_{k+1}},t_{k+1}) \big] \big[X_{t_{k+1}} - X_{t_k} \big] \, ,
\end{equation}
where $\Pi$ is a partition of $\mathcal{T}$ and $||\Pi||$ is its mesh. The second is called the \textit{It\^o integral} and is given by
\begin{equation}\label{eq:ItoF}
	\int_\mathcal{T} f(X_t,t) \, d_+ X_t
	:=
	\lim_{||\Pi||\rightarrow 0} \sum_k
	f(X_{t_k},t_k) \, \big[X_{t_{k+1}} - X_{t_k} \big] \, .
\end{equation}
A third is called a \textit{backward It\^o integral}, and is given by
\begin{equation}\label{eq:ItoB}
	\int_\mathcal{T} f(X_t,t) \, d_- X_t
	:=
	\lim_{||\Pi||\rightarrow 0} \sum_k
	f(X_{t_{k+1}},t_{k+1}) \, \big[X_{t_{k+1}} - X_{t_k} \big] \, .
\end{equation}
Finally, in contrast to deterministic processes, stochastic process have a non-vanishing quadratic variation.\footnote{More precisely, in a deterministic theory the quadratic variation is non-vanishing for discontinuous processes only, while it is non-vanishing for all processes in a stochastic theory.} This allows to define a fourth integral, which is the \textit{integral over the quadratic variation} given by
\begin{equation}\label{eq:Qvar}
	\int_\mathcal{T} f(X_t,t) \, d[X,X]_t
	:=
	\lim_{||\Pi||\rightarrow 0} \sum_k
	f(X_{t_k},t_k) \, \big[X_{t_{k+1}} - X_{t_k} \big]^2 \,.
\end{equation}
\par 

In the remainder of this appendix and throughout the book, we will assume that the semi-martingale $X$ is continuous.\footnote{All results can be generalized to processes that are c\`adl\`ag or c\`agl\`ad, but they obtain corrections due to the presence of jump discontinuities.} Starting from their respective definitions, one can derive a relation between the various integrals that is given by
\begin{align}
	\int_\mathcal{T} f(X_t,t) \circ dX_t
	&= \int_\mathcal{T} f(X_t,t) \, d_+ X_t 
	+ \frac{1}{2} \int_\mathcal{T} \frac{\p}{\p x} f(X_t,t) \, d[X,X]_t \nonumber\\
	&= \int_\mathcal{T} f(X_t,t) \, d_- X_t 
	- \frac{1}{2} \int_\mathcal{T} \frac{\p}{\p x} f(X_t,t) \, d[X,X]_t \, .
\end{align}
It is common in stochastic calculus to work in a differential notation, where the integral sign is dropped. In this notation, these relations are expressed as
\begin{align}\label{eq:ItovsStratFlat}
	f(X_t,t) \circ dX_t
	&= f(X_t,t) \, d_+ X_t 
	+ \frac{1}{2} \, \p_x f(X_t,t) \, d[X,X]_t \nonumber\\
	&= f(X_t,t) \, d_- X_t 
	- \frac{1}{2} \, \p_x f(X_t,t) \, d[X,X]_t \, .
\end{align}
In the remainder of this appendix and throughout the book, we will make use this differential notation. In addition, we denote Stratonovich integrals by $d_\circ$ instead of $\circ\, d\,$.
\par 

We will now discuss some properties of the various integrals that can be derived from their respective definitions.
The Stratonovich integral satisfies the ordinary chain rule and Leibniz rule, i.e for any function $f,g$ with the same properties as before we have
\begin{align}
	d_\circ f &= \p_t f \, dt + \p_x f \, d_\circ X ,\\
	d_\circ (f\,g) &= f \, d_\circ g +  g \, d_\circ f .
\end{align}
The It\^o integral, on the other hand, satisfies It\^o's lemma and a modified Leibniz rule given by
\begin{align}
	d_+f &= \p_t f \, dt + \p_x f \, d_+X + \frac{1}{2} \, \p_x^2 f \, d[X,X]  ,\\
	d_+(f\,g) &= f \, d_+g +  g \, d_+f + d[f,g]  .
\end{align}
Similarly, for the backward It\^o integral, one obtains
\begin{align}
	d_-f &= \p_t f \, dt + \p_x f \, d_-X - \frac{1}{2} \, \p_x^2 f \, d[X,X] ,\\
	d_-(f\,g) &= f \, d_-g +  g \, d_-f - d[f,g] .
\end{align}
Furthermore, the quadratic variation satisfies a symmetry property together with a chain and product rule. These are given by
\begin{align}
	d[f,g] &= d[g,f] , \\
	d[f,g] &= \p_x f \, \p_x g \, d[X,X] , \\
	d[f \, g, h] &= f \, d[g,h] + g \, d[f,h] .
\end{align}
\par 

It\^o integrals along local martingales are themselves local martingales. It\^o integrals thus satisfy the following martingale property:
\begin{align}\label{eq:MartProp}
	\E\left[ \int_{s}^{t} f(X_r,r) \, d_+ M_r \, \Big| \, \vec{\mathcal{F}}_s \right] 
	&=  0 \qquad \forall \, s<t\in\mathcal{T} , \nonumber\\
	\E\left[ \int_{s}^{t} f(X_r,r) \, d_- M_r \, \Big| \, \cev{\mathcal{F}}_t \right] 
	&=  0 \qquad \forall \, s<t\in\mathcal{T},
\end{align}
where it is assumed that $M$ is a local martingale with respect to the past $\vec{\mathcal{F}}$ in the first line and with respect to the future $\cev{\mathcal{F}}$ in the second line and that $X$ is adapted to these filtrations.
\par

All results from this appendix can be extended to higher dimensional semi-martingales. For a $n$-dimensional real continuous semi-martingale $X:\mathcal{T}\times(\Omega,\Sigma,\mathcal{F},\mathbb{P})\rightarrow(\R^n,\mathcal{B}(\R^n)$ the Stratonovich and It\^o integrals can be defined for every integrable and Borel measurable form $f\in T^\ast\R^d$ by
\begin{align}
	\int_\mathcal{T} f(d_\circ X_t) &= \int_\mathcal{T} f_i(X_t) \, d_\circ X^i_t \, , \nonumber\\
	\int_\mathcal{T} f(d_\pm X_t) &= \int_\mathcal{T} f_i(X_t) \, d_\pm X^i_t\, .
\end{align}
Similarly, the integral over the quadratic variation can be defined using bilinear forms $g\in T^2(T^\ast\R^d)$ such that
\begin{align}
	\int_\mathcal{T} g(d_\circ X_t,d_\circ X_t) &= \int_\mathcal{T} g_{ij}(X_t) \, d[X^i,X^j]_t \, , \nonumber\\
	\int_\mathcal{T} g(d_\pm X_t,d_\pm X_t) &= \int_\mathcal{T} g_{ij}(X_t) \, d[X^i,X^j]_t\, .
\end{align}
\par 

All the results can also be extended to complex valued continuous stochastic processes $Z$, using their decomposition $Z=X+\ri \, Y$ into real stochastic processes. For the Stratonovich and It\^o integrals one then obtains
\begin{align}
	\int_\mathcal{T} f(Z_t) \, d_\circ Z_t 
	&= 
	\int_\mathcal{T} f(Z_t) \, d_\circ X_t
	+ \ri \int_\mathcal{T} f(Z_t) \, \, d_\circ Y_t\\
	\int_\mathcal{T} f(Z_t) \, d_\pm Z_t 
	&= 
	\int_\mathcal{T} f(Z_t) \, d_\pm X_t
	+ \ri \int_\mathcal{T} f(Z_t) \, d_\pm Y_t	
\end{align}
and for the quadratic variation one obtains
\begin{equation}
	\int_\mathcal{T} f(Z_t) \, d[Z,Z]_t 
	= 
	\int_\mathcal{T} f(Z_t) \, d[X,X]_t
	- \int_\mathcal{T} f(Z_t) \, d[Y,Y]_t
	+ 2 \, \ri \int_\mathcal{T} f(Z_t) \, d[X,Y]_t \, ,
\end{equation}
where we used the symmetry of the quadratic variation.

\clearpage
\section{Second Order Geometry}\label{Ap:2Geometry}
Second order geometry is a geometrical framework that allows to extend stochastic calculus to manifolds. 
Stratonovich calculus is more adaptable for such an extension than It\^o calculus, since the Stratonovich formalism obeys the ordinary Leibniz' rule and chain rule: for any $f,g\in C^\infty(\M)$ and $h\in C^2(\R)$
\begin{align*}
	d_\circ(f \, g) &= f \, d_\circ g + g \, d_\circ f \, ,\\
	d_\circ(h \circ f) &= \left( h' \circ f \right) d_\circ f \, .
\end{align*}
In It\^o calculus, on the other hand, Leibniz rule is violated and the chain rule is replaced by It\^o's lemma:
\begin{align*}
	d_\pm(f \, g) &= f \, d_\pm g + g \, d_\pm f \pm d[f,g]\, , \\
	d(h \circ f) &= \left( h' \circ f \right) d_\pm f  \pm \frac{1}{2} \left( h'' \circ f \right) d[f,f] \, .
\end{align*}
Nevertheless, since It\^o calculus satisfies the martingale property \eqref{eq:MartProp}, it is worth extending the framework to manifolds. If the manifold is twice continuously differentiable and equipped with an affine connection, this can be done using second order geometry, which was developed by Schwartz and Meyer \cite{Schwartz,Meyer,Emery,Huang:2022}.
\par 

The idea of second order geometry is to incorporate the violation of Leibniz' rule, which is a property of functions defined on the manifold, into the underlying geometry. In this appendix, we review some aspects of this framework for smooth manifolds.
\par 

We will consider a smooth manifold equipped with affine connection $(\M,\Gamma)$ and smooth functions $f\in C^\infty(\M)$. In classical geometry, the differential of such a function along a trajectory $X:\mathcal{T}\rightarrow\M$ is given by
\begin{align}
	df(X_t,t) 
	&= \p_t f \, dt + \p_{i} f \, dX^i_t + \mathcal{O}(dt^2) \nonumber\\
	&= \left[ \p_t f + v^i \, \p_{i} f \right] dt + \mathcal{O}(dt^2) \,,
\end{align}
where $v(X_t,t)$ is a velocity field defined on the tangent bundle $T\M$.
\par 

In a stochastic theory, the trajectory $X$ becomes a stochastic process and a similar expression can be obtained for semi-martingales in the It\^o formulation:
\begin{align}
	\E\left[ d_\pm f(X_t,t) \, \Big| \, X_t \right] 
	&= \E \left[\p_t f \, dt + \p_{i} f \, d_\pm X^i_t \pm \frac{1}{2} \, \p_j \p_i f \, d[X^i,X^j]_t + o(dt) \, \Big| \, X_t \right] \nonumber\\
	&= \left[ \p_t f + v_\pm^i \, \p_{i} f \pm \frac{1}{2}  \,v^{ij}_2 \, \p_j \p_i f \right] dt + o(dt) \, ,
\end{align}
where, cf. appendix \ref{sec:L2StochProc},
\begin{align}
	v_\pm(X_t,t) &= \lim_{dt\rightarrow 0} \E \left[\frac{d_\pm X_t}{dt} \, \Big| \, X_t \right] ,\nonumber\\
	v_2(X_t,t) &= \lim_{dt\rightarrow 0} \E \left[\frac{d[X,X]_t}{dt} \, \Big| \, X_t \right] .
\end{align}
\par

In a deterministic theory, terms appearing at $\mathcal{O}(dt)$ are associated with velocities, terms appearing at $\mathcal{O}(dt^2)$ with accelerations etc. Second order geometry preserves this idea and thus interprets both the fields $v_\pm$ and $v_2$ as a velocity. Here, $v_\pm$ is a velocity field in the usual sense, while $v_2$ is a velocity field associated with the quadratic variation of $X$. Thus, by the L\'evy characterization, cf. Appendix \ref{ap:Wiener}, if $X$ is a Wiener process, $v_\pm$ determines a velocity associated to the expectation value $\E[X_t]$, while $v_2$ determines the velocity associated to the variance ${\rm Var}(X_t)=\E[X_t^2]-\E[X_t]^2$.
\par 

As a consequence, on a $n$-dimensional manifold, second order velocity fields $(v^i,v^{jk})$ have $\frac{n(n+3)}{2}$ degrees of freedom, where $n$ degrees of freedom are encoded in $v^i$ and the remaining $\frac{n(n+1)}{2}$ degrees of freedom in the symmetric object $v^{jk}$. It follows that, at every point $x\in\M$, the $n$-dimensional first order tangent space $T_x\M$ must be extended to a $\frac{n(n+3)}{2}$-dimensional second order tangent space $T_{2,x}\M$. 
Then, in a local coordinate chart, second order vectors $v\in T_{2,x}\M$ can be represented with respect to their canonical basis as\footnote{The factor $\frac{1}{2}$ is not universal in the definition of second order vectors. Some works, e.g.~Refs.~\cite{Kuipers:2021ylr,Huang:2022}, include it, while other works, e.g.~Refs.~\cite{Emery,Kuipers:2021jlh}, do not.}
\begin{equation}
	v = v^i \, \p_i + \frac{1}{2} \, v^{jk} \, \p_{jk} \, ,
\end{equation}
whereas first order vectors $v\in T_x\M$ are given by
\begin{equation}
	v = v^i \, \p_i \, .
\end{equation}
\par 

For the same reason, for any $x\in\M$, the first order cotangent space $T^\ast_x\M$ must be extended to a second order cotangent space $T^\ast_{2,x}\M$. In a local coordinate chart, first order forms $\omega\in T^\ast_x\M$ are given with respect to their canonical basis by
\begin{equation}
	\omega = \omega_i \, dx^i \, 
\end{equation}
and second order forms $\omega\in T^\ast_{2,x}\M$ are given by
\begin{equation}
	\omega = \omega_i \, d_2x^i + \frac{1}{2} \, \omega_{ij} \, d[x^i,x^j] \, .
\end{equation}
\par

The duality pairing of first order vectors $v\in T_x\M$ with forms $\omega\in T_x^\ast\M$ is given by
\begin{equation}
	\langle \omega, v\rangle = \omega_i \, v^i \, .
\end{equation}
Similarly, the duality pairing for second order vectors $v\in T_{2,x}\M$ with second order forms $\omega\in T_{2,x}^\ast\M$ is
\begin{equation}
	\langle \omega, v\rangle = \omega_i \, v^i + \frac{1}{2} \, \omega_{ij} \, v^{ij} \, ,
\end{equation}
and can be derived from the duality pairing of the basis elements
\begin{alignat}{2}
	\langle d_2 x^i, \p_j \rangle &= \p_j x^i &&= \delta^i_j \, , \nonumber\\
	\langle d_2 x^i, \p_{jk} \rangle &= \p_j \p_k x^i &&= 0 \, , \nonumber\\
	\langle d[x^i,x^j], \p_k \rangle &= \p_k ( x^i x^j) - x^i \p_k x^j - x^j \p_k x^i &&= 0 \, , \nonumber\\
	\langle d[x^i,x^j], \p_{kl} \rangle &= \p_k \p_l (x^i x^j) - x^i \p_k\p_l x^j - x^j \p_k \p_l x^i &&= \delta^i_k \delta^j_l + \delta^i_l \delta^j_k \, .
\end{alignat}
\par 

Second order vectors and forms do not transform in a covariant manner. Indeed, one can easily verify that the active coordinate transformation laws for vectors are
\begin{alignat}{2}
	v^i &\rightarrow \tilde{v}^i 
	&&= v^{k} \, \frac{\p \tilde{x}^i}{\p x^k} 
	+ \frac{1}{2} \, v^{kl} \, \frac{\p^2 \tilde{x}^i}{\p x^k \p x^l}\, ,\nonumber\\
	v^{ij} &\rightarrow \tilde{v}^{ij} 
	&&= v^{kl} \, \frac{\p \tilde{x}^i}{\p x^k} \frac{\p \tilde{x}^j}{\p x^l} \,  ,
\end{alignat}
and for forms they are given by
\begin{alignat}{2}
	\omega_i &\rightarrow \tilde{\omega}_i 
	&&= \omega_{k}  \, \frac{\p x^k}{\p \tilde{x}^i} \, ,\nonumber\\
	\omega_{ij} &\rightarrow \tilde{\omega}_{ij} 
	&&= \omega_{k} \, \frac{\p^2 x^k}{\p \tilde{x}^i \p \tilde{x}^j}
	+ \omega_{kl} \, \frac{\p x^k}{\p \tilde{x}^i} \frac{\p x^l}{\p \tilde{x}^j}\, .
\end{alignat}
Similarly, the passive transformation laws are given by
\begin{alignat}{2}
	\p_i &\rightarrow \tilde{\p}_i &&= \frac{\p x^k}{\p\tilde{x}^i} \, \p_k\,,\nonumber\\
	\p_{ij} &\rightarrow \tilde{\p}_{ij}
	&&= \frac{\p^2 x^k}{\p \tilde{x}^i \p \tilde{x}^j} \, \p_k
	+ \frac{\p x^k}{\p \tilde{x}^i}  \frac{\p x^l}{\p \tilde{x}^j} \, \p_{kl} \, ,
\end{alignat}
and
\begin{alignat}{2}
	d_2 x^i &\rightarrow d_2\tilde{x}^i 
	&&= 
	\frac{\p \tilde{x}^i}{\p x^k} \, d_2 x^k 
	+ 
	\frac{1}{2} \, \frac{\p^2 \tilde{x}^i}{\p x^k \p x^l} \,
	d[x^k, x^l] \, ,\nonumber\\
	d[x^i, x^j] &\rightarrow d[\tilde{x}^i , d\tilde{x}^j] 
	&&=
	\frac{\p \tilde{x}^i}{\p x^k} 
	\frac{\p \tilde{x}^j}{\p x^l} \,
	d[x^k, x^l] \, . 
\end{alignat}
\par 

However, if the manifold is equipped with an affine connection, one can construct  covariant representations for second order vectors and forms. These are given by
\begin{align}
	\hat{v}^i &= v^i + \frac{1}{2} \, \Gamma^i_{jk} \, v^{jk} \,, \nonumber\\
	\hat{v}^{ij} &= v^{ij} \, ,\nonumber\\
	\hat{\omega}_i &= \omega_i \, , \nonumber\\
	\hat{\omega}_{ij} &= \omega_{ij} - \Gamma^k_{ij} \, \omega_k \, .
\end{align}
Similarly, the covariant basis elements are
\begin{align}
	\hat{\p}_i &= \p_i \, ,\nonumber\\
	\hat{\p}_{ij} &= \p_{ij}  - \Gamma^k_{ij} \, \p_{k} \, ,\nonumber\\
	d_2\hat{x}^i &= d_2x^i + \frac{1}{2} \, \Gamma^i_{kl} \, d[x^k,x^l] \, , \nonumber\\
	d[\hat{x}^i,\hat{x}^j] &= d[x^i,x^j] \, .
\end{align}
These objects transform covariantly, i.e.
\begin{alignat}{2}
	\hat{v}^i &\rightarrow \tilde{\hat{v}}^i 
	&&= \hat{v}^{k} \, \frac{\p \tilde{x}^i}{\p x^k} \, ,\nonumber\\
	\hat{\omega}_{ij} &\rightarrow \tilde{\hat{\omega}}_{ij} 
	&&= \hat{\omega}_{kl} \, \frac{\p x^k}{\p \tilde{x}^i} \frac{\p x^l}{\p \tilde{x}^j}\, ,
\end{alignat}
and
\begin{alignat}{2}
	\hat{\p}_{ij} &\rightarrow \tilde{\hat{\p}}_{ij}
	&&= \frac{\p x^k}{\p \tilde{x}^i}  \frac{\p x^l}{\p \tilde{x}^j} \, \hat{\p}_{kl} \, ,\nonumber\\
	d_2 \hat{x}^i &\rightarrow d_2\tilde{\hat{x}}^i 
	&&= 
	\frac{\p \tilde{x}^i}{\p x^k} \, d_2 \hat{x}^k \, .
\end{alignat}
\par

We conclude this section by noting that second order geometry is a geometrical framework that generalizes structures from ordinary (first order) geometry to second order structures, as was done for vectors and forms in the preceding discussion. This can be done independently of any notions from stochastic analysis. However, the natural domain of application of second order geometry is stochastic calculus, as it provides an interpretation of the second order structures. It is straightforward to relate the previous general discussion on second order geometry to its application in It\^o calculus, as discussed in appendix \ref{ap:ReviewStochProc}, by replacing
\begin{align}
	d_2 x &\rightarrow d_\pm X \, ,\nonumber\\
	d[x,x] &\rightarrow \pm \, d[X,X] \,.
\end{align}

\subsection{Maps Between First and Second Order Geometry} 
The preceding discussion can be reformulated in a coordinate independent way by constructing unique mappings between first and second order geometry, cf. e.g. Ref.~\cite{Emery}.
\par 

For functions $f,g\in C^\infty(\M)$ and linear operators $L$ on $C^\infty(\M)$, one can define \textit{l'op\'erateur carr\'e du champ} or the \textit{squared field operator} by
\begin{equation}
	\Gamma_L(f,g) := L(f \, g) - f \, L(g) - g \, L(f) \, .
\end{equation}
Since $\Gamma_L(f,g)=0$, if $L$ is a derivation, this operator is a measure for how close the operator $L$ is to being a derivation. Moreover, it provides an alternative definition of the quadratic variation, since 
\begin{equation}
	\Gamma_{d_2}(f,g) 
	= d[f,g]\, .
\end{equation}
\par 

Starting from first order vectors $A , B\in T\M$, one can construct second order vectors by taking their product $A B \in T_2\M$. Similarly, starting from first order forms, one can construct second order forms, using a map
\begin{equation}
	\mathcal{H}: T^\ast\M \otimes T^\ast\M \rightarrow T^\ast_{2}\M
	\qquad
	{\rm s.t.}
	\qquad 
	\alpha \otimes \beta \mapsto \alpha \cdot \beta \, ,
\end{equation}
which satisfies the following properties:
\begin{alignat}{2}
	\langle\mathcal{H}(b) , A \, B \rangle 
	&= \frac{1}{2}\left[b(A,B) + b(B,A) \right] \qquad 
	&& \forall \, A,B\in T\M, \, b\in T^2(T^\ast\M) \, , \nonumber\\
	df \cdot dg 
	&= \frac{1}{2} \, d[f,g] \qquad
	&& \forall \, f,g\in C^\infty(\M) \, .
\end{alignat}
Moreover, the adjoint $\mathcal{H}^\ast:T_{2}\M \rightarrow T\M \otimes T\M$ defines a map from second order vectors to symmetric $(2,0)$-tensors. 
\par 

Vice versa, any second order form can be projected onto a first order form in an intrinsic way using a map $\mathcal{P}:T_2^\ast\M\rightarrow T^\ast\M$, such that
\begin{alignat}{2}
	\mathcal{P}(d_2 f) &= df \qquad &&\forall\, f\in C^\infty(\M) \, , \nonumber\\
	\mathcal{P}(\alpha \cdot \beta) &= 0 \qquad &&\forall\, \alpha,\beta \in T^\ast\M \, .
\end{alignat}
\par 

Alternatively, second order forms can be constructed using a linear map ${\underline{d}:T^\ast\M \rightarrow T^\ast_{2}\M}$, which satisfies the following properties
\begin{alignat}{2}
	\underline{d}(df) 
	&= d_2f \qquad 
	&& \forall \, f\in C^\infty(\M) \, , 
	\nonumber\\
	\underline{d}(f\, \alpha) 
	&= df \cdot \alpha + f \, \underline{d}\alpha  \qquad 
	&& \forall \, \alpha\in T^\ast\M, \, f\in C^\infty(\M) \, , 
	\nonumber\\
	\mathcal{P}(\underline{d}\alpha)
	&= \alpha \qquad 
	&& \forall \, \alpha\in T^\ast\M \, , 
	\nonumber\\
	\langle \underline{d}\alpha, A B - B A\rangle
	&= \langle \alpha, [A,B]\rangle \qquad 
	&& \forall \, \alpha\in T^\ast\M, \, A,B\in T\M \, , 
	\nonumber\\
	\langle \underline{d}\alpha, A B + B A \rangle
	&= A \langle \alpha, B\rangle + B \langle \alpha, A \rangle \qquad 
	&& \forall \, \alpha\in T^\ast\M, \, A,B\in T\M \, . 
\end{alignat}
\par 

Finally, using the affine connection $\Gamma: \mathfrak{X}(\M) \times \mathfrak{X}(\M)\rightarrow \mathfrak{X}(\M)$. One can construct mappings $\mathcal{F}:T_2\M\rightarrow T\M$ and $\mathcal{G}:T^\ast\M\rightarrow T_2^\ast\M$ through the following relations
\begin{alignat}{2}
	(\mathcal{F} \, V) \, f 
	&= V \, f - \langle \mathcal{H} \, \Gamma^\ast(df) , V \rangle \qquad
	&& \forall \, V \in T_2 \M, \, f\in C^\infty(\M) \, , 
	\nonumber\\
	\Gamma(A,B) 
	&= A \, B \, f - F( A \, B) \, f \qquad
	&& \forall \, A,B\in T\M, \, f\in C^\infty(\M) \, ,
\end{alignat}
and
\begin{alignat}{2}
	\mathcal{G}(df) 
	&= d_2 f - \mathcal{H} \, \Gamma^\ast(df) \qquad
	&& \forall \, f\in C^\infty(\M) \, ,
	\nonumber\\
	\Gamma(A,B) \, f 
	&= A \, B \, f - \langle \mathcal{G}(df), A B \rangle \qquad
	&& \forall \, A,B\in T\M,\, f\in C^\infty(\M) \, .
\end{alignat}
$\mathcal{F},\mathcal{G}$ then satisfy the following properties:
\begin{alignat}{2}
	\mathcal{F}(f \, V)
	&= f \, \mathcal{F}(V) \qquad
	&& \forall \, V\in T_2\M, \, f\in C^\infty(\M) \, , 
	\nonumber\\
	\mathcal{F}(A)
	&= A \qquad 
	&& \forall \, A\in T\M \subset T_2\M \, , 
	\nonumber\\
	\mathcal{G}(f\, \alpha) 
	&= f \, \mathcal{G}(\alpha) \qquad
	&& \forall \, \alpha\in T^\ast\M, \, f\in C^\infty(\M) \, , 
	\nonumber\\
	\mathcal{P}[\mathcal{G}(\alpha)] 
	&= \alpha \qquad
	&& \forall \, \alpha\in T^\ast\M \, , 
	\nonumber\\
	\langle \alpha , \mathcal{F}(V) \rangle 
	&= \langle \mathcal{G}(\alpha), V\rangle \qquad
	&& \forall \, \alpha\in T^\ast\M, \, V\in T_2\M \, .
\end{alignat}
\par 

We conclude this section by summarizing the action of the various maps in a local coordinate frame.
Given two first order vector fields $A,B\in T\M$ their product is a second order vector given by
\begin{equation}
	A \, B = \left[ A^i \, \p_i (B^j) \right] \p_j + \left[A^i \, B^j\right] \, \p_{ij} \, .
\end{equation}
Given a second order form $\omega\in T^\ast_{2}\M$, the projection map acts as
\begin{equation}
	\mathcal{P}(\omega) 
	= \mathcal{P}\left(\omega_i \, d_2 x^i + \frac{1}{2} \, \omega_{ij} \, d[x^i,x^j]\right) 
	= \omega_i \, dx^i \, .
\end{equation}
Given a first order form $\alpha\in T^\ast\M$, the map $\underline{d}$ acts as
\begin{equation}
	\underline{d}(\alpha) 
	= \underline{d}\left(\alpha_i \, d x^i \right) 
	= \alpha_i \, d_2 x^i + \frac{1}{2} \, \p_j \alpha_i \, d[x^i,x^j] \, .
\end{equation}
Given a first order form $\alpha\in T^\ast\M$, the map $\mathcal{G}$ acts as
\begin{equation}
	\mathcal{G}(\alpha) 
	= \mathcal{G}(\alpha_i \, dx^i)
	= \alpha_i \left(d_2 x^i + \frac{1}{2} \, \Gamma^i_{kl} \, d[x^k, x^l] \right)
	= \alpha_i \, d_2\hat{x}^i \, .
\end{equation}
Given two first order forms $\alpha,\beta\in T^\ast\M$, the map $\mathcal{H}$ acts as
\begin{equation}
	\mathcal{H}(\alpha \otimes \beta) 
	= \mathcal{H}(\alpha_i \, \beta_j \, dx^i \otimes dx^j)
	= \frac{1}{2} \, \alpha_i \, \beta_j\, d[x^i,x^j] \, .
\end{equation}
Given a second order vector field $V\in T^\ast_{2}\M$, the map $\mathcal{F}$ acts as
\begin{equation}
	\mathcal{F}(V) 
	= \mathcal{G}\left(V^i \, \p_i + \frac{1}{2} \, V^{ij} \, \p_{ij}\right)
	= \left( V^i + \frac{1}{2} \, \Gamma^i_{kl} \, V^{kl}  \right) \p_i
	= \hat{V}^i \, \p_i \, .
\end{equation}
Given a second order vector field $V\in T^\ast_{2}\M$, the map $\mathcal{H}^\ast$ acts as
\begin{equation}
	\mathcal{H}^\ast(V)
	= \mathcal{H}^\ast\left(V^i \, \p_i + \frac{1}{2} \, V^{ij} \, \p_{ij}\right)
	= \frac{1}{2} \, V^{ij} \, \p_i \otimes \p_j \, .
\end{equation}

\subsection{The Second Order Tangent Bundle}\label{sec:Structuregroup}

The (co)tangent bundle in first order geometry can be defined as follows.
\begin{mydef}{\rm (Tangent bundle)}\\
	The tangent bundle $(T\M,\tau_\M,\M)$ is the fiber bundle with base space $\M$, projection $\tau_\M:T\M\rightarrow\M$, typical fiber $\R^n$ and structure group ${\rm GL}(n,\R)$ acting from the left.
\end{mydef}
\begin{mydef}{\rm (Cotangent bundle)}\\
	The cotangent bundle $(T^\ast\M,\tau^{\ast}_\M,\M)$ is the fiber bundle dual to $(T\M,\tau_\M,\M)$ with base space $\M$, projection $\tau^{\ast}_\M:T^\ast\M\rightarrow\M$, typical fiber $(\R^n)^\ast$ and structure group ${\rm GL}(n,\R)$ acting from the right.
\end{mydef}
\noindent
These bundles can also be constructed as the bundle of tangent spaces at points $x\in\M$:
\begin{align}
	T\M &= \bigsqcup_{x\in\M} T_{x}\M \, , \nonumber\\
	T^\ast\M &= \bigsqcup_{x\in\M} T^\ast_{x}\M \, .
\end{align}
\par

Second order (co)tangent bundles can be defined in a similar way. Here, we quote the definitions given in Ref.~\cite{Huang:2022}.
\begin{mydef}{\rm (It\^o group)}\\
	The It\^o group $G_I^n$ is the Cartesian product ${\rm GL}(n,\R)\times{\rm Lin}(\R^n\otimes\R^n,\R^n)$ equipped with the binary operation
	\begin{equation}
		(g',\kappa') \, (g,\kappa) = (g' \, g, \, g'\circ\kappa + \kappa'\circ(g\otimes g))
	\end{equation}
	for all $g,g'\in{\rm GL}(n,\R)$ and $\kappa,\kappa'\in{\rm Lin}(\R^n\otimes\R^n,\R^n)$.
\end{mydef}
\begin{mydef}{\rm (Left action of the It\^o group)}\\
	The left group action of $G_I^n$ on $\R^n\times {\rm Sym}(T\R^n\otimes T \R^n)$ is defined by
	\begin{equation}
		(g,\kappa) \, (x,x_2) = (g \, x + \kappa \, x_2, \, (g\otimes g) \, x_2)
	\end{equation}
	for all $(g,\kappa)\in G^n_I$, $x\in\R^n$ and $x_2 \in {\rm Sym}(T\R^n\otimes T \R^n)$.
\end{mydef}
\begin{mydef}{\rm (Right action of the It\^o group)}\\
	The right group action of $G_I^n$ on $(\R^n\times {\rm Sym}(T\R^n\otimes T \R^n))^\ast$ is given by
	\begin{equation}
		(p,p_2) \, (g,\kappa) = (g^\ast \, p, \, \kappa^\ast \, p +  (g^\ast\otimes g^\ast) \, p_2)
	\end{equation}
	for all $(g,\kappa)\in G^n_I$, $p\in(\R^n)^\ast$ and $p_2 \in {\rm Sym}(T\R^n\otimes T \R^n)^\ast$.
\end{mydef}
\begin{mydef}{\rm (Second order tangent bundle)}\\
	The second order tangent bundle $(T_2\M,\tau^2_\M,\M)$ is the fiber bundle with base space $\M$, projection $\tau^2_\M:T_2\M\rightarrow\M$, typical fiber $\R^n\times{\rm Sym}(T\R^n\otimes T \R^n)$ and structure group $G^n_I$ acting from the left.
\end{mydef}
\begin{mydef}{\rm (Second order cotangent bundle)}\\
	The second order cotangent bundle $(T^\ast_2\M,\tau^{2\ast}_\M,\M)$ is the fiber bundle dual to $(T_2\M,\tau^2_\M,\M)$ with base space $\M$, projection $\tau^{2\ast}_\M:T^\ast_2\M\rightarrow\M$, typical fiber $(\R^n\times{\rm Sym}(T\R^n\otimes T \R^n))^\ast$ and structure group $G^n_I$ acting from the right.
\end{mydef}
\noindent
We note that these bundles can also be obtained from the second order (co)tangent spaces that were introduced earlier in this appendix, since
\begin{align}
	T_2\M &= \bigsqcup_{x\in\M} T_{2,x}\M \, , \nonumber\\
	T^\ast_2\M &= \bigsqcup_{x\in\M} T^\ast_{2,x}\M \, .
\end{align}

\subsection{Stochastic Integration on Manifolds}
Integration on manifolds along deterministic trajectories is performed by forms ${\omega\in T^\ast\M}$. Using second order order geometry, this can be extended to stochastic integration on manifolds by mapping these first order forms to second order forms.
\par

The Stratonovich integral is defined for any first order form $\omega\in T^\ast\M$ and is given by
\begin{equation}
	\dashint_{X_\mathcal{T}} \omega = \int_\mathcal{T} \omega_i(X_t) \, d_\circ X^i_t \, .
\end{equation}
The right hand side can then be evaluated in a local coordinate chart using the definition of the Stratonovich integral on $\R^n$ given in eq.~\eqref{eq:Stratonovich}. 
Alternatively, the Stratonovich integral can be defined as an integral over second order forms using the map $\underline{d}$ and the relation \eqref{eq:ItovsStratFlat} between It\^o and Stratonovich integrals on $\R^n$:
\begin{align}
	\dashint_{X_\mathcal{T}} \omega 
	&:= \int_{X_\mathcal{T}} \underline{d}_\pm(\omega) \nonumber\\
	&= \int_\mathcal{T} \omega_i(X_t)  \, d_\pm X^i_t
	\pm \frac{1}{2} \int_\mathcal{T} \p_j \omega_i(X_t)  \, d[X^i,X^j]_t 
	\nonumber\\
	&= \int_\mathcal{T} \omega_i(X_t) \, d_\circ X^i_t \, .
\end{align}
\par

The It\^o integral is defined using the map $\mathcal{G}$, such that the forward It\^o integral is given by
\begin{align}
	\lowint_{X_\mathcal{T}} \omega 
	&:=
	\int_{X_\mathcal{T}} \mathcal{G}_+(\omega) \nonumber\\
	&=
	\int_\mathcal{T} \omega_i(X_t) \, d_+ \hat{X}_t^i  \nonumber\\
	&=
	\int_\mathcal{T} \omega_i(X_t) \, d_+ X_t^i 
	+ \frac{1}{2} \int_\mathcal{T} \omega_{i}(X_t) \, \Gamma^i_{kl}(X_t) \, d[X^k,X^l]_t 
\end{align}
and the backward It\^o integral by
\begin{align}
	\upint_{X_\mathcal{T}} \omega 
	&:=
	\int_{X_\mathcal{T}} \mathcal{G}_-(\omega) \nonumber\\
	&=
	\int_\mathcal{T} \omega_i(X_t) \, d_- \hat{X}_t^i  \nonumber\\
	&=
	\int_\mathcal{T} \omega_i(X_t) \, d_- X_t^i 
	- \frac{1}{2} \int_\mathcal{T} \omega_{i}(X_t) \, \Gamma^i_{kl}(X_t) \, d[X^k,X^l]_t \, .
\end{align}
The right hand side of these expressions can be evaluated using the definitions of the It\^o integral and the integral over quadratic variation on $\R^n$, given in eqs.~\eqref{eq:ItoF}, \eqref{eq:ItoB} and \eqref{eq:Qvar}.
\par 

The integral over the quadratic variation is defined as an integral over a $(0,2)$-tensor $h\in T^2(T^\ast\M)$ and is given by
\begin{align}
	\int_{X_\mathcal{T}} h 	
	&=
	\int_{X_\mathcal{T}} \mathcal{H}(h) \nonumber\\
	&=
	\frac{1}{2} \, \int_{\mathcal{T}} h_{ij}(X_t) \, d[X^i,X^j]_t \, ,
\end{align}
where the right hand side can be evaluated using the definition of the integral over quadratic variation on $\R^n$, given in eq.~\eqref{eq:Qvar}. We note that this integral is defined for any bilinear form $h$, but only the symmetric part of $h$ contributes to the integral.
\par 

Finally, using the definition of the various integrals, one can derive a relation between the Stratonovich and the It\^o integral. In a local coordinate frame, it is given by
\begin{align}
	\int_\mathcal{T} \omega_i(X_t) \, d_\circ X^i_t
	&=
	\int_\mathcal{T} \omega_i(X_t) \, d_\pm X^i_t
	\pm \frac{1}{2} \int_\mathcal{T} \p_j\omega_i(X_t) \, d[X^i,X^j]_t
	\nonumber\\
	&=
	\int_\mathcal{T} \omega_i(X_t) \, d_\pm \hat{X}^i_t
	\pm \frac{1}{2} \int_\mathcal{T} \nabla_j\omega_i(X_t) \, d[X^i,X^j]_t \, ,
\end{align}
which can be written in a coordinate independent way as
\begin{align}
	\dashint_{X_\mathcal{T}} \omega
	&=
	\lowint_{X_\mathcal{T}} \omega 
	+ \int_{X_\mathcal{T}} \nabla\omega \nonumber\\
	&=
	\upint_{X_\mathcal{T}} \omega 
	- \int_{X_\mathcal{T}} \nabla\omega \, .
\end{align}

\clearpage
\section{Construction of the It\^o Lagrangian}\label{sec:ItoLagrangian}
In this appendix, we calculate the It\^o Lagrangian $L^\pm(x,v_\pm,v_2,t)$ corresponding to the Stratonovich Lagrangian \eqref{eq:StratLag} on a pseudo-Riemannian manifold:
\begin{equation}
	L^\circ(x,v_\circ,t) = \frac{m}{2} \, g_{ij}(x,t) \, v_\circ^i v_\circ^j + q \, A_i(x,t) \, v_\circ^i - \mathfrak{U}(x,t) \, .
\end{equation}
We do this by writing down the action for this Lagrangian
\begin{align}
	S_\circ(X)
	&=
	\E\left[ \int_{t_0}^{t_f} L^\circ(X_t,V_{\circ,t},t) \, dt \right]
	\nonumber\\
	&= \E\left[ 
	\int_{t_0}^{t_f} \left(
	\frac{m}{2} \, g_{ij}(X_t,t) \, V^i_{\circ,t} V^j_{\circ,t} 
	+ q \, A_i(X_t,t) \, V^i_{\circ,t} 
	- \mathfrak{U}(X_t,t)
	\right) dt
	\right] \, \label{eq:ApStratAction}
\end{align}
and imposing that
\begin{align}\label{eq:itoDerImpose}
	S_\circ(X) = S_\pm(X) = \E\left[ \int_{t_0}^{t_f} L^\pm(X_t,V_{\pm,t},V_{2,t},t) \, dt \right].
\end{align}
Thus, we have to rewrite the Stratonovich velocity as an It\^o velocity process. This can be done by evaluating the three terms in the action separately.
For the third term, involving the scalar potential, this is straightforward, as it does not depend on the velocity process. Therefore, the third term is the same in the It\^o and Stratonovich formulation.
\par 

For the second term, which is linear in velocity, we find
\begin{align}
	\E\left[ \int_{t_0}^{t_f} A_i(X_t,t) V^i_{\circ,t} \, dt  \right]
	&=
	\E\left[ \int_{t_0}^{t_f} A_i(X_t,t) \, d_\circ X^i_t  \right]
	\nonumber\\
	&=
	\E\left[ 
	\int_{t_0}^{t_f} \left(
	A_i(X_t,t) \, d_\pm X^i_t 
	\pm \frac{1}{2} \, \p_j A_i(X_t,t) \, d[X^i,X^j]_t
	\right)
	\right]
	\nonumber\\
	&=
	\E\left[ 
	\int_{t_0}^{t_f} \left(
	A_i(X_t,t) \, V^i_{\pm,t} 
	\pm \frac{1}{2} \, \p_j A_i(X_t,t) \, V^{ij}_{2,t}
	\right) dt
	\right]
	\nonumber\\
	&=
	\E\left[ 
	\int_{t_0}^{t_f} \left(
	A_i(X_t,t) \, \hat{V}^i_{\pm,t} 
	\pm \frac{1}{2} \, \nabla_j A_i(X_t,t) \, V^{ij}_{2,t}
	\right) dt
	\right] \, ,
\end{align}
where $\hat{V}_\pm^i = V_\pm^i \pm \frac{1}{2} \, \Gamma^i_{jk} \, V_2^{jk}$ is the covariant velocity process.
\par 

For the first term, which is quadratic in velocity, we find
\begin{align}\label{eq:Itoder}
	\E\left[ \int g_{ij}  \, V^i_{\circ} V^j_{\circ} \, dt \right]
	&=
	\E\left[ \int g_{ij}  \, V^i_{\circ} \, d_\circ X^j \right] 
	\nonumber\\
	&=
	\E\left[ \int \delta_{ab} \, e^b_j \, V_{\circ}^a \, d_\circ X^j  \right] 
	\nonumber\\
	&=
	\E\left[ \int \left(
	\delta_{ab} \, e^b_j \, V_{\circ}^a \, d_\pm X^j 
	\pm \frac{1}{2} \, \frac{\p}{\p x^k} \left( \delta_{ab} \, e^b_j \, V_\circ^a \right) d[X^j,X^k]
	\right. \right.
	\nonumber\\
	&\qquad \qquad \left.\left.
	\pm \frac{1}{2} \, \frac{\p}{\p v^c} \left( \delta_{ab} \, e^b_j \, V_\circ^a \right) d[X^j,V_\circ^c]
	\right) \right] 
	\nonumber\\
	&=
	\E\left[ \int \left(
		g_{ij} V_{\circ}^i \, d_\pm X^j 
		\pm \frac{1}{2} \, \delta_{ab} \, \Gamma^l_{kj} e^b_l \, V_\circ^a \, d[X^j,X^k]
		\pm \frac{1}{2} \, \delta_{ab} \, e^b_j \, d[X^j,V_\circ^a]
	\right) \right] 
	\nonumber\\
	&=
	\E\left[ \int \left(
		g_{ij} V_{\circ}^i \, d_\pm X^j 
		\pm \frac{1}{2} \, g_{ij}  \, \Gamma^j_{kl} \, V_\circ^i \, d[X^k,X^l]
		\pm \frac{1}{2} \, g_{ij} \, d[X^j,V_\circ^i]
	\right) \right] 
	\nonumber\\
	&=
	\E\left[ \int 
	\left( 
	g_{ij} V_{\circ}^i \, \hat{V}_\pm^j \, dt 
	\pm \frac{1}{2} \, g_{ij} \, d[X^i,V_\circ^j]
	\right)
	\right]. 
\end{align}
We can then repeat the same calculation for $V_\circ^i$, which yields
\begin{equation}\label{eq:ItoderEarly}
	\E\left[ \int g_{ij}  \, V^i_{\circ} V^j_{\circ} \, dt \right]
	=
	\E\left[ \int 
	\left( 
	g_{ij} \hat{V}_{\pm}^i \, \hat{V}_\pm^j \, dt 
	\pm \frac{1}{2} \, g_{ij} \, d[X^i,V_\circ^j+ \hat{V}_\pm^j]
	\right)
	\right] .
\end{equation}
An expression for the quadratic variation $[X,V_\circ]$ is given in section \ref{sec:MomentumProcess}, but we do not yet have an expression for $[X,\hat{V}_\pm]$. In order to obtain this expression, we rewrite the result of eq.~\eqref{eq:Itoder} using
\begin{align}
	V_\circ = \frac{1}{2} \Big( \hat{V}_+ + \hat{V}_- \Big) \, ,\nonumber\\
	V_\perp = \frac{1}{2} \Big( \hat{V}_+ - \hat{V}_- \Big) \, .
\end{align}
The first expression follows from the definition of the Stratonovich velocity and the second defines a velocity perpendicular to the Stratonovich velocity. We recalculate the quadratic term starting from the result in eq.~\eqref{eq:Itoder} and find
\begin{align}\label{eq:Itodera}
	\E\left[ \int g_{ij}  \, V^i_{\circ} V^j_{\circ} \, dt \right]
	&=
	\E\left[ \int 
		\left( 
			g_{ij} \, V_{\circ}^i \, ( V_\circ^j \pm \hat{V}_\perp^j ) \, dt 
			\pm \frac{1}{2} \, g_{ij} \, d[X^i,V_\circ^j]
		\right)
	\right] \, \nonumber\\
	&=
	\E\left[ \int 
		\left( 
		g_{ij} \, V_\circ^i \, \hat{V}_{\pm}^j \, dt 
		\pm g_{ij} \, V_{\circ}^i  \hat{V}_\perp^j  \, dt 
		\pm g_{ij} \, d[X^i,V_\circ^j]
		\right)
	\right] ,
\end{align}
where we repeated the calculation of eq.~\eqref{eq:Itoder} in the second line. Now, we must calculate the second term involving the perpendicular velocity. Repeating the calculation of eq.~\eqref{eq:Itoder}, this term can be written as
\begin{equation}\label{eq:Itoderb}
	\E\left[ \int g_{ij} \, V_{\circ}^i \, \hat{V}_\perp^j  \, dt \right]
	=
	\E\left[ \int 
	\left( 
	g_{ij} \hat{V}_{\pm}^i \, \hat{V}_\perp^j \, dt 
	\pm \frac{1}{2} \, g_{ij} \, d[X^i,\hat{V}_\perp^j]
	\right)
	\right] .
\end{equation}
When plugged into eq.~\eqref{eq:Itodera}, this immediately yields our earlier result \eqref{eq:ItoderEarly}. 
However, this expression can also be evaluated in a different way, since, unlike the other velocities, the perpendicular velocity represents an acceleration. This can be seen by writing the expressions of the velocity in differential notation:
\begin{align}
	V_+ \, dt &= X(t+dt) - X(t) \, , \nonumber\\
	V_- \, dt &= X(t) - X(t-dt) \, , \nonumber\\
	V_\circ \, dt &= \frac{1}{2} \, \Big[ X(t+dt) - X(t-dt) \Big] , \nonumber\\
	V_\perp \, dt &= \frac{1}{2} \, \Big[ X(t+dt) - 2 \, X(t) + X(t-dt) \Big] .
\end{align}
Therefore, $d[X,V_\perp]$ can be reduced to the quadratic variation $d[X,X]$ by improving It\^o's lemma by one order in the Taylor expansion. This yields
\begin{align}
	\E\left[ \int g_{ij} \, V_{\circ}^i \, \hat{V}_\perp^j  \, dt \right]
	&=
	\E\left[ \int 
		\left( 
			g_{ij} \, \hat{V}_\perp^j \, d_\circ X^i 
		\right)
	\right]
	\nonumber\\
	&=
	\E\left[ \int 
		\left( 
			\delta_{ab} \, e^a_i \, V_\perp^b \, d_\circ X^i 
		\right)
	\right]
	\nonumber\\
	&=
	\E\left[ \int 
		\left( 
			\delta_{ab} \, e^a_i \, V_\perp^b \, d_\pm X^i 
			\pm \frac{1}{2} \frac{\p}{\p x^k} \Big(\delta_{ab} \, e^a_i \Big) \, V_\perp^b \, d[X^i,X^k]
	\right. \right.	\nonumber\\
	&\qquad \qquad
			\pm \frac{1}{6} \frac{\p^2}{\p x^l \p x^k} \Big(\delta_{ab} \, e^a_i \Big) \, e^b_j \, d[X^j,X^l] \, d[X^i,X^k]
	\nonumber\\
	&\qquad \qquad \left. \left.
			\mp \frac{1}{6} \, \delta_{ab} \, e^a_i \, \frac{\p^2}{\p x^l \p x^k} \Big(e^b_j \Big) \, d[X^j,X^l] \, d[X^i,X^k]
		\right)
	\right]
	\nonumber\\
	&=
	\E\left[ \int 
		\left( 
			g_{ij} \hat{V}_\perp^j \, d_\pm X^i 
			\pm \frac{1}{2} \, g_{ij} \hat{V}_\perp^j \, \Gamma^i_{kl} \, d[X^k,X^l]
	\right. \right.	\nonumber\\
	&\qquad \qquad
			\pm \frac{1}{6} \, g_{mj} \, \big(\p_l \Gamma^m_{ki} + \Gamma^m_{ln} \Gamma^n_{ki} \big) \, d[X^j,X^l] \, d[X^i,X^k]
	\nonumber\\
	&\qquad \qquad \left. \left.
			\mp \frac{1}{6} \, g_{mi} \, \big(\p_l \Gamma^m_{kj} + \Gamma^m_{ln} \Gamma^n_{kj} \big) \, d[X^j,X^l] \, d[X^i,X^k]
	\right)
	\right]
	\nonumber\\
	&=
	\E\left[ \int 
		\left( 
			g_{ij} \hat{V}_\perp^j \, d_\pm\hat{X}^i
			\pm \frac{1}{6} \, g_{im} \, \mathcal{R}^m_{\;\; j k l} \, d[X^i,X^k] \, d[X^j,X^l]
		\right)
	\right] 
	\nonumber\\
	&=
	\E\left[ \int 
		\left( 
			g_{ij} \hat{V}_\pm^i \hat{V}_\perp^j 
			\pm \frac{1}{6} \, \mathcal{R}_{i j k l} \, V_2^{ik} \, V_2^{jl}
		\right) dt
	\right] .
\end{align}
This result can be plugged into eq.~\eqref{eq:Itodera}, which yields the result for the quadratic term
\begin{align}
	\E\left[ \int g_{ij}  \, V^i_{\circ} V^j_{\circ} \, dt \right]
	&=
	\E\left[ \int 
	\left( 
	g_{ij} \, \hat{V}_{\pm}^i \, \hat{V}_\pm^j \, dt 
	+ \frac{1}{6} \, \mathcal{R}_{ijkl} V_2^{ik} V_2^{jl} \, dt
	\pm g_{ij} \, d[X^i,V_\circ^j]
	\right)
	\right] .
\end{align}
Moreover, by comparing to eqs.~\eqref{eq:ItoderEarly} and \eqref{eq:Itoderb}, we find
\begin{align}
	d[X^i,\hat{V}_\pm^j] 
	&= d[X^i,V_\circ^j] \pm d[X^i,\hat{V}_\perp^j] \nonumber\\
	&= d[X^i,V_\circ^j] \pm \frac{1}{3} \, \mathcal{R}_{ijkl} \, V_2^{jl} \, d[X^i,X^k] \, .
\end{align}
\par 

Since all terms appearing in the Stratonovich Lagrangian have been rewritten into an It\^o formulation, we can finally impose eq.~\eqref{eq:itoDerImpose}.
This yields
\begin{equation}
	L^\pm(x,v_\pm,v_2,t) = L^\pm_0(x,v_\pm,v_2,t) \pm \, L_\infty (x,v_\circ) \, ,
\end{equation}
where the finite part is given by
\begin{align}
	L^\pm_0(x,v_\pm,v_2,t) 
	&= 
	\frac{m}{2} \, g_{ij} \left( v_\pm^i \pm \frac{1}{2} \Gamma^i_{kl} v_2^{kl} \right)  \left( v_\pm^j \pm \frac{1}{2} \Gamma^j_{kl} v_2^{kl} \right)
	+ \frac{m}{12} \, \mathcal{R}_{ijkl} v_2^{ik} v_2^{jl} 
	\nonumber\\
	&\quad
	+ q \, A_i v_\pm^i \pm \frac{q}{2} \, v_2^{ij} \p_j A_i 
	- \mathfrak{U} \, ,
\end{align}
which can be rewritten into an explicitly covariant form as
\begin{align}
	L^\pm_0(x,\hat{v}_\pm,v_2,t) 
	&= 
	\frac{m}{2} \, g_{ij} \hat{v}_\pm^i \hat{v}_\pm^j 
	+ \frac{m}{12} \, \mathcal{R}_{ijkl} v_2^{ik} v_2^{jl} 
	+ q \, A_i \hat{v}_\pm^i \pm \frac{q}{2} \, v_2^{ij}  \nabla_j A_i 
	- \mathfrak{U} \, .
\end{align}
Furthermore, the divergent part $L_\infty$ is defined by the integral condition
\begin{equation}
	\E\left[\int L_\infty (x,v_{\circ}) \, dt\right] 
	=
	\E\left[ \int \frac{m}{2} \, g_{ij} \, d[x^i,v_\circ^j] \right] .
\end{equation}

\clearpage
\section{Stochastic Variational Calculus}\label{ap:VariationalCalculus}

\subsection{Stratonovich-Euler-Lagrange Equations}\label{Ap:StratELEqs}
We consider a set $\mathcal{T}=[t_0,t_f]$, a pseudo-Riemannian manifold $\M$, a Stratonovich tangent bundle $T_\circ\M$, a Stratonovich Lagrangian $L^\circ(x,v_\circ,t):T_\circ\M\times \mathcal{T} \rightarrow \R$ and the action
\begin{equation}
	S(X) = \E\left[\int_{t_0}^{t_f} L^\circ(X_t,V_{\circ,t},t) \, dt \right],
\end{equation}
where $X$ is a continuous semi-martingale process on $\M$ with a Doob-Meyer decomposition of the form
\begin{equation}
	X^i_t = C^i_t + e^i_a(X_t) \, M^a.
\end{equation}
We can vary this action with respect to another process $\delta X$ that is stochastically independent, i.e. $\delta M\indep M$, and satisfies the boundary conditions $\delta X_{t_0}=\delta X_{t_f} =0$. We find
\begin{align}
	\delta S(X) 
	&= S(X+\delta X) - S(X) 
	\nonumber\\
	&= 
	\E\left[ \int_{t_0}^{t_f} \Big( 
		L^\circ(X_t+\delta X_t,V_{\circ,t} + \delta V_{\circ,t},t) 
		- L^\circ(X_t,V_{\circ,t},t) 
	\Big) \, dt \right] 
	\nonumber\\
	&= 
	\E\left[ \int_{t_0}^{t_f} \left(
		\frac{\p L^\circ}{\p x^i} \, \delta X^i_t 
		+ \frac{\p L^\circ}{\p v_\circ^i} \, \delta V^i_{\circ,t} 
		+ \mathcal{O}(\delta X_t^2)
	\right) dt \right] 
	\nonumber\\
	&= 
	\E\left[ \int_{t_0}^{t_f} \left( 
		\frac{\p L^\circ}{\p x^i} \, \delta X^i_t \, dt
		+ \frac{\p L^\circ}{\p v_\circ^i}\, d_\circ \delta X^i_t
	\right) \right]
	+ \mathcal{O}||\delta X||^2 
	\nonumber\\
	&= 
	\E\left[ 
	\frac{\p L^\circ}{\p v_\circ^i} \, \delta X^i_{t} \Big|_{t_0}^{t_f} 
	+ \int_{t_0}^{t_f} \delta X^i_t \left(
	\frac{\p L^\circ}{\p x^i} \, dt 
	- d_\circ \frac{\p L^\circ}{\p v_\circ^i}
	\right) \right]
	+ \mathcal{O}||\delta X||^2 \nonumber\\
	&= 
	\E\left[ \int_{t_0}^{t_f} \delta X^i_t \left(
		\frac{\p L^\circ}{\p x^i} \, dt 
		- d_\circ \frac{\p L^\circ}{\p v_\circ^i} 
	\right) \right]
	+ \mathcal{O}||\delta X||^2\, ,
\end{align}
where we used the Stratonovich integration by parts formula in the fifth line and ${\delta X_{t_0}=\delta X_{t_f} =0}$ in the sixth line.
\par 

By taking the limit $||\delta X||\rightarrow 0$, we find
\begin{equation}
	\frac{\delta S(X)}{\delta{X}^i} 
	= 
	\E\left[ \int_{t_0}^{t_f} \left(
		\frac{\p L^\circ}{\p x^i} \, dt
		- d_\circ \frac{\p L^\circ}{\p v_\circ^i} 
	\right) \right] .
\end{equation}
If we impose $\frac{\delta S}{\delta X} = 0$, this yields the Stratonovich-Euler-Lagrange equations
\begin{equation}
	\E\left[ \int_{t_0}^{t_f} d_\circ \frac{\p}{\p v_\circ^i} L^\circ(X_t,V_{\circ,t},t) \right]
	=
	\E\left[ \int_{t_0}^{t_f} \frac{\p}{\p x^i} L^\circ(X_t,V_{\circ,t},t) \, dt \right] ,
\end{equation}
which can be written in a differential notation as
\begin{equation}
	d_\circ \frac{\p L^\circ}{\p v_\circ^i}
	=
	\frac{\p L^\circ}{\p x^i} \, dt \, .
\end{equation}

\subsection{It\^o-Euler-Lagrange Equations}\label{Ap:ItoELEqs}
We consider a set $\mathcal{T}=[t_0,t_f]$, a pseudo-Riemannian manifold $\M$, It\^o tangent bundles $T_\pm\M$, It\^o Lagrangians $L^\pm(x,v_\pm,v_2,t):T_\pm\M\times \mathcal{T} \rightarrow \R$ and the action
\begin{equation}
	S(X) = \E\left[\int_{t_0}^{t_f} L^\pm(X_t,V_{\pm,t},V_{2,t},t) \, dt \right].
\end{equation}
where $X$ is a continuous semi-martingale process on $\M$ with a Doob-Meyer decomposition of the form
\begin{equation}\label{eq:ApDoobMeyer}
	X^i_t = C^i_t + e^i_a(X_t) \, M_t^a.
\end{equation}
We can vary this action with respect to another process $\delta X$ that is stochastically independent, i.e. $\delta M\indep M$, and satisfies the boundary conditions $\delta X_{t_0}=\delta X_{t_f} =0$. We find
\begin{align}
	\delta S(X) 
	&= S(X+\delta X) - S(X) 
	\nonumber\\
	&= 
	\E\left[ \int_{t_0}^{t_f} \Big( 
	L^\pm(X_t+\delta X_t,V_{\pm,t} + \delta V_{\pm,t}, V_{2,t}+\delta V_{2,t},t) 
	- L^\pm(X_t,V_{\pm,t},V_{2,t},t) 
	\Big) \, dt \right] 
	\nonumber\\
	&= 
	\E\left[ \int_{t_0}^{t_f} \left(
		\frac{\p L^\pm}{\p x^i} \, \delta X^i_t 
		+ \frac{\p L^\pm}{\p v_\pm^i} \, \delta V^i_{\pm,t} 
		+ \frac{\p L^\pm}{\p v_2^{ij}} \, \delta V^{ij}_{2,t} 
		+ \mathcal{O}(\delta X_t^2)
	\right) dt \right] 
	\nonumber\\
	&= 
	\E\left[ \int_{t_0}^{t_f} \left( 
		\frac{\p L^\pm}{\p x^i} \, \delta X^i_t \, dt
		+ \frac{\p L^\pm}{\p v_\pm^i} \, d_\pm \delta X^i_t
		+ \frac{\p L^\pm}{\p v_2^{ij}} \, d[\delta X^i, X^j]_t 
		+ \frac{\p L^\pm}{\p v_2^{ij}} \, d[X^i, \delta X^j]_t
	\right) \right]
	\nonumber\\
	&\quad
	+ \mathcal{O}||\delta X||^2 
	\nonumber\\
	&=
	\E\left[ 
	\frac{\p L^\pm}{\p v_\pm^i} \, \delta X^i_{t} \Big|_{t_0}^{t_f}
	+ \int_{t_0}^{t_f} \delta X^i_t \left(
	\frac{\p L^\pm}{\p x^i} \, dt 
	- d_\pm \frac{\p L^\pm}{\p v_\pm^i} 
	\right)
	\mp d\left[ \frac{\p L^\pm}{\p v_\pm^i} , \delta X^i\right]_t
	\right]
	\nonumber\\
	&\quad 
	+ \E\left[ \int_{t_0}^{t_f} \left( \frac{\p L^\pm}{\p v_2^{ij}} + \frac{\p L^\pm}{\p v_2^{ji}}  \right)  d[X^j, \delta X^i]_t
	\right]
	+ \mathcal{O}||\delta X||^2 \nonumber\\
	&= 
	\E\left[ \int_{t_0}^{t_f} \left\{ \delta X^i_t \left(
		\frac{\p L^\pm}{\p x^i} \, dt 
		- d_\pm \frac{\p L^\pm}{\p v_\pm^i}
	\right) 
	+ \left( \frac{\p L^\pm}{\p v_2^{ij}} + \frac{\p L^\pm}{\p v_2^{ji}}  \right) d[X^j, \delta X^i]_t
	\right. \right. \nonumber\\
	&\qquad \left. \left.
	\mp \frac{\p^2 L^\pm}{\p x^j \p v_\pm^i} \, d\left[ X^j , \delta X^i \right]_t
	\mp \frac{\p^2 L^\pm}{\p v_\pm^j \p v_\pm^i} \, d\left[ V_\pm^j , \delta X^i\right]_t
	\mp \frac{\p^2 L^\pm}{\p v_\pm^{jk} \p v_\pm^i} \, d\left[ V_2^{jk} , \delta X^i\right]_t
	\right\}
	\right]
	\nonumber\\
	&\quad
	+ \mathcal{O}||\delta X||^2\, ,
\end{align}
where we used the It\^o integraton by parts formula in the fifth line and ${\delta X_{t_0}=\delta X_{t_f} =0}$ in the sixth line.
\par 

Since we have not yet factorized $\delta X^i$, we must further evaluate the quadratic variation containing $\delta X^i$. We find
\begin{align}
	d[X^j,\delta X^i] 
	&= e^j_a \,  \delta e^i_b \, d[M^a, M^b] + e^j_a \, e^i_b \, d[M^a, \delta M^b] + \mathcal{O}(\delta X^2)  + o(dt) \nonumber\\
	&= - e^j_a \, \Gamma^i_{kl} \, e^l_b \, \delta X^k \, d[M^a, M^b] + \mathcal{O}(\delta X^2)  + o(dt)
	\nonumber\\
	&= - \Gamma^i_{kl} \, \delta X^k \, d[X^j, X^l] + \mathcal{O}(\delta X^2)  + o(dt)  \, ,
\end{align}
where we used the Doob-Meyer decomposition \eqref{eq:ApDoobMeyer} in the first and last line, and the stochastic independence $\delta M\indep M$, which implies $d[M^a,\delta M^b]=0$, in the second line.
By a similar calculation we find
\begin{align}
	d[V_\pm^j,\delta X^i] 
	&= - \Gamma^i_{kl} \, \delta X^k  \, d[V_\pm^j,X^l] + \mathcal{O}(\delta X^2)  + o(dt) \, , \\
	d[V_2^{jk},\delta X^i] 
	&= - \Gamma^i_{lm} \, \delta X^m \, d[V_2^{jk},X^l] + \mathcal{O}(\delta X^2)  + o(dt) \, .
\end{align}
These results can be plugged into the variation of the action. Then, in the limit ${||\delta X||\rightarrow 0}$, we obtain
\begin{align}
	\frac{\delta S(X)}{\delta{X}^i} 
	&= 
	\E\left[ 
		\int_{t_0}^{t_f} \left(
			\frac{\p L^\pm}{\p x^i} \, dt
			- d_\pm \frac{\p L^\pm}{\p v_\pm^i}
			- \Gamma^k_{ij} \left( \frac{\p L^\pm}{\p v_2^{kl}} + \frac{\p L^\pm}{\p v_2^{lk}} \right) d[X^j,X^l]_t
		\right. \right. \nonumber\\
		&\qquad
			\pm \Gamma^k_{ij} \, \frac{\p^2 L^\pm}{\p x^l \p v_\pm^k} \, d[X^j,X^l]_t 
			\pm \Gamma^k_{ij} \, \frac{\p^2 L^\pm}{\p v_\pm^l \p v_\pm^k} \, d[X^j,V_\pm^l]_t
		\nonumber\\
		&\qquad \left. \left.		
			\pm \Gamma^k_{ij} \, \frac{\p^2 L^\pm}{\p v_2^{lm} \p v_\pm^k} \, d[X^j,V_2^{lm}]_t
		\right) 
	\right] .
\end{align}
If we impose $\frac{\delta S}{\delta X} = 0$, this yields the It\^o-Euler-Lagrange equations.
These can be written in differential notation as
\begin{align}
	d_\pm \frac{\p L^\pm}{\p v_\pm^i} 
	&=
	\frac{\p L^\pm}{\p x^i} \, dt 
	- \Gamma^k_{ij} \left( \frac{\p L^\pm}{\p v_2^{kl}} + \frac{\p L^\pm}{\p v_2^{lk}} \right) d[X^j,X^l]_t
	\pm \Gamma^k_{ij} \, \frac{\p^2 L^\pm}{\p x^l \p v_\pm^k} \, d[X^j,X^l]_t 
	\nonumber\\
	&\quad
	\pm \Gamma^k_{ij} \, \frac{\p^2 L^\pm}{\p v_\pm^l \p v_\pm^k} \, d[X^j,V_\pm^l]_t
	\pm \Gamma^k_{ij} \, \frac{\p^2 L^\pm}{\p v_2^{lm} \p v_\pm^k} \, d[X^j,V_2^{lm}]_t \, .
\end{align}

\subsection{Hamilton-Jacobi-Bellman Equations}\label{ap:HamJac}
We consider a set $\mathcal{T}=[t_0,t_f]$, a pseudo-Riemannian manifold $\M$, It\^o tangent bundles $T_\pm\M$ and It\^o Lagrangians $L^\pm(x,v_\pm,v_2,\varepsilon,t):T_\pm\M \times \R \times \mathcal{T} \rightarrow \R$, where $\varepsilon \in \R$ is a Lagrange multiplier. For $L^+$, we construct the principal function
\begin{align}\label{eq:PFnctFwrdA}
	S^+(x,\varepsilon,t) 
	&= S^+(x,\varepsilon,t;x_f,\varepsilon_f,t_f) \nonumber\\
	&= -\E\left[\int_{t}^{t_f} L^+(X_s,V_{+,s},V_{2,s},\mathcal{E}_s,s)\, ds \Big| X_t = x, X_{t_f}=x_f, \mathcal{E}_t = \varepsilon, \mathcal{E}_{t_f} = \varepsilon_f \right] ,
\end{align}
where $(X_s,V_s,\mathcal{E}_s)$ is a solution of the It\^o-Euler-Lagrange equations passing through $(x,\varepsilon,t)$ and $(x_f,\varepsilon_f,t_f)$. For $L^-$, we construct the principal function
\begin{align}\label{eq:PFnctBwrdA}
	S^-(x,\varepsilon,t) 
	&= S^-(x,\varepsilon,t;x_0,\varepsilon_0,t_0) \nonumber\\
	&= \E\left[\int_{t_0}^{t} L^-(X_s,V_{-,s},V_{2,s},\mathcal{E}_s,s)\, ds \Big| X_t = x, X_{t_0}=x_0, \mathcal{E}_t = \varepsilon, \mathcal{E}_{t_0} = \varepsilon_0 \right] ,
\end{align}
where $(X_s,V_s,\mathcal{E}_s)$ is a solution of the It\^o-Euler-Lagrange equations passing through $(x_0,\varepsilon_0,t_0)$ and $(x,\varepsilon,t)$.
\par

We can vary these principal functions with respect to their end point $(x,\varepsilon,t)$. For $S^+$, we obtain
\begin{align}
	\delta S^+(x,\varepsilon,t) 
	&= 
	S^+(x+\delta x,\varepsilon, t) - S^+(x,\varepsilon, t)\nonumber\\
	&= 
	- \E\left\{ 
		\E\left[ \int_{t}^{t_f} L^+ \, ds \, \Big| \, X_t= x + \delta x \right]
		- \E\left[ \int_{t}^{t_f} L^+ \, ds \, \Big| \, X_t= x  \right]
	\Big| X_{t_f} = x_f
	\right\}
	\nonumber\\
	&= 
	- \E\left[ 
		\int_{t}^{t_f} \left( L_{X+\delta X}^+ - L^+_X \right) ds \, 
	\Big| \,  
		X_t = x, X_{t_f} = x_f, \delta X_t = \delta x, \delta X_{t_f} = 0
	\right]
	\nonumber\\
	&= 
	- \E\left[ 
		\int_{t}^{t_f}  \left(
			\frac{\p L^+ }{\p x^i}  \, \delta X^i_s 
			+ \frac{\p L^+}{\p v_+^i}  \, \delta V^i_{+,s}
		\right. \right. \nonumber\\
		&\qquad \qquad \qquad \left. \left. 
			+ \frac{\p L^+}{\p v_2^{ij}}  \, \delta V^{ij}_{2,s}
			+ \mathcal{O}(\delta X_s^2)
		\right) ds \, 
	\Big| \, 
		X_t, X_{t_f}, \delta X_{t_f}, \delta X_t
	\right]
	\nonumber\\
	&=
	- \E\left[ 
		\int_{t}^{t_f}  \left(
			 \delta X^i_s \, d_+ \frac{\p L^+}{\p v_+^i}
			+ \frac{\p L^+}{\p v_+^i}  \, d_+ \delta X^i_{s}
		\right. \right. \nonumber\\
		&\qquad \qquad \qquad \left. \left. 		
			+ d\left[\frac{\p L^+}{\p v_+^i} , \delta X^i \right]_s
		\right) \, 
	\Big| \, 
		X_t, X_{t_f}, \delta X_{t_f}, \delta X_t
	\right]
	+ \mathcal{O}||\delta x||^2
	\nonumber\\
	&=
	- \E\left[ 
		\int_{t}^{t_f}  d_+ \left(\frac{\p L^+}{\p v_+^i} \, \delta X_s^i \right)
	\Big| \, 
		X_t, X_{t_f}, \delta X_{t_f}, \delta X_t
	\right]
	+ \mathcal{O}||\delta x||^2
	\nonumber\\
	&=
	- \E\left[ 
		\frac{\p L^+}{\p v_+^i} \, \delta X_{s}^i \, \Big|_{t}^{t_f} \,
	\Big| \, 
		X_{t} = x , \delta X_{t} = \delta x, X_{t_f} = x_f , \delta X_{t_f} = 0
	\right]
	+ \mathcal{O}||\delta x||^2
	\nonumber\\
	&=
	\E\left[ 
	\frac{\p}{\p v_+^i} L^+(X_{t},V_{+,t},V_{2,t},\mathcal{E}_t,t) \,
	\Big| \, 
	X_{t} = x, \mathcal{E}_t = \varepsilon
	\right]
	\delta x^i
	+ \mathcal{O}||\delta x||^2\,,
\end{align}
where we suppressed the dependence on the initial and final condition of the Lagrange multiplier $\mathcal{E}_t$ in the first seven lines.
Moreover, we used the It\^o-Euler-Lagrange equation in the fifth line and the It\^o integration by parts formula in the sixth line.
\par

By a similar calculation, we find for $S^-$
\begin{equation}
	\delta S^-(x,\varepsilon,t) 
	=
	\E\left[  \frac{\p}{\p v_-^i} L^-(X_t,V_{-,t},V_{2,t},\mathcal{E}_t,t) \, \Big| \, X_t=x, \mathcal{E}_t = \varepsilon \right] \delta x^i 
	+ \mathcal{O}||\delta x||^2\, .
\end{equation}
After taking the limit $||\delta x||\rightarrow 0$, we obtain the first Hamilton-Jacobi equation
\begin{equation}
	\frac{\p}{\p x^i} S^\pm (X_t,\mathcal{E}_t,t) = p^\pm_i(X_t,\mathcal{E}_t,t)
\end{equation}
with 
\begin{equation}
	p^\pm_i(X_t,\mathcal{E}_t,t) = \E\left[  \frac{\p L^\pm }{\p v_\pm^i} \, \Big| \, X_t, \mathcal{E}_t\right] .
\end{equation}
\par

The second Hamilton-Jacobi equation can be obtained by applying It\^o's lemma to $S^\pm$, which yields
\begin{align}
	d_\pm S^\pm(X_t,\mathcal{E}_t, t) 
	&= 
	\frac{\p S^\pm}{\p t}  \, dt 
	+ \frac{\p S^\pm}{\p \varepsilon}  \, d \mathcal{E}_t
	+ \frac{\p S^\pm}{\p x^i} \, d_\pm X^i_t 
	\pm \frac{1}{2} \, \frac{\p^2 S^\pm}{\p x^j \p x^i}  \, d[X^i,X^j]_t \nonumber\\
	&=
	\frac{\p S^\pm}{\p t}  \, dt 
	+ \frac{\p S^\pm}{\p \varepsilon} \, \frac{d \mathcal{E}_t}{dt} \, dt
	+ p^\pm_i \, d_\pm X^i_t 
	\pm \frac{1}{2} \, \frac{\p \, p^\pm_i}{\p x^j}  \, d[X^i,X^j]_t \, ,
\end{align}
where we used that $\mathcal{E}_t$ is a deterministic process and the first Hamilton-Jacobi equation in the second line. By taking a conditional expectation of this expression, we find
\begin{align}
	\E \left[ d_\pm S^\pm (X_t,\mathcal{E}_t, t) \Big| X_t, \mathcal{E}_t\right]
	&= 
	\left[ \frac{\p S^\pm}{\p t}
	+ \frac{\p S^\pm}{\p \varepsilon}  \, \frac{d \mathcal{E}_t}{dt}
	+ p^\pm_i v_\pm^i
	\pm \frac{1}{2} \, v_2^{ij} \, \frac{\p \, p^\pm_i}{\p x^j} \right] dt \, .\label{eq:ApHJ21}
\end{align}
Moreover, using the definition of Hamilton's principal function, given in eqs.~\eqref{eq:PFnctFwrdA} and \eqref{eq:PFnctBwrdA}, we obtain 
\begin{align}\label{eq:ApHJ22}
	\E \left[ d_\pm S^\pm \Big| X_t, \mathcal{E}_t \right] 
	&= 
	L^\pm\left[X_t,v_{\pm}(X_t,\mathcal{E}_t,t),v_{2}(X_t,\mathcal{E}_t,t),\mathcal{E}_t,t\right] \, dt
	\nonumber\\
	&=
	\Big\{ L_0^\pm\left[X_t,v_{\pm},v_{2},\mathcal{E}_t,t\right]  
	 \pm L_\infty\left[X_t,v_{\circ},\mathcal{E}_t\right] \Big\} \, dt\, .
\end{align}
Then, comparing eqs.~\eqref{eq:ApHJ21} and \eqref{eq:ApHJ22} at finite order yields
\begin{equation}
	\frac{\p}{\p t} S(X_t,\mathcal{E}_t,t) + \frac{d \mathcal{E}_t}{dt} \, \frac{\p}{\p \varepsilon} S(X_t,\mathcal{E}_t,t)
	= - H_0^\pm\left[X_t, p^\pm(X_t,\mathcal{E}_t,t),\p p^\pm(X_t,\mathcal{E}_t,t),\mathcal{E}_t,t\right]
\end{equation}
with the Hamiltonian given by
\begin{align}
	H_0^\pm(x, p^\pm,\p p^\pm,\varepsilon,t) 
	&= p_i^\pm \, v^i_\pm 
	\pm \frac{1}{2} v^{ij}_2 \, \p_j p_i^\pm 
	- L_0^\pm(x,v_\pm,v_2,\varepsilon,t) \nonumber\\
	&= p_i^\pm \, \hat{v}^i_\pm 
	\pm \frac{1}{2} v^{ij}_2 \, \nabla_j p_i^\pm 
	- L_0^\pm(x,v_\pm,v_2,\varepsilon,t) \, .
\end{align}
Furthermore, comparing the divergent parts of eqs.~\eqref{eq:ApHJ21} and \eqref{eq:ApHJ22} yields an integral constraint
\begin{equation}
	\E\left[ \oint L_\infty(X_s,V_{\circ,s},\mathcal{E}_s) \, ds \, \Big| \, X_t, \mathcal{E}_t \right] 
	=
	\pm \oint \left(  p^\pm_i \, \hat{v}_\pm^i \pm \frac{1}{2} \, v_2^{ij} \, \nabla_j p^\pm_i \right) dt \, .
\end{equation}

\vspace{3.0cm}
\fbox{%
	\parbox{\textwidth}{ 
		This is a preprint of the following work:\\
		Folkert Kuipers, ``Stochastic Mechanics: the Unification of Quantum Mechanics with Brownian Motion'', SpringerBriefs in Physics, Springer (2023),\\[0.1cm]
		reproduced with permission of Springer Nature Switzerland AG. \\[0.2cm]
		The final authenticated version is available online at:\\
		\href{https://link.springer.com/book/10.1007/978-3-031-31448-3}{http://dx.doi.org/10.1007/978-3-031-31448-3}
	}
}
\end{document}